%% file: article.tex
\documentclass{amsart}
\usepackage[margin=1in]{geometry}
\usepackage{amsmath, amsfonts, amssymb}
\usepackage{color}
\usepackage{algpseudocode}
\usepackage{algorithm}
\usepackage{siunitx}
\usepackage{graphicx}
\usepackage{subfig}
\graphicspath{{images/}}

\newcommand\bx{\boldsymbol{x}}
\newcommand\bv{\boldsymbol{v}}

\newcommand\bxi{\boldsymbol{\xi}}
\newcommand\bw{\boldsymbol{w}}
\newcommand\bc{\boldsymbol{c}}
\newcommand\bn{\boldsymbol{n}}
\newcommand\bcH{\boldsymbol{\mathcal{H}}}
\newcommand\bomega{\boldsymbol{\omega}}

\newcommand\He{\mathit{He}}

\theoremstyle{remark}

\title{Moment method as a numerical solver: Challenge from shock structure problems}

\author{Zhenning Cai}
\address[Zhenning Cai]{Department of Mathematics, National University of Singapore,
  Level 4, Block S17, 10 Lower Kent Ridge Road, Singapore 119076}
\email{matcz@nus.edu.sg}

\thanks{Zhenning Cai was supported by the Academic Research Fund of the
Ministry of Education of Singapore under grant Nos. R-146-000-305-114 and
R-146-000-326-112.}


\begin{document}
\maketitle

\begin{abstract}
  We survey a number of moment hierarchies and test their performances in
  computing one-dimensional shock structures. It is found that for high Mach
  numbers, the moment hierarchies are either computationally expensive or hard
  to converge, making these methods questionable for the simulation of
  highly nonequilibrium flows. By examining the convergence issue of Grad's
  moment methods, we propose a new moment hierarchy to bridge the hydrodynamic
  models and the kinetic equation, allowing nonlinear moment methods to be used
  as a numerical tool to discretize the velocity space for high-speed flows.
  For the case of one-dimensional velocity, the method is formulated
  for odd number of moments, and it can be extended seamlessly to the
  three-dimensional case. Numerical tests show that the method is capable of
  predicting shock structures with high Mach numbers accurately, and the
  results converge to the solution of the Boltzmann equation as the number of
  moments increases. Some applications beyond the shock structure problem are
  also considered, indicating that the proposed method is suitable for
  computation of transitional flows.
\end{abstract}

\input{article_introduction.tex}
\input{article_review.tex}
\input{article_num_test.tex}
\input{article_hmb.tex}
\input{article_application.tex}
\input{article_conclusion.tex}
\input{article_appendix.tex}

\bibliographystyle{amsplain}
\bibliography{../article}
\end{document}

%% file: article_introduction.tex
\section{Introduction}
Rarefied gas dynamics is an important branch of fluid mechanics with wide
applications in astronautics and microelectromechanical systems. The Boltzmann
equation, devised nearly 150 years ago in \cite{Boltzmann1872}, is one of the
most basic models for rarefied gases. Due to its high dimensionality and
complicated collision term, the numerical simulation of this equation is highly
challenging. Despite the fast development of the computational infrastructure,
solving the Boltzmann equation in the three-dimensional case is still rather
resource-demanding \cite{Dimarco2018}, mainly due to its complicated collision
term. While researchers are still improving the numerical solver \cite{Wu2013,
Hu2016, Liu2016, Gamba2017, Kitzler2019}, an alternative is to consider model
reduction to reduce the numerical difficulty. One important technique for
kinetic model reduction is the moment method first introduced by Grad
\cite{Grad1949}. This methodology is the main topic to be studied in this
paper.

The basic idea of the moment method is to take all the moments of the Boltzmann
equation, and then consider only a subset of these equations. To make the
subsystem self-contained, moment closure needs to be applied. By including
increasing numbers of moments to the systems, a sequence of models can be
established, which bridges the hydrodynamic models (such as the Euler equations
and the Navier-Stokes equations) and the Boltzmann equation. Due to the nested
structure of these models, we call such a model sequence a \emph{moment
hierarchy}. Different moment closure can result in different moment
hierarchies, while all of them are expected to converge to the Boltzmann
equation as the number of moments tends to infinity. With such convergence, the
moment methods can also be regarded as the discretization of the velocity space
in the Boltzmann equation, as have been studied recently in multiple works
\cite{Hu2020, Sarna2020, Fan2020}.

While such convergence holds formally according to the derivation of the moment
hierarchies, numerical experiments sometimes suggest otherwise. The convergence
of Grad's moment method has been demonstrated in \cite{Au2001} for a shock tube
problem. In the linear regime, the convergence has been shown both numerically
and theoretically for boundary value problems \cite{Torrilhon2015, Sarna2018}.
Despite these positive results, some pessimistic observations have been
reported in other references, especially when the moment method is nonlinear
and the number of moments in the system is large. In \cite{Cai2012}, it is
found that Grad's moment method fails to work for the shock tube problem with a
large density ratio due to loss of hyperbolicity \cite{Muller1998}. Even if the
hyperbolicity is fixed by the framework proposed in \cite{Cai2015a}, the method
still fails for the Fourier flow problem due to the convergence issue raised in
\cite{Cai2020}. Another moment hierarchy, called regularized moment equations
\cite{Struchtrup2005}, is unable to produce shock structure with large Mach
numbers when the number of moments is large \cite{Cai2012}. For the moment
hierarchy based on the maximum-entropy closure \cite{Levermore1996}, the
equilibrium state turns out to be a singularity \cite{Junk1998}, making the
numerical validation extremely difficult. It seems that these bridges from
hydrodynamic to kinetic turn fragile when non-equilibrium effect gets strong.
It is unpredictable when and where the nonequilibrium will cause them to
collapse. Such a property makes it difficult to use the nonlinear moment
methods as a numerical solver of the Boltzmann equation.

In this work, we will survey a number of moment hierarchies, and test their
performances using the shock structure problem, which is a frequently-used
benchmark problem involving obvious non-equilibrium effects. Observing the
difficulty in computing high-speed flows, we introduce a novel moment
hierarchy, named as ``highest-moment-based moment method'', to realize an
efficient and stable connection between the fluid models and the Boltzmann
equation, whose properties are verified by shock structure problems with both
one-dimensional and three-dimensional physics. The moment equations
with odd number of moments are formulated for one-dimensional physics, and for
the three-dimensional physics, based on spherical harmonics, the moments are
chosen to preserve the rotational invariance. Experiments are also carried out
for some applications beyond the shock structure problems.

The rest of the paper is organized as follows. The Boltzmann equation and some
existing moment hierarchies are reviewed in Section \ref{sec:review}. In
Section \ref{sec:num_test}, we pick some moment hierarchies with less
computational difficulties to test their capabilities to simulate shock
structures. Our novel moment hierarchy is introduced in Section \ref{sec:1d}
and \ref{sec:3d} for one-dimensional and three-dimensional cases, respectively.
Some numerical study beyond the shock structure problem is carried out in
Section \ref{sec:application}. Finally, we conclude the paper and present some
discussion on the proposed method in Section \ref{sec:conclusion}.

%% file: article_review.tex
\section{Review of the Boltzmann equation and some moment hierarchies}
\label{sec:review}
The Boltzmann equation gives the time evolution of the distribution function
$f(\bx,\bxi,t)$, which is related to the macroscopic quantities such as the
number density of molecules $\rho(\bx,t)$, velocity $\bv(\bx,t)$ and the
specific internal energy $e(\bx,t)$ by 
\begin{displaymath}
\rho(\bx,t) = \langle f(\bx,\bxi,t) \rangle, \qquad
\bv(\bx,t) = \frac{1}{\rho(\bx,t)} \langle \bxi f(\bx,\bxi,t) \rangle, \qquad
e(\bx,t) = \frac{1}{\rho(\bx,t)}
  \left\langle \frac{1}{2} |\bxi - \bv(\bx,t)|^2 f(\bx,\bxi,t) \right\rangle,
\end{displaymath}
where $\langle \cdot \rangle$ denotes the integral of $\cdot$ with respect to
$\bxi$. Considering the $d$-dimensional space and assuming that $\bx =
(x_1, \cdots x_{d})^T$ and $\bxi = (\xi_1, \cdots, \xi_d)^T$, we can write the
$d$-dimensional Boltzmann equation as
\begin{equation} \label{eq:Boltzmann}
\frac{\partial f}{\partial t} +
  \sum_{k=1}^d \xi_k \frac{\partial f}{\partial x_k} = \mathcal{S}[f],
\end{equation}
where the right-hand side $\mathcal{S}[f]$ is the collision term describing the
interaction between gas molecules. Below we are going to consider two types of
collision terms:
\begin{itemize}
\item Quadratic collision operator (only for $d \geqslant 2$):
  \begin{equation} \label{eq:quadratic}
  \mathcal{S}[f](\bx,\bxi,t) =
  \int_{\mathbb{R}^d} \int_{\mathbb{S}^{d-1}} B(\bxi - \bxi_*, \bomega)
    [f(\bx,\bxi',t) f(\bx,\bxi_*',t) - f(\bx,\bxi,t) f(\bx,\bxi_*,t)] \,\mathrm{d}\bomega \,\mathrm{d}\bxi_*,
  \end{equation}
  where $B(\cdot,\cdot)$ is the collision kernel, and
  \begin{displaymath}
  \bxi' = \frac{\bxi + \bxi_*}{2} + \frac{|\bxi - \bxi_*|}{2} \bomega, \qquad
  \bxi_*' = \frac{\bxi + \bxi_*}{2} - \frac{|\bxi - \bxi_*|}{2} \bomega.
  \end{displaymath}
\item Bhatnagar-Gross-Krook (BGK) operator:
  \begin{displaymath}
  \mathcal{S}[f](\bx,\bxi,t) =
    \frac{1}{\tau(\bx,t)} [\mathcal{M}(\bx,\bxi,t) - f(\bx,\bxi,t)],
  \end{displaymath}
  where $\tau(\bx,t)$ denotes the mean relaxation time, which depends usually on
  the density and the temperature, and $\mathcal{M}(\bx,\bxi,t)$ is the local
  equilibrium defined by
  \begin{equation} \label{eq:Maxwellian}
  \mathcal{M}(\bx,\bxi,t) = \frac{\rho(\bx,t)}{(2\pi \theta(\bx,t))^{d/2}}
    \exp \left( -\frac{|\bxi - \bv(\bx,t)|^2}{2\theta(\bx,t)} \right),
  \qquad \theta(\bx,t) = \frac{1}{d} e(\bx,t).
  \end{equation}
\end{itemize}
The BGK operator will be used when discussing one-dimensional model problems,
while the quadratic operator will be used in the more realistic
three-dimensional cases.

Below we are going to review some existing moment hierarchies in the
literature. For simplicity, we will present these equations only based on the
one-dimensional dynamics, for which the spatial and velocity variables will be
written as $x$ and $\xi$.

\subsection{Grad's moment methods}
Grad's work \cite{Grad1949} introduced the moment method to the kinetic theory,
which is based on the following series expansion of the distribution function:
\begin{equation} \label{eq:ansatz}
f(x,\xi,t) = \sum_{\alpha=0}^{+\infty}
  f_{\alpha}(x,t) [\theta(x,t)]^{-\alpha/2}
  \He_{\alpha} \left( \frac{\xi - v(x,t)}{\sqrt{\theta(x,t)}} \right)
  \mathcal{M}_0(x,\xi,t),
\end{equation}
where $\mathcal{M}_0(x,\xi,t) = \mathcal{M}(x,\xi,t) / \rho(x,t)$, and
$\He_{\alpha}(\cdot)$ is the Hermite polynomial defined by
\begin{displaymath}
\He_{\alpha}(\xi) = (-1)^{\alpha} \exp \left( \frac{\xi^2}{2} \right)
  \frac{\mathrm{d}^k}{\mathrm{d}\xi^k} \exp \left( -\frac{\xi^2}{2} \right),
\end{displaymath}
which are orthogonal polynomials defined on $\mathbb{R}$ with weight function
$\omega(\xi) = \exp(-\xi^2/2)$. Due to such orthogonality, each coefficient
$f_{\alpha}$ can be viewed as a ``moment'' since it can be represented by the
integral of the distribution function times a polynomial of $\xi$. Due to the
nonlinearity incorporated into \eqref{eq:ansatz} by the parameters $v(x,t)$ and
$\theta(x,t)$, the coefficients satisfies $f_1(x,t) = f_2(x,t) \equiv 0$, and
any Maxwellian can be exactly expressed by this series, which actually has only
one nonzero term with index $\alpha = 0$, so that the resulting moment system
can be considered as a natural generalization of Euler equations. To derive the
moment system, we plug \eqref{eq:ansatz} into the Boltzmann equation with BGK
collision term, and obtain
\begin{equation} \label{eq:f_alpha}
\frac{\mathrm{d} f_{\alpha}}{\mathrm{d} t} +
  \frac{\mathrm{d} v}{\mathrm{d} t} f_{\alpha-1} +
  \frac{1}{2} \frac{\mathrm{d} \theta}{\mathrm{d} t} f_{\alpha-2}
+ \theta F_{\alpha-1} + (\alpha+1) F_{\alpha+1}
  = -\frac{1-\delta_{\alpha0}}{\tau} f_{\alpha},
\qquad \forall \alpha \in \mathbb{N},
\end{equation}
where $F_{\alpha} = \partial_x f_{\alpha} + f_{\alpha-1} \partial_x v +
\frac{1}{2} f_{\alpha-2} \partial_x \theta$, and
$\dfrac{\mathrm{d}}{\mathrm{d}t} = \partial_t + v \partial_x$. In the above
equations, whenever a negative subscript appears, the quantity is regarded as
zero, e.g. $F_{-1} = 0$, $F_0 = \partial_x f_0$.

Clearly the equations \eqref{eq:f_alpha} need to be closed if we aim at a
finite system. Grad created the moment hierarchies by the simple closure
\begin{equation} \label{eq:Grad_closure}
f_{\alpha} = 0, \qquad \forall \alpha \geqslant N, 
\end{equation}
where $N \geqslant 3$ is the number of equations (or the number of moments) in
the finite system, which is usually called Grad's $N$-moment system. By
choosing different values of $N$, a moment hierarchy can be generated. In
particular, when $N = 3$, the system is identical to the Euler equations. The
Navier-Stokes equations are not directly included in the hierarchy, but it can
be derived from the moment system with $N=4$ by Chapman-Enskog expansion. In
fact, Grad's 4-moment system contains one more order than Navier-Stokes,
meaning that the Burnett equations can also be derived by Chapman-Enskog
expansion.

The three-dimensional case is more complicated. In general, the 13-moment
equations include the Navier-Stokes limit, but it includes the Burnett limit
only in some special cases such as Maxwell molecules. We refer the readers to
\cite{Struchtrup2005} for more details. Nevertheless, the moment hierarchy
generated by Grad's moment system still formally connects the Euler equations
and the Boltzmann equation.

However, when nonequilibrium is strong, the computation based on Grad's moment
equations often breaks down due to the loss of hyperbolicity of Grad's system,
which has been pointed out in a number existing works \cite{Muller1998,
Torrilhon2000, Cai2013, Cai2013g}, and its numerical effect has been observed
\cite{Torrilhon2010} by a shock tube test. Such a drawback highly restricts the
application of Grad's moment equations. It is to be shown in Section
\ref{sec:num_test} that Grad's method does not work for shock structures with a
high Mach number.

\subsection{Linearized Grad's moment equations}
Despite the loss of hyperbolicity, Grad's method is known to be linearly stable
around the global Maxwellian. Therefore when Grad's equations are linearized,
one can expect that the hyperbolicity can be restored. The linearized moment
equations are
\begin{align*}
& \frac{\partial f_0}{\partial t} + \frac{\partial v}{\partial x} = 0, \\
& \frac{\partial v}{\partial t} + \frac{\partial f_0}{\partial x}
  + \frac{\partial \theta}{\partial x} = 0, \\
& \frac{\partial \theta}{\partial t} + \frac{\partial v}{\partial x}
  + 3 \frac{\partial f_3}{\partial x} = 0, \\
& \frac{\partial f_3}{\partial t} +
  \frac{1}{2} \frac{\partial \theta}{\partial x} +
  4 \frac{\partial f_4}{\partial x} = -\frac{1}{\tau_0} f_3, \\
& \frac{\partial f_{\alpha}}{\partial t} +
  \frac{\partial f_{\alpha-1}}{\partial x} +
  (\alpha + 1) \frac{\partial f_{\alpha+1}}{\partial x}
= -\frac{1}{\tau_0} f_{\alpha}, \qquad \alpha \geqslant 4.
\end{align*}
The moment closure is again given by \eqref{eq:Grad_closure}. By increasing
$N$, we again obtain a moment hierarchy. The convergence of such systems as
$N$ approaches infinity has been studied in \cite{Torrilhon2015}, and it is
seen that the limit equation is the linearized Boltzmann equation about the
same global equilibrium state. Manifestly, such a system is applicable only
when the fluid states are close to a global equilibrium state. It cannot be
applied to shock structure computations for a large Mach number. Furthermore,
it does not include the nonlinear Euler equations, let alone the higher-order
Navier-Stokes or Burnett equations.  Therefore we do not take these systems
into account in this work.

\subsection{Grad's moment equations with linearized ansatz} \label{sec:lin_ansatz}
A related method is to linearize Grad's expansion \eqref{eq:ansatz} instead of
the moment equations:
\begin{equation} \label{eq:linear_ansatz}
f(x,\xi,t) = \sum_{\alpha=0}^{+\infty}
  f_{\alpha}(x,t) \overline{\theta}^{-\alpha/2}
  \He_{\alpha} \left( \frac{\xi - \overline{v}}{\sqrt{\overline{\theta}}} \right)
  \cdot \frac{1}{\sqrt{2\pi \overline{\theta}}}
  \exp \left( -\frac{|\xi - \overline{v}|^2}{2\overline{\theta}} \right),
\end{equation}
where $\overline{v}$ and $\overline{\theta}$ are preset constant parameters,
which do not vary with $x$ and $t$. Such linearization essentially turns Grad's
method to the Hermite spectral method, and the coefficients $f_1(x,t)$ and
$f_2(x,t)$ are no longer zero. The equations for the coefficients are
\begin{equation} \label{eq:linear_equation}
\frac{\partial f_{\alpha}}{\partial t} +
  \overline{\theta} \frac{\partial f_{\alpha-1}}{\partial x} +
  \overline{v} \frac{\partial f_{\alpha}}{\partial x} +
  (\alpha + 1) \frac{\partial f_{\alpha+1}}{\partial x} =
\frac{1}{\tau} (\mathcal{M}_{\alpha} - f_{\alpha}), \qquad \alpha \geqslant 0,
\end{equation}
where
\begin{displaymath}
\mathcal{M}_{\alpha}(x,t) = \frac{\bar{\theta}^{\alpha/2}}{\alpha!}
  \int_{\mathbb{R}} \He_{\alpha}
  \left( \frac{\xi - \overline{v}}{\sqrt{\overline{\theta}}} \right)
  \mathcal{M}(x,\xi,t) \,\mathrm{d}\xi.
\end{displaymath}
The moment closure is again given by \eqref{eq:Grad_closure}. For such a method,
the hyperbolicity can be guaranteed since all the characteristic speeds are
constant. For this moment hierarchy, the 3-moment equations do not correspond
to Euler equations; however, when $N \geq 4$, Euler equations can be derived by
Hilbert expansion. More concretely, when $\tau$ is small in
\eqref{eq:linear_equation}, the leading order term of $f_{\alpha}$ is the same
as $\mathcal{M}_{\alpha}$, yielding Euler equations. The general order of
accuracy is yet to be further studied.

This method is numerically studied in \cite{Sarna2018, Hu2019, Hu2020}, and some
computation for shock structures has been carried out in \cite{Hu2020}, which
shows good agreement with the DSMC results. However, there are two parameters
$\overline{u}$ and $\overline{\theta}$ in the ansatz \eqref{eq:linear_equation},
which have to be chosen appropriately to ensure the convergence of the sequence
\eqref{eq:linear_ansatz}. These two parameters have to be chosen manually; they
cannot be determined by the distribution function itself or physical
experiments. Different choices of these parameters will lead to completely
different models, as is undesired from the modelling point of view.

\subsection{Hyperbolic moment equations} \label{sec:hyp}
To gain hyperbolicity without changing the ansatz used by Grad \eqref{eq:ansatz},
another moment hierarchy is introduced in \cite{Cai2013g, Cai2014}. In these
works, the closure \eqref{eq:Grad_closure} is supplemented by removing the term
$F_N$ in the system, so that the resulting equations are
\begin{equation}
\frac{\mathrm{d} f_{\alpha}}{\mathrm{d} t} +
  \frac{\mathrm{d} v}{\mathrm{d} t} f_{\alpha-1} +
  \frac{1}{2} \frac{\mathrm{d} \theta}{\mathrm{d} t} f_{\alpha-2}
+ \theta F_{\alpha-1} + (1-\delta_{N-1,\alpha})(\alpha+1) F_{\alpha+1}
  = -\frac{1-\delta_{\alpha0}}{\tau} f_{\alpha},
\quad \forall \alpha = 0,1,\cdots,N-1.
\end{equation}
The resulting system is hyperbolic if and only if $\theta > 0$, and all the
characteristic speeds are determined by $v$ and $\theta$. When $N = 3$, the
above system is again the same as Euler equations. The convergence test of such
equations on benchmark microflow problems have been carried out in
\cite{Cai2013, Cai2018}. Despite the hyperbolicity, for the one-dimensional
five-moment system, it is shown to be linearly unstable at nonequilibrium
states in \cite{Zhao2017}. Therefore these systems are never used in the
simulation of high-Mach number flows.

In a recent work \cite{Cai2020}, it is found that the computation using
hyperbolic moment equations may fail when the maximum temperature ratio is
larger than two in the problem, due to the divergence of Grad's series expansion
\eqref{eq:ansatz}. The convergence of \eqref{eq:ansatz} requires
\begin{equation} \label{eq:convergence}
\int_{\mathbb{R}} \frac{[f(x,\xi,t)]^2}{\mathcal{M}_0(x,\xi,t)} \,\mathrm{d}\xi
  < +\infty, \qquad \forall x,t.
\end{equation}
However, it is shown in \cite{Cai2020} that for the BGK collision term, if there
exists $x_1$ and $x_2$ such that $\theta(x_1) > 2\theta(x_2)$ for the exact
solution of a steady-state problem, then the convergence condition
\eqref{eq:convergence} will fail, leading to the divergence of
\eqref{eq:ansatz}. In this case, even if the solution of the moment equations
exists, the convergence to the Boltzmann equation is questionable. Both shock
structure and the Fourier flow computation are considered in \cite{Cai2020},
where it is observed that when the maximum temperature ratio is larger than $2$
in the exact solution, the computation using hyperbolic moment equations fails
when $N$ is sufficiently large. Such a deficiency, which prohibits the
simulation for high-speed flows using many moments, exists in Grad's moment
method, hyperbolic moment method and any other methods based on Grad's ansatz
such as \cite{Koellermeier2014}.

\subsection{Regularized moment equations}
Another method to overcome the loss of hyperbolicity is to introduce stabilizing
diffusive terms, which can be done by mimicking the derivation of
Navier-Stokes equations as extensions of Euler equations. The most well-known
model in this category is the regularized 13-moment equations (``13 moments''
refers to the 3D case), which are derived in \cite{Struchtrup2003} for Maxwell
molecules with the super-Burnett order. The regularized 26-moment equations are
derived in \cite{Gu2009}.  For the BGK collision model with one-dimensional
physics, the closure is
\begin{displaymath}
f_N = \tau \left(
  \frac{\theta}{\rho} f_{N-1} \frac{\partial \rho}{\partial x} -
  \theta \frac{\partial f_{N-1}}{\partial x} -
  \frac{1}{2} \theta f_{N-3} \frac{\partial \theta}{\partial x} -
  \frac{N-1}{2} f_{N-1} \frac{\partial \theta}{\partial x}
\right).
\end{displaymath}
When $N = 3$, the system is identical to the Navier-Stokes equations. The
numerical tests in \cite{Torrilhon2004, Cai2012, Timokhin2015, Timokhin2017}
show that a small number of moments (13 or 20 moments in the three-dimensional
case) can well describe the shock structure for equilibrium variables (density,
velocity and temperature) up to Mach number $9.0$. However, a significant
underestimate of the heat flux can be observed in these numerical results for
Mach number larger than $3.0$.

Despite the stabilization, this moment hierarchy still suffers from the
instability of the system when the number of moments is large, as reported in
\cite{Cai2012}. As far as we know, the regularized 26-moment equations are
applied only to microflow problems \cite{Gu2009}. The numerical experiment will
be redone in the one-dimensional case in Section \ref{sec:num_test}. Moreover,
the derivation of the equations in the three-dimensional case is extremely
tedious for non-Maxwell molecules, even for the 13-moment case with locally
linearized collision term \cite{Cai2020r}. The derivation of regularized moment
equations may be impracticable for larger moment system.

\subsection{Method of maximum entropy}
This moment hierarchy takes a different ansatz
\begin{equation} \label{eq:max_entropy}
f(x,\xi,t) =
  \exp \left( \sum_{\beta = 0}^{N-1} k_{\beta}(x,t) \xi^{\beta} \right),
\end{equation}
where we require that $N$ is odd and $k_{N-1}(x,t) \leqslant 0$ so that the
distribution function is integrable. Using $M_{\alpha}(x,t)$ to denote the
moment $\int_{\mathbb{R}} \xi^{\alpha} f(x,\xi,t) \,\mathrm{d}\xi$, we can
write the moment system as
\begin{equation} \label{eq:mom_sys}
\frac{\partial M_{\alpha}}{\partial t} + \frac{\partial M_{\alpha+1}}{\partial x}
  = \frac{1}{\tau} \left(
    \int_{\mathbb{R}} \xi^{\alpha} \mathcal{M}(x,\xi,t) \,\mathrm{d}\xi - M_{\alpha}
  \right), \qquad \alpha = 0,1,\cdots,N-1,
\end{equation}
where
\begin{equation} \label{eq:M_N}
M_N(x,t) = \int_{\mathbb{R}} \xi^N 
  \exp \left( \sum_{\beta = 0}^{N-1} k_{\beta}(x,t) \xi^{\beta} \right) \,\mathrm{d}\xi,
\end{equation}
and the coefficients $k_{\beta}$ can be determined from $M_{\alpha}$ by solving
the moment inversion problem:
\begin{equation} \label{eq:M_alpha}
M_{\alpha}(x,t) = \int_{\mathbb{R}} \xi^{\alpha}
  \exp \left( \sum_{\beta = 0}^{N-1} k_{\beta}(x,t) \xi^{\beta} \right) \,\mathrm{d}\xi,
\qquad \alpha = 0,1,\cdots,N-1.
\end{equation}
Such a method has been studied in \cite{Dreyer1987, Levermore1996, Muller1998}.
Theoretically, the system is hyperbolic, entropic, and preserves positivity.
It preserves at least the Navier-Stokes limit when the Knudsen number is small.
However, the numerical computation of such a system is extremely difficult when
$N > 3$, due to the large characteristic speed and the lack of explicit
formulas for the map between moments and the variables $k_{\beta}$. The first
numerical results are reported in \cite{Tallec1997}, where shock structures
with Mach numbers $1.2$ and $2.0$ for the 3D 14-moment theory is computed. The
density and temperature profiles well agree with the DSMC results. When the
Mach number increases to $4.0$, the results in \cite{Schaerer2017} shows the
inadequacy of 14 moments, while the 35-moment theory can provide much better
results \cite{Schaerer2017, Schaerer2017e}. In these methods, the velocity
domain is truncated to avoid infinite characteristic speed, so that CFL
conditions can be imposed more easily.

We believe that such a moment hierarchy is suitable for shock structure
computation with larger Mach numbers. However, current numerical methods seem
insufficient to make the system competitive due to the significant difficulty
in solving the moment inversion problem. Research works are still ongoing
to improve the efficiency \cite{Sadr2019}.

One relevant method is the moment closure based on $\varphi$-divergences
proposed in \cite{Abdelmalik2016}. It replaces the ansatz
\eqref{eq:max_entropy} by
\begin{equation}
f(x,\xi,t) = \mathcal{M}(x,\xi,t)
  \left( 1 + \frac{1}{n} \sum_{\beta = 0}^{N-1} k_{\beta}(x,t) \xi^{\beta} \right)_+^n,
\end{equation}
where $(\cdot)_+$ means $\max(\cdot,0)$. Such an ansatz can be viewed as an
approximation of \eqref{eq:max_entropy}, due to the fact that
\begin{displaymath}
\exp(x) = \lim_{n\rightarrow \infty} \left( 1 + \frac{x}{n} \right)_+^n.
\end{displaymath}
The moment inversion problem again needs to be solved for this ansatz, which
has been implemented in \cite{Abdelmalik2016a, Abdelmalik2017}. The shock
structure problem with Mach number $1.4$ for one-dimensional physics is
considered in \cite{Abdelmalik2017}, but the paper does not show the accuracy
of the model by comparing with other methods.

More studies on both the numerical methods and the model verification are
needed for these entropy stable models. Due to the complicated algorithm used
for the moment inversion, these methods are not studied in this paper.

\subsection{Quadrature-based moment methods} \label{sec:QBMM}
Another method requiring moment inversion is the quadrature-based moment
methods \cite{Fox2008, Fox2009}. In the one-dimensional case, the ansatz of the
distribution function for even $N$ is
\begin{equation} \label{eq:QBMM}
f(x,\xi,t) = \sum_{n = 0}^{N/2} \varrho_n(x,t) \delta(\xi - \xi_n(x,t)).
\end{equation}
The moment system can again be written as \eqref{eq:mom_sys}, and the last flux
$M_N$ can be defined similarly to \eqref{eq:M_N}\eqref{eq:M_alpha}, with the
ansatz of $f$ correspondingly replaced. Since both $\varrho_n$ and $\xi_n$ need
to be determined, the moment inversion problem is again highly nontrivial,
especially in the high-dimensional case. For one-dimensional velocity, some
efficient algorithms have been developed \cite{Wheeler1974, Yuan2011}, and in
the multi-dimensional case, the conditional quadrature method of moments
\cite{Yuan2011} is developed to generate quadrature points and weights by
making use of the one-dimensional algorithm. We have not seen the discussion of
the asymptotic limits of such systems to the best of our knowledge.

Problems with shocks with Mach number up to $2.05$ have been considered in
\cite{Fox2009} for three-dimensional physics. The simulation appears to be
quite stable, whereas the accuracy of the shock structure is not carefully
checked. Such a method will also be taken into account in our numerical
experiments. Note that the hyperbolic version of the quadrature-based moment
method has been studied in \cite{Fox2018}, where it is mentioned that such
method currently works only for $N=3$ and $N=5$, and therefore it will not be
considered in this work.

\subsection{Other moment hierarchies}
Besides the aforementioned moment hierarchies, some other moment hierarchies
have also appeared in a relatively smaller amount of literature, which we do
not study in this work. For example, in \cite{Fan2015}, the authors proposed a
variation of the hyperbolic moment system by replacing the Maxwellian in Grad's
ansatz with an anisotropic Gaussian. The method is identical to the hyperbolic
moment method in the one-dimensional case, while it makes some difference in
the multi-dimensional case. The extended quadrature-based moment method
\cite{Yuan2012} replaces the Dirac delta functions in \eqref{eq:QBMM} by
Gaussians, but it is rarely applied to the gas kinetic theory. Some encouraging
results are shown in \cite{Laplante2016} for one-dimensional shock structure
computation with two Gaussians. However, the generalization to multiple
Gaussians and higher-dimensional cases has intrinsic difficulties. More
recently, the entropic quadrature closure is proposed in \cite{Bohmer2020},
which combines the idea of maximum entropy and quadrature-based moment methods.
The numerical examples in \cite{Bohmer2020} also shows its great potential in
tackling high-speed flows. We expect further development of this method.

%% file: article_num_test.tex
\section{Numerical computation of shock structures for one-dimensional physics}
\label{sec:num_test}
Despite a variety of different moment methods proposed in the history,
experiments showing the convergence of the moment hierarchies have hardly been
done for high-speed flows. The review of the methods in the previous section
shows three problems in such simulations: numerics, hyperbolicity, and
convergence. In this section, we are going to consider the one-dimensional
physics and use numerical tests to demonstrate such problems. For simplicity,
we only pick the methods which are easier to implement in our experiements.
The one-dimensional shock structure problem will be used as the benchmark.
Below we first provide the definition of the problem.

For the Boltzmann equation \eqref{eq:Boltzmann}, the steady shock structure of
Mach number $\mathit{Ma}$ can be solved by setting the initial data to be
\begin{equation} \label{eq:init}
f(x,\xi,0) = \frac{\rho_0(x)}{\sqrt{2\pi \theta_0(x)}}
  \exp \left( -\frac{|\xi - v_0(x)|^2}{2\theta_0(x)} \right),
\end{equation}
where 
\begin{gather*}
\rho_0(x) = \left\{ \begin{array}{@{}ll}
  1, & \text{if } x < 0, \\
  \dfrac{2 \mathit{Ma}^2}{\mathit{Ma}^2 + 1}, & \text{if } x > 0,
\end{array} \right. \qquad
v_0(x) = \left\{ \begin{array}{@{}ll}
  \sqrt{3} \mathit{Ma}, & \text{if } x < 0, \\
  \dfrac{\sqrt{3}}{2} \dfrac{\mathit{Ma}^2+1}{\mathit{Ma}}, & \text{if } x > 0,
\end{array} \right. \\[5pt]
\theta_0(x) = \left\{ \begin{array}{@{}ll}
  1, & \text{if } x < 0, \\
  \dfrac{(3\mathit{Ma}^2-1)(\mathit{Ma}^2+1)}{4\mathit{Ma}^2}, & \text{if } x > 0.
\end{array} \right.
\end{gather*}
The steady state solution is the structure of the shock wave. For moment
equations, the initial condition can be obtained by computing the moments from
the initial distribution function. The five moment methods we are going to test 
in our work are Grad's moment methods, Grad's moment methods with linearized
ansatz, hyperbolic moment methods, regularized moment methods and
quadrature-based moment methods. Two Mach numbers $\mathit{Ma} = 1.4$ and
$\mathit{Ma} = 2.0$ will be taken into account. In all the following numerical
tests, the finite volume method with Lax-Friedrichs numerical flux and forward
Euler method is used, which is known to be dissipative but numerically stable.
The number of spatial grid cells is $10,000$ unless otherwise specified. We
terminate the simulation at $t = 50$. When showing the numerical results, we
are mainly interested in the density, velocity and temperature, and their
normalized values are to be plotted, which are defined by
\begin{displaymath}
\widehat{\rho}(x) = \frac{\rho(x) - \rho_0(-\infty)}{\rho_0(+\infty) - \rho_0(-\infty)}, \qquad
\widehat{v}(x) = \frac{v(x) - v_0(+\infty)}{v_0(-\infty) - v_0(+\infty)}, \qquad
\widehat{\theta}(x) = \frac{\theta(x) - \theta_0(-\infty)}{\theta_0(+\infty) - \theta_0(-\infty)}.
\end{displaymath}
Below the results of the two Mach numbers are to be presented separately in the
following subsections.

\subsection{Mach number $1.4$}
In this case, the density ratio is $1.32$ and the temperature ratio is $1.84$.
We consider this case as the ``small Mach number'', and we mainly use this
example to show the convergence of the moment hierarchies and compare the
qualities of these models. The computational domain is set to be $[-20, 40]$.
When we show the numerical results, only a region $[-10,10]$ is plotted, since
the curves outside this region are nearly flat. The results of the five
methods will be presented in the following subsections.

\subsubsection{Grad's moment method with linearized ansatz}
\begin{figure}[!ht]
\centering
\includegraphics[width=.45\textwidth,clip]{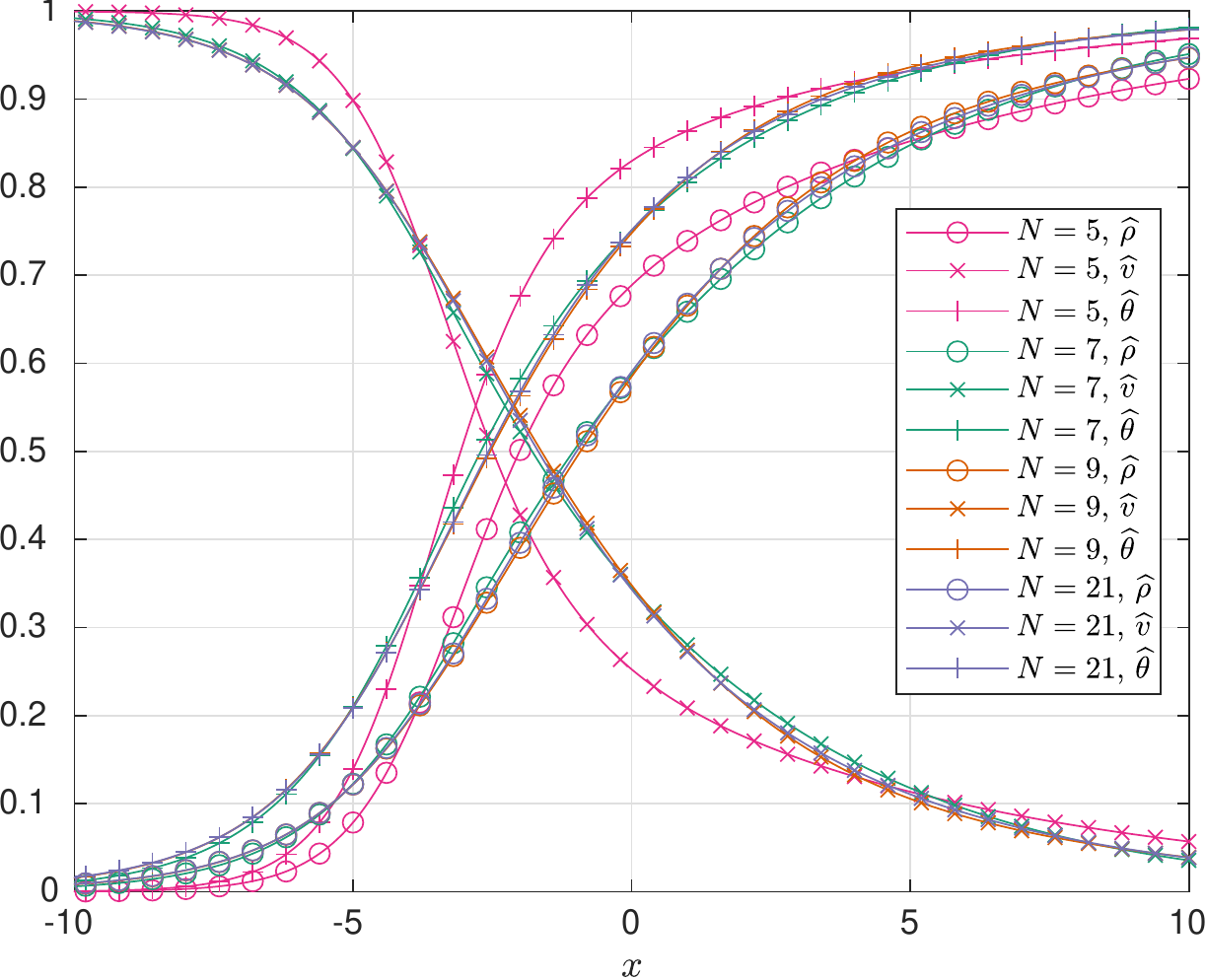}
\caption{Numerical results for $\mathit{Ma} = 1.4$ using Grad's moment method with
linearized ansatz}
\label{fig:linear_Ma1.4}
\end{figure}

Since this method is essentially the spectral method for the Boltzmann
equation, the convergence as $N \rightarrow \infty$ can be guaranteed if the
parameters $\overline{u}$ and $\overline{\theta}$ in \eqref{eq:linear_ansatz}
are chosen properly so that the series \eqref{eq:linear_ansatz} converges for
the exact solution of the Boltzmann equation. Here we choose $\overline{u} =
\sqrt{3} \mathit{Ma}$ and $\overline{\theta} = 1$, and the results for
$N=5,7,9,21$ are shown in Figure \ref{fig:linear_Ma1.4}.

As a spectral method, its fast convergence can be clearly observed. Taking the
results of $N=21$ as the reference solution, we find that the profiles computed
by $N=7$ already have good agreement with the reference solution
quantitatively. However, due to the lack of nonlinearity, the results of the
five-moment system $N=5$ shows a significantly underestimated shock thickness.
Below, in our experiments for the other four methods, the results of this
method with $N=21$ will be used as the references solutions.

\subsubsection{Grad's moment method and hyperbolic moment method}
\begin{figure}[!ht]
\centering
\subfloat[Grad's moment method]{%
  \label{fig:Grad_Ma1.4}
  \includegraphics[width=.45\textwidth]{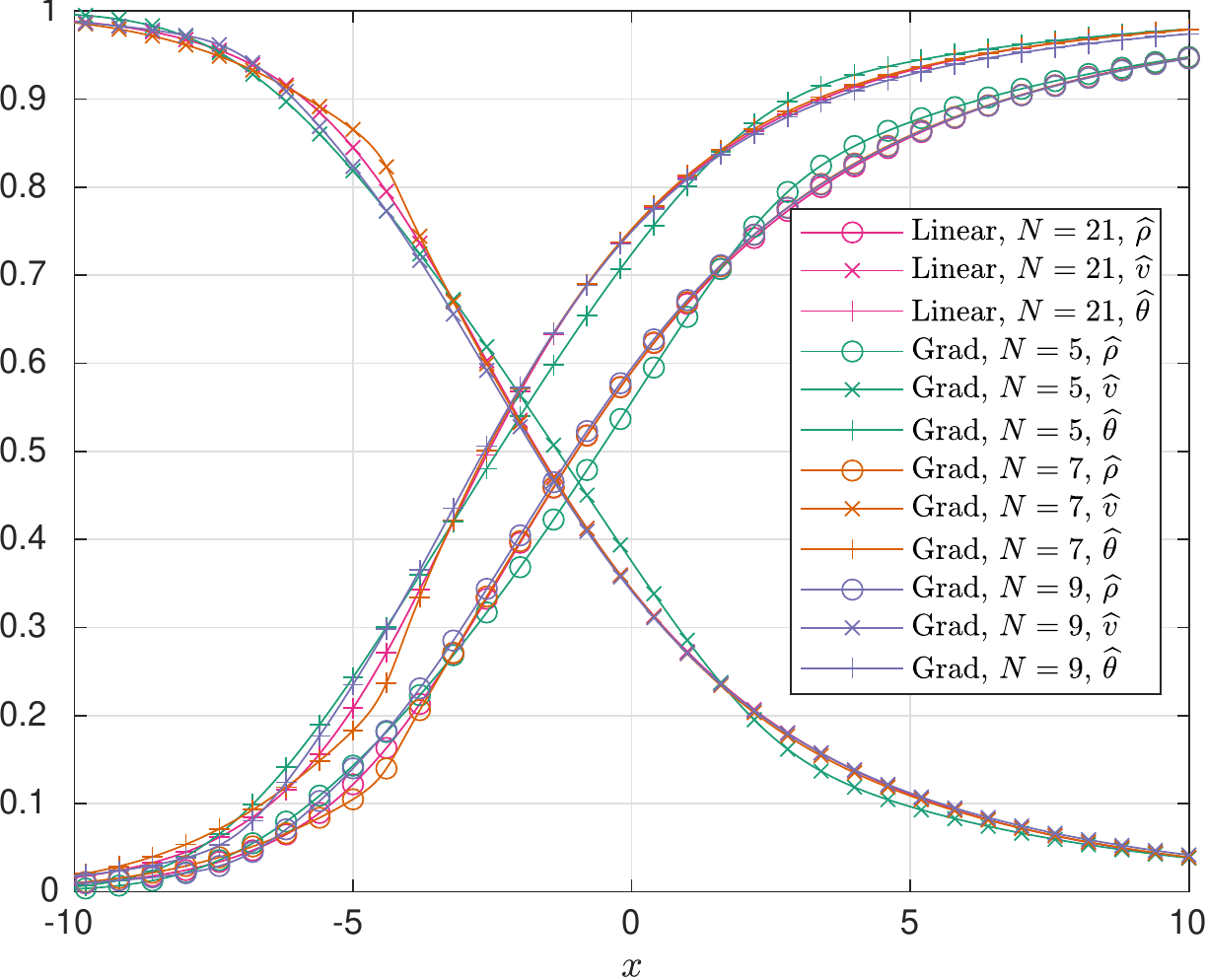}}
\qquad
\subfloat[Hyperbolic moment method]{%
  \label{fig:Hyp_Ma1.4}
  \includegraphics[width=.45\textwidth,clip]{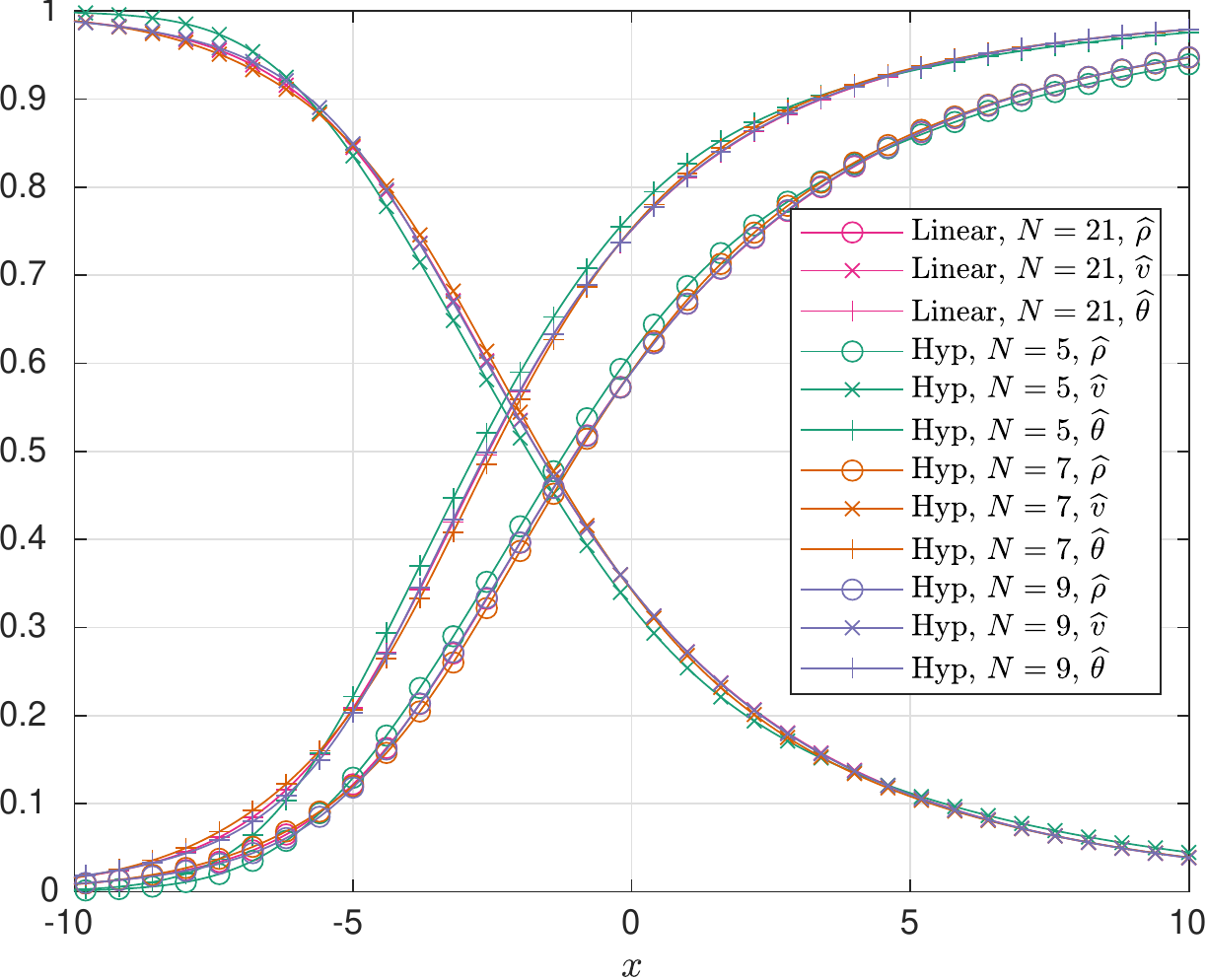}}
\caption{Numerical results for $\mathit{Ma} = 1.4$ using Grad's moment method and the hyperbolic moment method}
\label{fig:GradHyp_Ma1.4}
\end{figure}

For Mach number $1.4$, the temperature ratio is below $2.0$. Therefore
convergence of both Grad's moment method and the hyperbolic moment method can
be expected. The numerical solutions for $N = 5,7,9$ are provided in Figure
\ref{fig:GradHyp_Ma1.4}. Two methods provide results with similar quality, due
to the same ansatz \eqref{eq:ansatz} for the distribution function. Compared
with Figure \ref{fig:linear_Ma1.4}, the results for $N=5$ have been greatly
improved, showing the superiority of a nonlinear ansatz for the distribution
function. Comparing Figure \ref{fig:Grad_Ma1.4} and Figure \ref{fig:Hyp_Ma1.4},
one can find that the hyperbolic moment method gives slightly better results
than Grad's moment method, as agrees with the observation in \cite{Cai2020h}.
The underlying reason remains to be further studied. The simulation for the
hyperbolic moment method is also faster than Grad's moment method, since at
each time step, we need to find the roots of a polynomial on each spatial grid
to determine the time step according to the CFL condition.

\subsubsection{Regularized moment method}
\begin{figure}[!ht]
\centering
\includegraphics[width=.45\textwidth,clip]{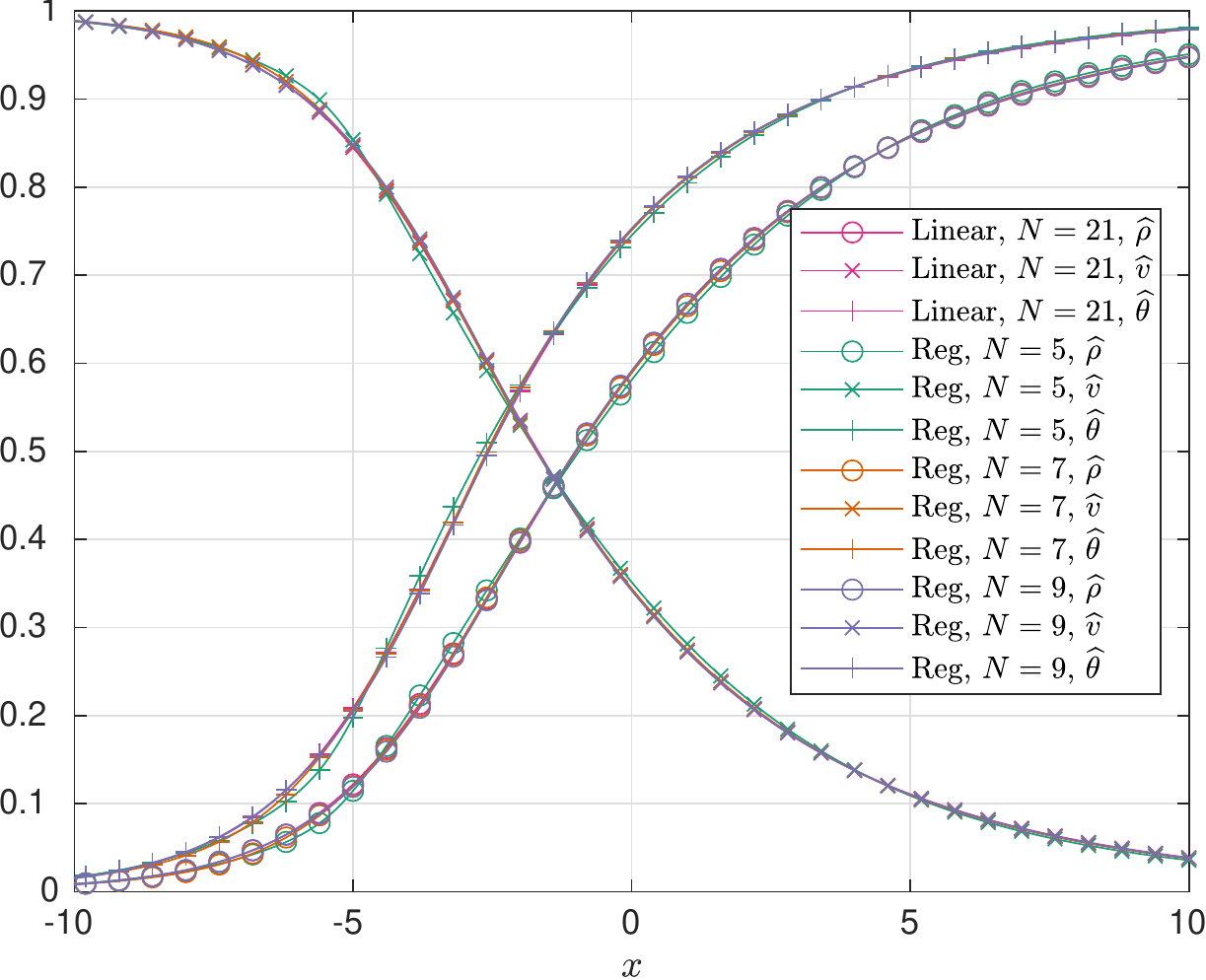}
\caption{Numerical results for $\mathit{Ma} = 1.4$ using regularized moment method}
\label{fig:Reg_Ma1.4}
\end{figure}

The results of the regularized moment method with $N = 5,7,9$ are shown in
Figure \ref{fig:Reg_Ma1.4}, which also shows convergence as $N$ increases.
Compared with Grad's moment method and the hyperbolic moment method, the
regularized moment method shows better results for $N=5$, since it essentially
includes part of the information from the fifth moment. However, due to the
second-order derivatives in the equations, much smaller time steps are required
in the simulation, causing significantly longer computational time. Although
this can be improved by using implicit treatment of these second-order
derivatives, it can still be expected that longer computational time is needed
than Grad's moment method. The advantage of the regularized moment method is
clearer in the high-dimensional case, where the number of moments is
proportional to $N^3$, so that the regularized moment method can effectively
reduce the degrees of freedom.

\subsubsection{Quadrature-based moment method}
\begin{figure}[!ht]
\centering
\includegraphics[width=.45\textwidth]{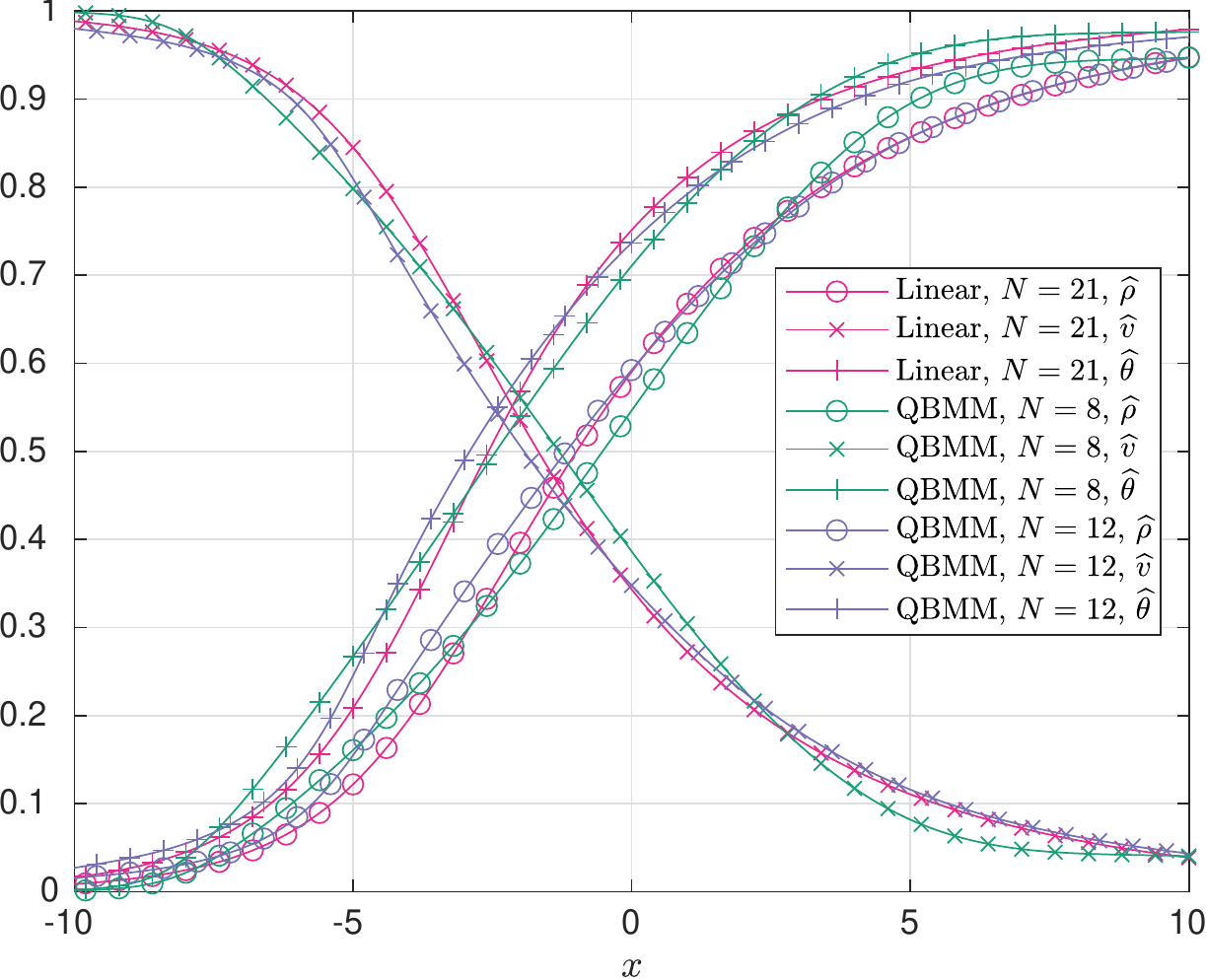}
\caption{Numerical results for $\mathit{Ma} = 1.4$ using quadrature-based moment method}
\label{fig:QBMM_Ma1.4}
\end{figure}

For this method, we choose to use even number of moments in our tests as
mentioned in Section \ref{sec:QBMM}. The cases tested in our simulation include
$N=8$ and $12$, and we show the results in Figure \ref{fig:QBMM_Ma1.4}. In
general, QBMM can provide stable results, while the quality of approximation is
not as good as previous methods. When $N = 12$, we can still observe obvious
difference between the QBMM result and the reference solution. The reason is
likely to be the ansatz of QBMM, which consists of several Dirac delta
functions that do not match the form of distribution functions in the shock
structure.

\subsection{Mach number $2.0$}
The above numerical experiments show that the moment methods generally work
well for Mach number $1.4$, despite the failure in some cases. However, when we
increase the Mach number to $2.0$, the situation becomes much less optimistic.
In this case, the density ratio is $1.6$ and temperature ratio is $3.44$. Since
the temperature ratio exceeds $2$, the convergence issue of Grad-type methods
(see \cite{Cai2020}) will surface, resulting breakdown of the computation. The
computational domain is set to be $[-30, 30]$ for this much number. Our
numerical results will be detailed in the following subsections.

\subsubsection{Grad's moment method with linearized ansatz}
\begin{figure}[!ht]
\centering
\includegraphics[width=.45\textwidth]{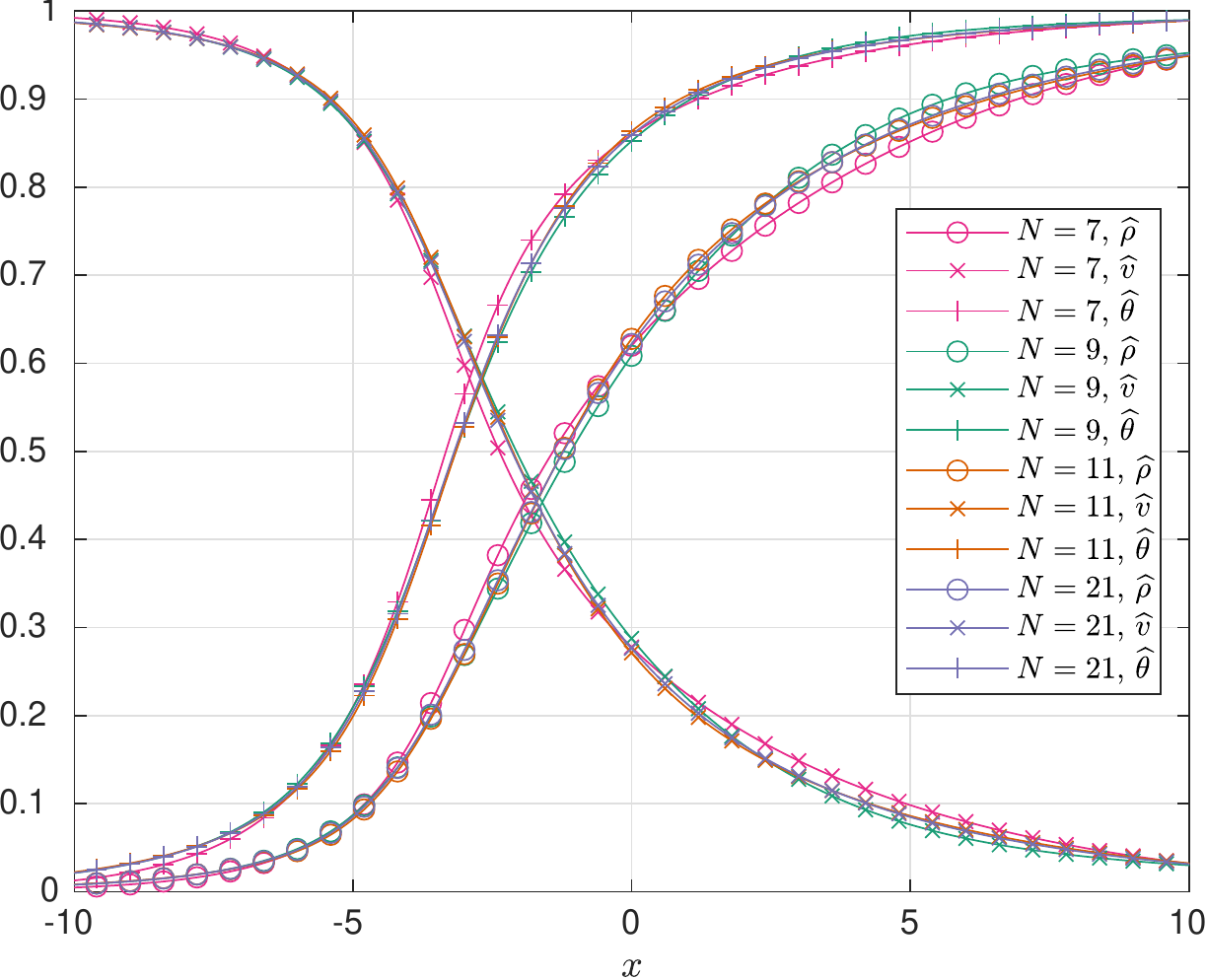}
\caption{Numerical results for $\mathit{Ma} = 2.0$ using Grad's moment method with
linearized ansatz}
\label{fig:linear_Ma2}
\end{figure}

For Mach number $2.0$, choosing $\overline{\theta} = 1$ quickly leads to
breakdown of computation due to the divergence of \eqref{eq:linear_ansatz} for
fluid states behind the shock wave. This indicates that the manual selection of
different parameters for different problems is essential in this method. By
choosing $\overline{\theta} = 2$ and $\overline{v} = \sqrt{3}\mathit{Ma}$, we
can get convergent results, and the spectral method still shows its high
efficiency (see Figure \ref{fig:linear_Ma2}), and we will again use the result
of $N=21$ as reference solution in the following numerical tests.

\subsubsection{Grad's moment method and hyperbolic moment method}
\begin{figure}[!ht]
\centering
\subfloat[Grad's moment method]{%
  \label{fig:Grad_Ma2}
  \includegraphics[width=.45\textwidth]{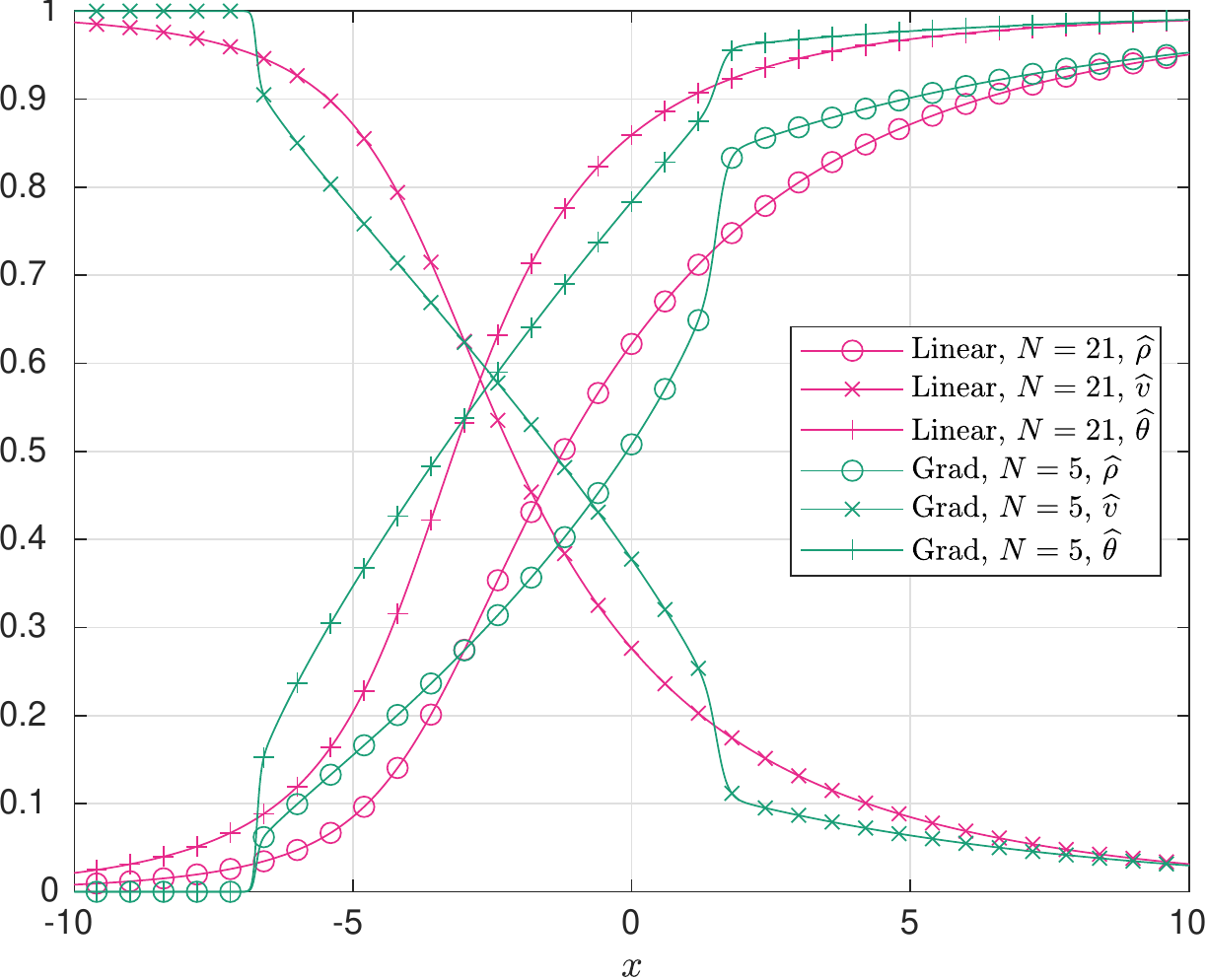}}
\qquad
\subfloat[Hyperbolic moment method]{%
  \label{fig:Hyp_Ma2}
  \includegraphics[width=.45\textwidth,clip]{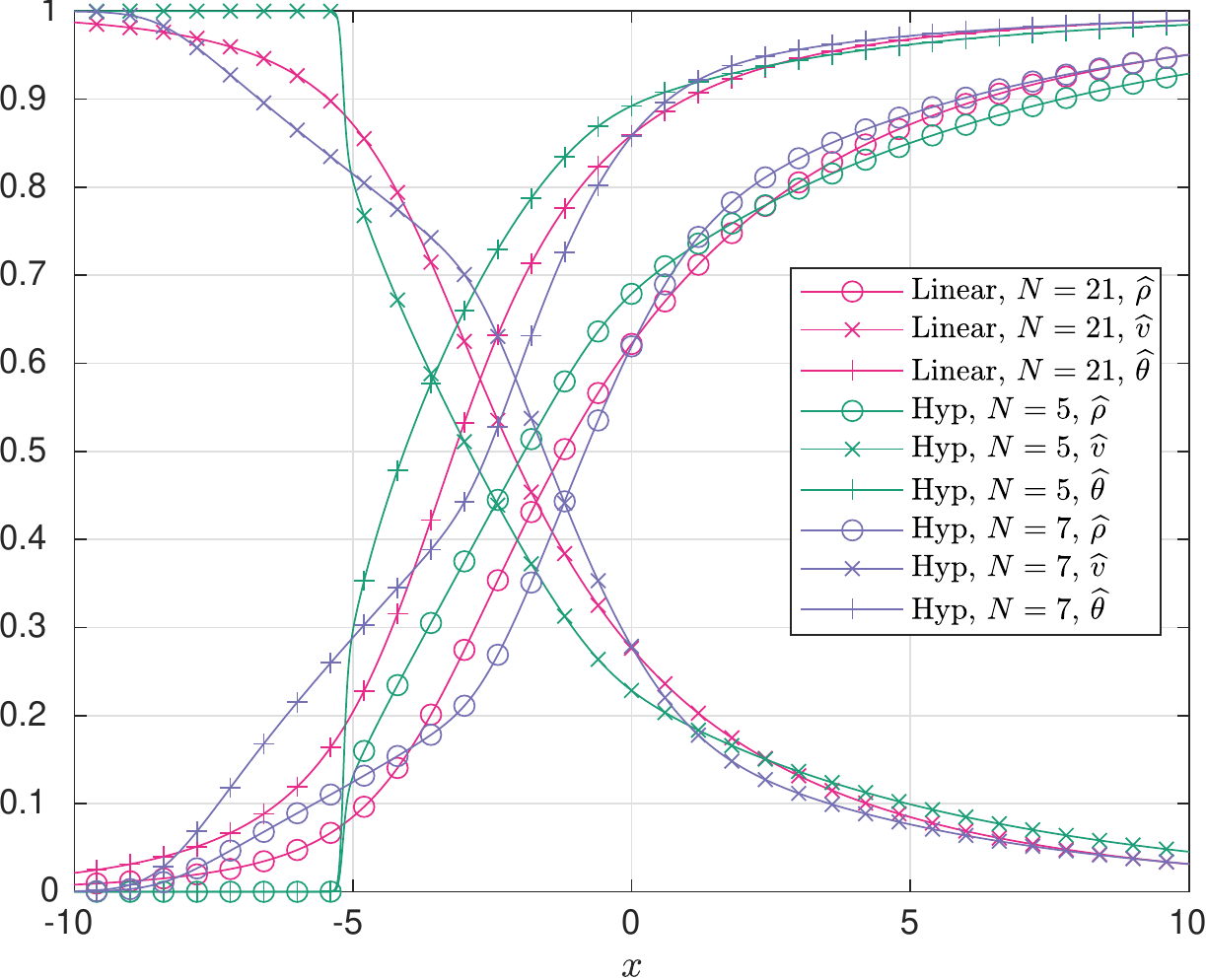}}
\caption{Numerical results for $\mathit{Ma} = 2.0$ using Grad's moment method and the hyperbolic moment method}
\label{fig:GradHyp_Ma2}
\end{figure}

Now Grad's moment method faces problems coming from both hyperbolicity and
convergence, and the instability of Grad's moment method emerges. We tested $N
= 5,7,9,11$, and it turns out that only for $N=5$, we can achieve the steady
state numerically from the discontinuous initial data. All other choices of $N$
end up with negative temperature in the computational process, which forces the
simulation to terminate. In the numerical result for $N=5$ (Figure
\ref{fig:Grad_Ma2}), a discontinuity can be observed near $x = -7$, although it
is smeared due to the dissipative numerical scheme, as is the notorious
subshock phenomenon. Note that here we do not introduce any high-order schemes
since these methods may introduce numerical instability. This discontinuity is
known as an unphysical ``subshock'', which is caused by the insufficient
characteristic speed of the moment equations \cite{Grad1952, Torrilhon2000}.
Although using more moments can increase the characteristic speed, such a
strategy is voided by the convergence problem, which will be detailed in
Section \ref{sec:ansatz}.

As for the hyperbolic moment method, we can obtain stable numerical results for
$N = 5,7$, while for $N=9,11$, the computation breaks down due to the
appearance of negative temperature, which agrees with the observation in
\cite{Cai2020h}. This shows some advantages of the hyperbolic moment method due
to the hyperbolicity fix. However, this does not overcome the convergence
issue. The numerical results for $N=5,7$ are plotted in Figure
\ref{fig:Hyp_Ma2}. Again, we can observe a subshock for $N=5$ near $x = -5$;
when $N = 7$, although the subshock no longer appears in the solution, the
approximation of the shock structure is still unsatisfactory. In general, the
performance of these moment methods is poorer than the case of $\mathit{Ma} =
1.4$, owing to the stronger nonequilibrium inside the shock wave.
Unfortunately, we are not able to further improve the result by increasing the
number of moments with such an ansatz.

\subsubsection{Regularized moment method}
\begin{figure}[!ht]
\centering
\includegraphics[width=.45\textwidth,clip]{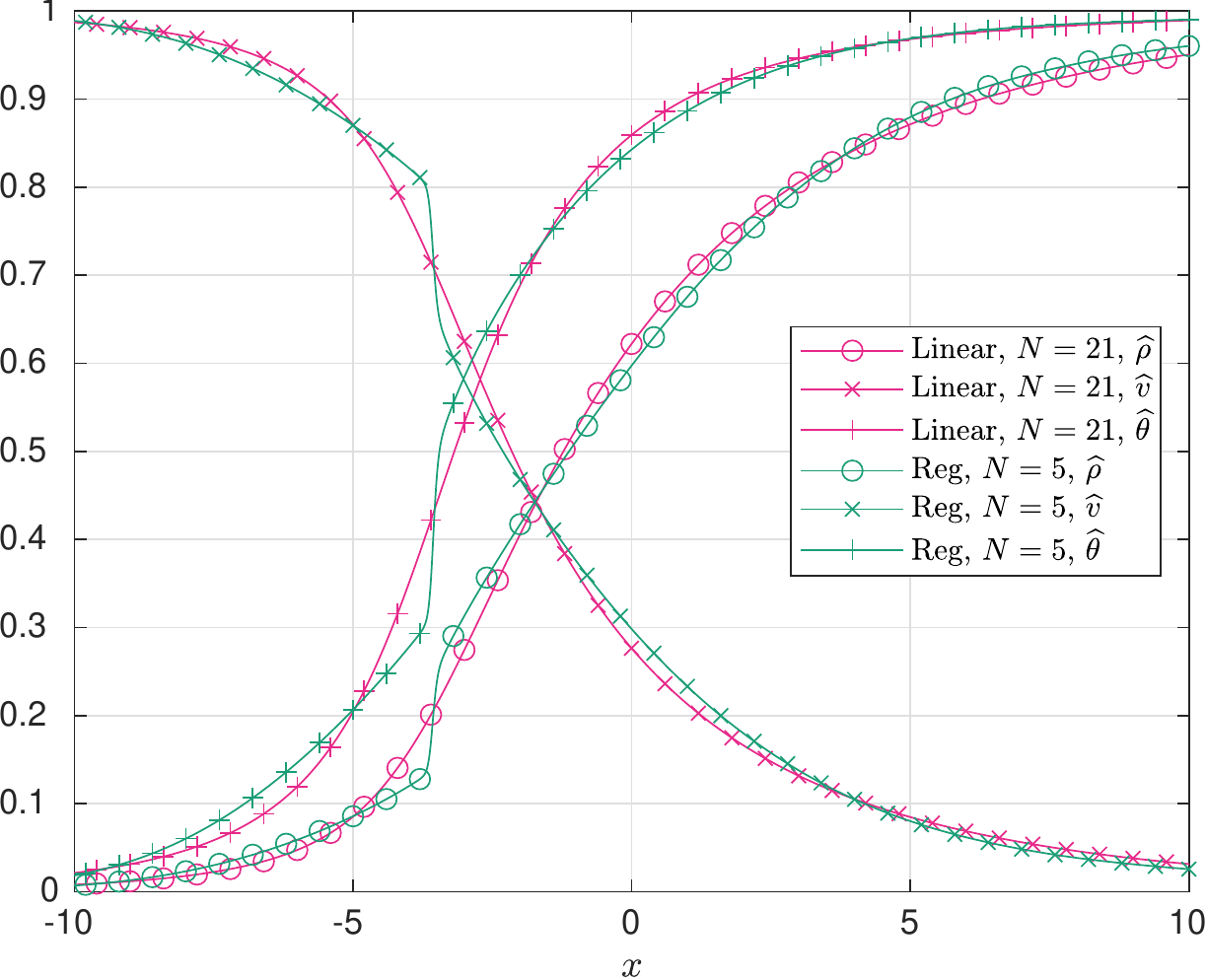}
\caption{Numerical results for $\mathit{Ma} = 2.0$ using regularized moment method}
\label{fig:Reg_Ma2}
\end{figure}

The regularized moment method suffers from the same issue as Grad's moment
method and the hyperbolic moment method. We tested $N=5,7,9,11$, and the only
case that works is $N=5$, and we plot the results in Figure \ref{fig:Reg_Ma2}.
Due to the regularization, the quality of approximation is better than Grad's
moment method, but the deviation is still obvious. Unfortunately, the
regularization is not sufficient to stabilize the moment method when $N
\geqslant 7$, and the simulation still fails by the ocurrence of negative
temperature during the evolution.

\subsubsection{Quadrature-based moment method}
\begin{figure}[!ht]
\centering
\includegraphics[width=.45\textwidth,clip]{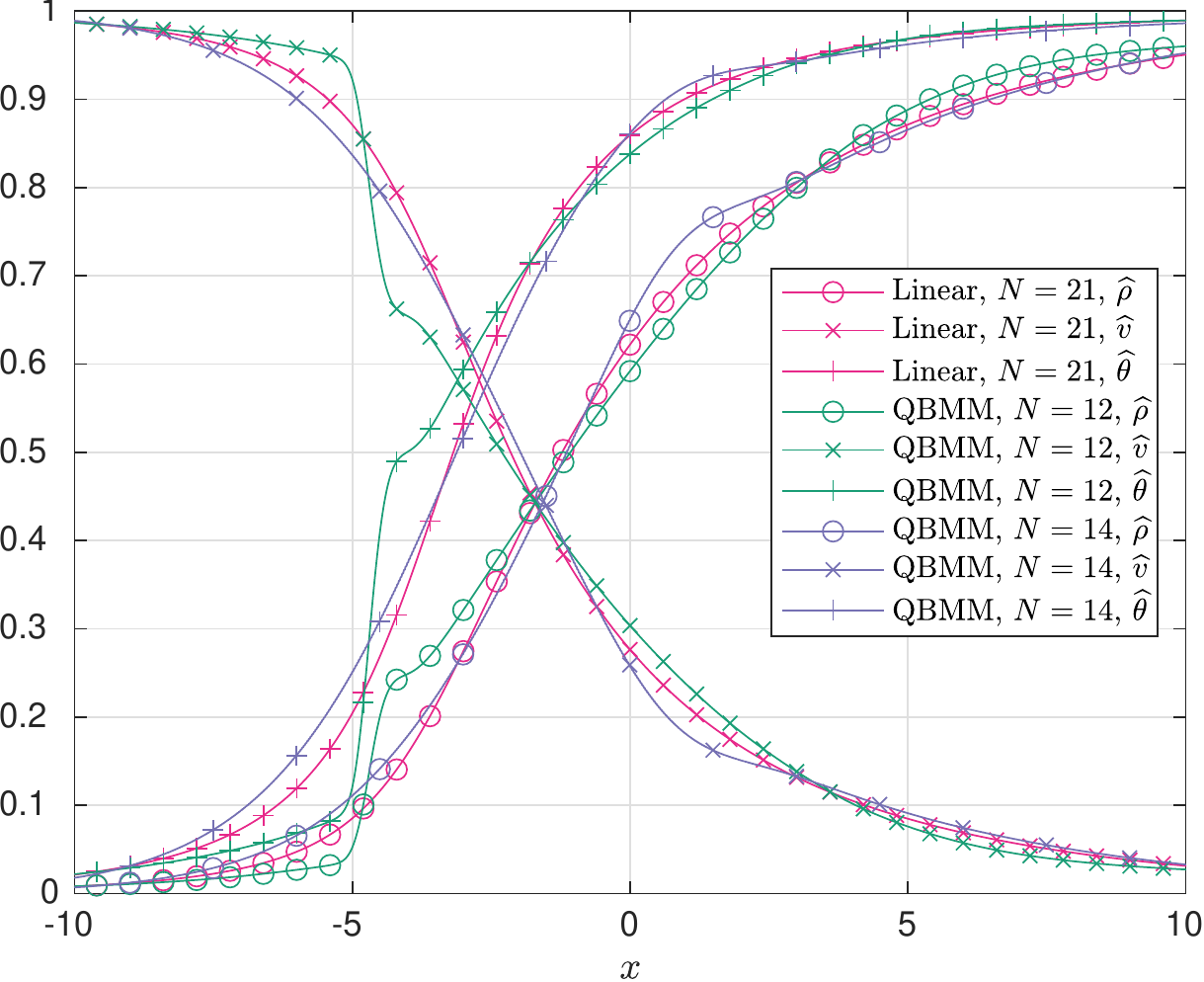}
\caption{Numerical results for $\mathit{Ma} = 2.0$ using quadrature-based moment method}
\label{fig:QBMM_Ma2}
\end{figure}

The numerical results for the quadrature-based moment method are plotted in
Figure \ref{fig:QBMM_Ma2}, where the curves for $N=12$ and $14$ are provided.
Figure \ref{fig:QBMM_Ma2} shows that the method shows the trend of convergence,
but the efficiency is unsatisfactory, especially when compared with Figure
\ref{fig:linear_Ma2}. Even for $N = 14$, the deviation from the reference
solution is still obvious. Note that $N=14$ looks like a small number of
degrees of freedom in the one-dimensional case, while in the three-dimensional
case, the corresponding number of moments is $N(N+1)(N+2)/6 = 560$, and this
number grows in the cubic manner as $N$ increases. Here we note that QBMM and
related methods have achieved significant success in simulating disperse
multiphase flows \cite{Fox2018}.

%% file: article_hmb.tex
\section{Highest-moment-based moment method: One-dimensional case} \label{sec:1d}
The above numerical tests show intrinsic difficulties in computing shock
structures using moment methods even for Mach number $2.0$ in the
one-dimensional case, which does not look very high. The quadrature-based
moment method looks like a possible solution, but its efficiency is lower than
the Grad-type methods, and its generalization to three-dimensional case is
nontrivial. Some untested methods such as the maximum entropy method may work
better for high Mach numbers. Unfortunately, significant numerical difficulties
exist in such computations. Is it possible to find alternative moment
hierarchies as simple as Grad's moment methods to deal with high-speed flows?
We are going to explore such possibilities in this section. For simplicity, we
will only address the one-dimensional case in this section. The more realistic
three-dimensional case will be discussed later in Section \ref{sec:3d}.

\subsection{Derivation of the moment equations}
The derivation of the novel moment hierarchy in the one-dimensional case can be
demonstrated in two steps. We will first propose an ansatz for the distribution
function, and then apply the model-reduction technique in \cite{Cai2015a} to
obtain moment systems.

\subsubsection{Ansatz for the distribution function} \label{sec:ansatz}
\textcolor{red}{Before} writing down our ansatz of the distribution function, we
would like to propose several principles that the ansatz must follow. In order
to build reliable moment models to compute the shock structure for large Mach
numbers, the ansatz must satisfy:
\begin{description}
\item[C1] Any Maxwellian can be represented exactly by the ansatz.
\item[C2] The system does not require any parameters that cannot be determined
  by the moments in the system.
\item[C3] \label{cond:convergence} The ansatz converges to any linear
  combination of any Gaussians as the number of moments tends to infinity.
\end{description}
The first condition means that the Euler limit must be preserved. The second
condition ensures that no additional parameters are introduced into the ansatz.
The third condition takes into account the possible shapes of the distribution
functions appearing in the shock structure. According to the Mott-Smith theory,
\cite{MottSmith1951}, the distribution functions inside the shock wave can be
well approximated by the linear combination of two Gaussians representing the
fluid state in front of and behind the shock wave. Condition C3 is the one that
Grad's ansatz violates, especially when the temperature ratio exceeds $2$. We
refer the readers to \cite{Cai2020} for details on this convergence issue. Here
we just briefly mention that given $\theta_1$ and $\theta_2$ satisfying
$\theta_1 > 2\theta_2 > 0$, for any $v_1, v_2 \in \mathbb{R}$ and $\rho_1,
\rho_2 \in \mathbb{R}_+$, we can always find a sufficiently small $\kappa > 0$
such that Grad's series \eqref{eq:ansatz} for the distribution function
\begin{equation} \label{eq:Mott-Smith}
f(\xi) = (1-\kappa) \frac{\rho_1}{\sqrt{2\pi\theta_1}}
  \exp \left( -\frac{(\xi - v_1)^2}{2\theta_1} \right) +
\kappa \frac{\rho_2}{\sqrt{2\pi\theta_2}}
  \exp \left( -\frac{(\xi - v_2)^2}{2\theta_2} \right)
\end{equation}
diverges, since when $\kappa$ is small, the temperature $\theta$ of the above
distribution function is close to $\theta_2$, so that $[f(\xi)]^2 /
\mathcal{M}_0(\xi)$ tends to infinity as $\xi \rightarrow \infty$, which
violates the condition \eqref{eq:convergence}.

Despite the this drawback of Grad's method, its simplicity and good
approximability inspire us to use a similar form of the ansatz:
\begin{equation} \label{eq:new_ansatz}
f(x,\xi,t) = \sum_{\alpha=0}^{N-1}
  f_{\alpha}(x,t) \Theta(x,t)^{-\alpha/2}
  \He_{\alpha} \left( \frac{\xi - V(x,t)}{\sqrt{\Theta(x,t)}} \right)
  \cdot \frac{1}{\sqrt{2\pi \Theta(x,t)}}
  \exp \left( -\frac{|\xi - V(x,t)|^2}{2\Theta(x,t)} \right).
\end{equation}
The value of $V(x,t)$ will still be chosen as the velocity $v(x,t)$, which
describes the center of the distribution function, so that $f_1 \equiv 0$.
However, to get convergence, we need to choose $\Theta(x,t)$ which can better
reflect the decay rate in the tail of the distribution function, so that
Condition C3 can be respected. By Condition C2, the quantity $\Theta(x,t)$ can
only be constructed from the first $N$ moments. While Grad's moment method only
uses the second moment to find $\Theta$, we would like to consider using the
highest moment to define $\Theta$, which better describes the behavior of the
tail. Based on this idea, we can figure out the precise definition of $\Theta$
by Condition C1, meaning that $\Theta$ must equal $\theta$ for the equilibrium
distribution function. For Maxwellians defined in \eqref{eq:Maxwellian}, the
highest moment ($(N-1)$th moment) is
\begin{displaymath}
\int_{\mathbb{R}} (\xi - v)^{N-1} \mathcal{M}(\xi) \,\mathrm{d}\xi =
\left\{ \begin{array}{ll}
  (N-2)!! \rho \theta^{(N-1)/2}, & \text{if } N \text{ is odd}, \\
  0, & \text{if } N \text{ is even}.
\end{array} \right.
\end{displaymath}
When $N$ is odd, we see that for Maxwellians,
\begin{displaymath}
\theta = \left( \frac{1}{(N-2)!!\rho} \int_{\mathbb{R}}
  (\xi - v)^{N-1} \mathcal{M}(\xi) \,\mathrm{d}\xi \right)^{2/(N-1)}.
\end{displaymath}
Therefore, for any distribution function $f(\xi)$, we choose to define
\begin{equation} \label{eq:Theta}
\Theta = \left( \frac{1}{(N-2)!!\rho} \int_{\mathbb{R}}
  (\xi - v)^{N-1} f(\xi) \,\mathrm{d}\xi \right)^{2/(N-1)},
\end{equation}
so that when $f(\xi)$ happens to be a Maxwellian, the value of $\Theta$
coincides with $\theta$, and this Maxwellian can be exactly represented by the
ansatz \eqref{eq:new_ansatz} with $f_1 = f_2 = \cdots = f_{N-1} = 0$. When $N$
is even, the highest moment for Maxwellians equal zero, which does not provide
any information for us to define $\Theta$. In this case, we can take the
$(N-2)$th moment instead and define $\Theta$ by
\begin{displaymath}
\Theta = \left( \frac{1}{(N-3)!!\rho} \int_{\mathbb{R}}
  (\xi - v)^{N-2} f(\xi) \,\mathrm{d}\xi \right)^{2/(N-2)},
\end{displaymath}
In this paper, we focus only on the case of odd $N$.

By choosing \eqref{eq:Theta} in \eqref{eq:ansatz}, we can get another
constraint of the coefficients other than $f_1 = 0$. Taking the $(N-1)$th
moment for \eqref{eq:ansatz}, we have
\begin{displaymath}
\int_{\mathbb{R}} (\xi - v)^{N-1} f(\xi) \,\mathrm{d}\xi
  = \sum_{\substack{\alpha = 0\\ \alpha \text{ is even}}}^{N-1}
  \frac{(N-1)!}{(N-1-\alpha)!!} f_{\alpha} \Theta^{(N-1-\alpha)/2}.
\end{displaymath}
Combining the above equation and \eqref{eq:Theta}, one finds that
\begin{equation} \label{eq:constraint}
\sum_{\substack{\alpha = 2\\ \alpha \text{ is even}}}^{N-1}
  \frac{f_{\alpha} \Theta^{(N-1-\alpha)/2}}{(N-1-\alpha)!!} = 0,
\end{equation}
where we have used the fact that $\rho = f_0$. This also confirms that
Condition C2 holds for this ansatz, since there are essentially only $N$
parameters in the ansatz \eqref{eq:new_ansatz}, which can be fully determined
by the $N$ moments.

The verification of Condition C3 is also straightforward. Given the sum of a
number of Gaussians
\begin{displaymath}
f(\xi) = \sum_{k=1}^K \frac{\rho_k}{\sqrt{2\pi \theta_k}}
  \exp \left( -\frac{(\xi - v_k)^2}{2\theta_k} \right),
\end{displaymath}
we have
\begin{displaymath}
\int_{\mathbb{R}} (\xi-v)^{N-1} f(\xi) \,\mathrm{d}\xi
  = (N-2)!! \sum_{k=1}^K \rho_k Q_k(\theta_k),
\end{displaymath}
where $Q_k(\cdot)$ are monic polynomials of degree $(N-1)/2$:
\begin{displaymath}
Q_k(x) = \sum_{\substack{\alpha=0\\ \alpha \text{ is even}}}^{N-1} 
  \begin{pmatrix} N-1 \\ \alpha \end{pmatrix} (v_k - v)^{\alpha}
  \frac{(N-2-\alpha)!!}{(N-2)!!} x^{(N-1-\alpha)/2}.
\end{displaymath}
Without loss of generality, we assume $\theta_1 \le \theta_2 \le \cdots \le
\theta_K$. Then it can be directly verified that
\begin{displaymath}
\lim_{N \rightarrow +\infty}
  \left( \sum_{k=1}^K \frac{\rho_k}{\rho} P_k(\theta_k) \right)^{2/(N-1)}
= \theta_K.
\end{displaymath}
This means for sufficiently large $N$, the slowest decay of the Gaussian can be
captured, as ensures the convergence of the ansatz.

It is worth mentioning that such a choice of $\Theta$ can capture the tail of a
much wider range of distribution functions. It can be shown that if the slowest
decay of the tail behaves like $\exp\big({-}\xi^2/(2\vartheta) \big)$, then
when $N$ tends to infinity, the value of $\Theta$ always tends to $\vartheta$.
This indicates that such a method is not specially designed for the shock
structure problem. Rigorous analysis of the convergence and wider applications
of this expansion are included in our ongoing works.

Based on this ansatz, we will apply the methodology introduced in
\cite{Cai2015a} to obtain a hyperbolic system. The details will be introduced
in the following section.

\subsubsection{System for the coefficients} \label{sec:system}
The reference \cite{Cai2015a} provides a general framework to derive hyperbolic
reduced models from the kinetic equations. Our ansatz of the distribution
function \eqref{eq:ansatz} fits this framework perfectly, so that we can
directly follow the steps therein the derive the equations. The details are
listed below:
\begin{itemize}
\item Compute time and spatial derivatives:
  \begin{align*}
  \frac{\partial f}{\partial t} &= \sum_{\alpha=0}^{N+1} \left(
    \frac{\partial f_{\alpha}}{\partial t} +
    \frac{\partial u}{\partial t} f_{\alpha-1} +
    \frac{1}{2} \frac{\partial \Theta}{\partial t} f_{\alpha-2}
  \right) \mathcal{H}_{\alpha}^{[\Theta(x,t)]}
    \left( \frac{\xi - v(x,t)}{\sqrt{\Theta(x,t)}} \right), \\
  \frac{\partial f}{\partial x} &= \sum_{\alpha=0}^{N+1} \left(
    \frac{\partial f_{\alpha}}{\partial x} +
    \frac{\partial u}{\partial x} f_{\alpha-1} +
    \frac{1}{2} \frac{\partial \Theta}{\partial x} f_{\alpha-2}
  \right) \mathcal{H}_{\alpha}^{[\Theta(x,t)]}
    \left( \frac{\xi - v(x,t)}{\sqrt{\Theta(x,t)}} \right),
  \end{align*}
  where $f_{\alpha}$ is regarded as zero if $\alpha < 0$ or $\alpha \geqslant
  N$, and
  \begin{displaymath}
  \mathcal{H}_{\alpha}^{[\Theta(x,t)]}(c) = \Theta(x,t)^{-\alpha/2}
  \He_{\alpha}(c)
  \cdot \frac{1}{\sqrt{2\pi \Theta(x,t)}} \exp \left( -\frac{c^2}{2} \right).
  \end{displaymath}
\item Truncate the series:
  \begin{align}
  \label{eq:dfdt}
  \frac{\partial f}{\partial t} & \approx \sum_{\alpha=0}^{N-1} \left(
    \frac{\partial f_{\alpha}}{\partial t} +
    \frac{\partial v}{\partial t} f_{\alpha-1} +
    \frac{1}{2} \frac{\partial \Theta}{\partial t} f_{\alpha-2}
  \right) \mathcal{H}_{\alpha}^{[\Theta(x,t)]}
    \left( \frac{\xi - v(x,t)}{\sqrt{\Theta(x,t)}} \right), \\
  \label{eq:dfdx}
  \frac{\partial f}{\partial x} & \approx \sum_{\alpha=0}^{N-1} \left(
    \frac{\partial f_{\alpha}}{\partial x} +
    \frac{\partial v}{\partial x} f_{\alpha-1} +
    \frac{1}{2} \frac{\partial \Theta}{\partial x} f_{\alpha-2}
  \right) \mathcal{H}_{\alpha}^{[\Theta(x,t)]}
    \left( \frac{\xi - v(x,t)}{\sqrt{\Theta(x,t)}} \right).
  \end{align}
  Here the step \eqref{eq:dfdx} does not exist in Grad's moment method, as is
  the reason for the loss of hyperbolicity.
\item Approximate the advection term: Define
  \begin{displaymath}
  \mathcal{F}_{\alpha} = 
    \frac{\partial f_{\alpha}}{\partial x} +
    \frac{\partial v}{\partial x} f_{\alpha-1} +
    \frac{1}{2} \frac{\partial \Theta}{\partial x} f_{\alpha-2}.
  \end{displaymath}
  Then
  \begin{equation} \label{eq:xi_dfdx}
  \begin{split}
  \xi \frac{\partial f}{\partial x} & \approx \sum_{\alpha=0}^{N-1}
    \mathcal{F}_{\alpha} \xi \mathcal{H}_{\alpha}^{[\Theta(x,t)]}
    \left( \frac{\xi - v(x,t)}{\sqrt{\Theta(x,t)}} \right) \\
  & = \sum_{\alpha=0}^N
    [\Theta \mathcal{F}_{\alpha-1} + v \mathcal{F}_{\alpha}
      + (\alpha+1) \mathcal{F}_{\alpha + 1}]
    \mathcal{H}_{\alpha}^{[\Theta(x,t)]}
      \left( \frac{\xi - v(x,t)}{\sqrt{\Theta(x,t)}} \right)
    \quad [\mathcal{F}_{-1} = \mathcal{F}_N = \mathcal{F}_{N+1} = 0] \\
  & \approx \sum_{\alpha=0}^{N-1}
    [\Theta \mathcal{F}_{\alpha-1} + v \mathcal{F}_{\alpha}
      + (\alpha+1) \mathcal{F}_{\alpha + 1}]
    \mathcal{H}_{\alpha}^{[\Theta(x,t)]}
      \left( \frac{\xi - v(x,t)}{\sqrt{\Theta(x,t)}} \right)
    \quad [\mathcal{F}_{-1} = \mathcal{F}_N = 0]
  \end{split}
  \end{equation}
\item Approximate the collision term:
  \begin{displaymath}
  \begin{split}
  \frac{1}{\tau} (\mathcal{M} - f) &= \frac{1}{\tau} 
    \sum_{\alpha=0}^{+\infty}
    \left[ \frac{1+(-1)^{\alpha}}{2} \frac{\rho}{\alpha!!}(\theta - \Theta)^{\alpha/2} - f_{\alpha} \right]
    \mathcal{H}_{\alpha}^{[\Theta(x,t)]}
      \left( \frac{\xi - v(x,t)}{\sqrt{\Theta(x,t)}} \right) \\
  &\approx \frac{1}{\tau} \sum_{\alpha=0}^{N-1}
    \left[ \frac{1+(-1)^{\alpha}}{2} \frac{\rho}{\alpha!!}(\theta - \Theta)^{\alpha/2} - f_{\alpha} \right]
    \mathcal{H}_{\alpha}^{[\Theta(x,t)]}
      \left( \frac{\xi - v(x,t)}{\sqrt{\Theta(x,t)}} \right),
  \end{split}
  \end{displaymath}
  where the temperature $\theta$ can be computed by
  \begin{equation} \label{eq:temperature}
  \theta = \frac{1}{\rho} \int_{\mathbb{R}} |\xi - v|^2 f(\xi) \,\mathrm{d}\xi
    = \Theta + \frac{2f_2}{\rho}.
  \end{equation}
\item Assemble all the terms and formulate the equations:
  \begin{equation} \label{eq:form1}
  \begin{split}
  \frac{\mathrm{d} f_{\alpha}}{\mathrm{d} t} +
    \frac{\mathrm{d} v}{\mathrm{d} t} f_{\alpha-1} +
    \frac{1}{2} \frac{\mathrm{d} \Theta}{\mathrm{d} t} f_{\alpha-2}
  + \Theta \mathcal{F}_{\alpha-1}
  + (1-\delta_{N-1,\alpha})(\alpha+1) \mathcal{F}_{\alpha+1} \\
    = \frac{1}{\tau} \left[ \frac{1+(-1)^{\alpha}}{2} 
      \frac{\rho}{\alpha!!}(\theta - \Theta)^{\alpha/2} - f_{\alpha} \right],
  \quad \forall \alpha = 0,1,\cdots,N-1.
  \end{split}
  \end{equation}
\end{itemize}
The final moment equations are given in \eqref{eq:form1}, together with the
relations $f_0 = \rho$, $f_1 = 0$ and
\eqref{eq:constraint}\eqref{eq:temperature}. For simplicity, we will refer to
this method as the \emph{highest-moment-based moment method} (HMBMM). In the
following section, several properties of the HMBMM will be discussed.

\subsection{Properties of the moment systems}
By the derivation of the moment system, it can be expected that this moment
hierarchy is more promising to get convergence to the Boltzmann equation for a
wider class of problems. Below we will focus on some other properties of the
equations, including its hyperbolicity, asymptotic limit, and the balance laws
of the moments.

\subsubsection{Hyperbolicity of moment equations}
The hyperbolicity of HMBMM can be observed by rewriting the equations using
matrices and vectors.  Let $\bw = (f_0, v, f_2, \cdots, f_{N-2}, \Theta)^T$ be
the vector of unknowns, and
\begin{displaymath}
\bcH(\xi) = \left(
  \mathcal{H}_0^{[\Theta]} \left( \frac{\xi - v}{\sqrt{\Theta}} \right),
  \mathcal{H}_1^{[\Theta]} \left( \frac{\xi - v}{\sqrt{\Theta}} \right), \cdots,
  \mathcal{H}_{N-1}^{[\Theta]} \left( \frac{\xi - v}{\sqrt{\Theta}} \right)
\right)^T
\end{displaymath}
be the vector of basis functions. Then \eqref{eq:dfdt}\eqref{eq:dfdx} can be
rewritten as
\begin{displaymath}
  \frac{\partial f}{\partial t} \approx
    \bcH(\xi)^T \mathbf{D} \frac{\partial \bw}{\partial t}, \qquad
  \frac{\partial f}{\partial x} \approx
    \bcH(\xi)^T \mathbf{D} \frac{\partial \bw}{\partial x},
\end{displaymath}
where $\mathbf{D}$ is a certain square matrix depending on $\bw$. Here the
vector $\mathbf{D} \partial_x \bw$ equals $(\mathcal{F}_0, \cdots,
\mathcal{F}_{N-1})$. Therefore \eqref{eq:xi_dfdx} can be rewritten as
\begin{displaymath}
\xi \frac{\partial f}{\partial x} \approx
    \bcH(\xi)^T \mathbf{M} \mathbf{D} \frac{\partial \bw}{\partial x},
\end{displaymath}
where the matrix $\mathbf{M}$ is tridiagonal:
\begin{displaymath}
\mathbf{M} = \begin{pmatrix}
  v & 1 \\
  \Theta & v & 2 \\
  & \Theta & v & 3 \\
  & & \ddots & \ddots & \ddots \\
  & & & \Theta & v & N-1 \\
  & & & & \Theta & v
\end{pmatrix}.
\end{displaymath}
Such a derivation shows that the moment system can be written as
\begin{displaymath}
\mathbf{D} \frac{\partial \bw}{\partial t} + 
  \mathbf{M} \mathbf{D} \frac{\partial \bw}{\partial x}
= \boldsymbol{\mathcal{S}},
\end{displaymath}
where the right-hand side $\boldsymbol{\mathcal{S}}$ represents the collision
term, which does not affect the hyperbolicity. This formula shows that the
system is hyperbolic if $\mathbf{M}$ is real diagonalizable, which can be
easily observed since the characteristic polynomial of $\mathbf{M}$ is
\begin{displaymath}
\det(\lambda \mathbf{I} - \mathbf{M}) =
  \He_N \left( \frac{\lambda - v}{\sqrt{\Theta}} \right),
\end{displaymath}
and all the roots of the Hermite polynomial are real and distinct if $\Theta$
is positive, as implies the hyperbolicity of the moment equations. Compared
with the hyperbolic moment equations introduced in Section \ref{sec:hyp}, HMBMM
is more nonlinear since the characteristic speeds involve $\Theta$, which is
expressed by the highest moment.

\subsubsection{Asymptotic limits}
We would now like to check whether the Euler and Navier-Stokes limits can be
preserved by this moment hierarchy. When $N=3$, it is easy to see that the
equations are identical to Euler equations of one-dimensional velocity.  Below
we only consider the case $N=5$. The equations with more moments can be
analyzed using the same technique.

For $N=5$, the moment equations \eqref{eq:form1} can be reformulated as
\begin{equation} \label{eq:1dsystem}
\begin{aligned}
\frac{\mathrm{d} f_0}{\mathrm{d} t} + f_0 \frac{\partial v}{\partial x} &= 0, \\
f_0 \frac{\mathrm{d} v}{\mathrm{d} t} + \Theta \frac{\partial f_0}{\partial x}
  + f_0 \frac{\partial \Theta}{\partial x} + 2\frac{\partial f_2}{\partial x} &= 0, \\
\frac{\mathrm{d} f_2}{\mathrm{d} t} +
  \frac{1}{2} f_0 \frac{\mathrm{d}\Theta}{\mathrm{d}t} +
  (3f_2 + \Theta f_0) \frac{\partial v}{\partial x} +
  3\frac{\partial f_3}{\partial x} &= 0, \\
\frac{\mathrm{d} f_3}{\mathrm{d} t} +
  f_2 \frac{\mathrm{d} v}{\mathrm{d} t} +
  \frac{1}{2} \Theta f_0 \frac{\partial \Theta}{\partial x} -
  \Theta \frac{\partial f_2}{\partial x} +
  4f_3\frac{\partial v}{\partial x} &= -\frac{f_3}{\tau}, \\
\frac{1}{2} \frac{\mathrm{d} f_2}{\mathrm{d} t} -
  \frac{f_3}{\Theta} \frac{\mathrm{d}v}{\mathrm{d}t} -
  f_2 \frac{\partial v}{\partial x} -
  \frac{\partial f_3}{\partial x} &=
-\frac{f_2}{2\tau} \left( 1 + \frac{f_2}{\Theta f_0} \right).
\end{aligned}
\end{equation}
Note that we have used the relation \eqref{eq:constraint} to eliminate the
variable $f_4$. We would like to first show that the system preserves Euler and
Navier-Stokes limits when $\tau$ is small. For any quantity $\varphi \in \{f_0,
v, f_2, f_3, \Theta\}$, we expand it in terms of $\tau$ as
\begin{displaymath}
  \varphi = \varphi^{(0)} + \tau \varphi^{(1)} + \tau^2 \varphi^{(2)} + \cdots.
\end{displaymath}
By the fourth equation of \eqref{eq:1dsystem}, it is clear that $f_3^{(0)} =
0$; and it can be seen from the last equation of \eqref{eq:1dsystem} that
\begin{displaymath}
  f_2^{(0)} \left( 1 + \frac{f_2^{(0)}}{\Theta^{(0)} f_0^{(0)}} \right) = 0.
\end{displaymath}
If $f_2^{(0)} = -f_0^{(0)} \Theta^{(0)}$, the equation \eqref{eq:temperature}
shows that the leading order of temperature is negative, as is a non-physical
state. Therefore $f_2^{(0)} = 0$. Thus the zeroth-order approximation of the
above system can be written from the first three equations of
\eqref{eq:1dsystem}:
\begin{equation} \label{eq:1dEuler}
\begin{aligned}
\frac{\mathrm{d} f_0}{\mathrm{d} t} + f_0 \frac{\partial v}{\partial x} &= 0, \\
f_0 \frac{\mathrm{d} v}{\mathrm{d} t} + \Theta \frac{\partial f_0}{\partial x}
  + f_0 \frac{\partial \Theta}{\partial x} &= 0, \\
\frac{\mathrm{d}\Theta}{\mathrm{d}t} +
  2\Theta \frac{\partial v}{\partial x} &= 0,
\end{aligned}
\end{equation}
which is exactly the Euler equations for one-dimensional physics.

To get the first-order limit, we adopt Chapman-Enskog expansion, so that
\begin{displaymath}
f_0^{(k)} = 0, \quad v^{(k)} = 0, \quad \Theta^{(k)} = 0,
  \qquad \text{for all } k \geqslant 1.
\end{displaymath}
By the fourth equation of \eqref{eq:1dsystem}, we see that
\begin{displaymath}
f_3^{(1)} = -\frac{1}{2} \Theta f_0 \frac{\partial \Theta}{\partial x},
\end{displaymath}
which gives the Fourier law for heat conduction. The last equation in
\eqref{eq:1dsystem} shows that $f_2^{(1)} = 0$. Therefore the first-order
asymptotic limit can be obtained by changing the last equation of
\eqref{eq:1dEuler} to 
\begin{equation} \label{eq:1dNS}
\frac{\mathrm{d}\Theta}{\mathrm{d}t} +
  2\Theta \frac{\partial v}{\partial x} =
  \frac{3}{f_0} \frac{\partial}{\partial x} \left(
    \tau f_0 \Theta \frac{\partial \Theta}{\partial x}
  \right),
\end{equation}
which corresponds to the Navier-Stokes equations for one-dimensional physics.

When we solve Euler equations or Navier-Stokes equations numerically, most of
the time, we do not use the forms of equations like \eqref{eq:1dEuler} or
\eqref{eq:1dNS}. Instead, we prefer writing the equations in the form of
conservation laws so that one can easily apply conservative numerical schemes.
Such a form for this moment hierarchy will be explored in the next section.

\subsubsection{Balance laws of the moments}
For the convenience of numerical computation, we would like to write the moment
systems in the form similar to \eqref{eq:mom_sys}, which requires us to define
the moments
\begin{equation} \label{eq:mnts}
M_k = \langle \xi^k f \rangle, \qquad k = 0,1,\cdots,N-1.
\end{equation}
For simplicity, again we only write down the equations for $N = 4$. By the
ansatz \eqref{eq:new_ansatz}, we can establish the relation between the moments
\eqref{eq:mnts} and the variables $f_0, v, f_2, f_3, \Theta$:
\begin{gather*}
M_0 = f_0, \qquad M_1 = v f_0, \qquad M_2 = f_0(v^2 + \Theta) + 2f_2, \\
M_3 = v f_0(v^2 + 3\Theta) + 6(v f_2 + f_3), \qquad
M_4 = f_0(v^4 + 6v^2 \Theta + 3\Theta^2) + 12v(vf_2 + 2f_3).
\end{gather*}
Thus the moment system \eqref{eq:1dsystem} can be written as
\begin{align*}
\frac{\partial M_0}{\partial t} + \frac{\partial M_1}{\partial x} &= 0, \\
\frac{\partial M_1}{\partial t} + \frac{\partial M_2}{\partial x} &= 0, \\
\frac{\partial M_2}{\partial t} + \frac{\partial M_3}{\partial x} &= 0, \\
\frac{\partial M_3}{\partial t} + \frac{\partial M_4}{\partial x}
  &= -\frac{1}{\tau} \frac{2M_1^3-3M_0 M_1 M_2 + M_0^2 M_3}{M_0^2}, \\
\frac{\partial M_4}{\partial t} + \frac{\partial M_5}{\partial x}
  +60 \left(\Theta f_2 \frac{\partial v}{\partial x} -
    f_3 \frac{\partial \Theta}{\partial x} \right)
  &= -\frac{1}{\tau} \frac{2M_1^4-3M_0^2 M_2^2 + M_0^3 M_4}{M_0^3},
\end{align*}
where
\begin{displaymath}
M_5 = \langle \xi^5 f \rangle
  = vf_0 (v^4 + 10 v^2 \Theta + 15 \Theta^2)
    + 20 v^2 (v f_2 + 3f_3) + 60 \Theta f_3.
\end{displaymath}
It is straightforward to verify that this system is equivalent to
\eqref{eq:1dsystem}. The first three equations are the conservation laws of
mass, momentum and energy, and the fourth equation is a balance law for the
third moment. These four equations can be considered as ``exact'' since they
can be obtained directly by taking moments of the Boltzmann-BGK equation. The
last equation contains 1) the moment closure given by the defintion of $M_5$;
2) a non-conservative product $\Theta f_2 \partial_x v - f_3 \partial_x
\Theta$, which helps maintain the hyperbolicity of the system. The
non-conservative product ruins the form of balance law in the last equation, as
can be considered as the compromise to obtain a hyperbolic structure.

Such a formulation can be generalized to the case of a larger $N$. The
resulting $N$-moment system always includes three conservation laws, $N-4$
balance laws, and one equation that involves the moment closure and the
hyperbolicity fix. Due to the exactness of the first $N-1$ equations, it can
also be expect that the asymptotic limit can be well preserved if the last
equation does not spoil the order of magnitude. Also, such a form is more
suitable for numerical simulation. In our numerical method, a standard finite
volume method with Lax-Friedrichs numerical flux is applied, and the
non-conservative product in the last equation is discretized by simple central
differences. The results will be reported in the next subsection.

\subsection{Numerical results}
\begin{figure}[!ht]
\centering
\subfloat[$\mathit{Ma} = 1.4$]{%
  \label{fig:New_Ma1.4}
  \includegraphics[width=.45\textwidth]{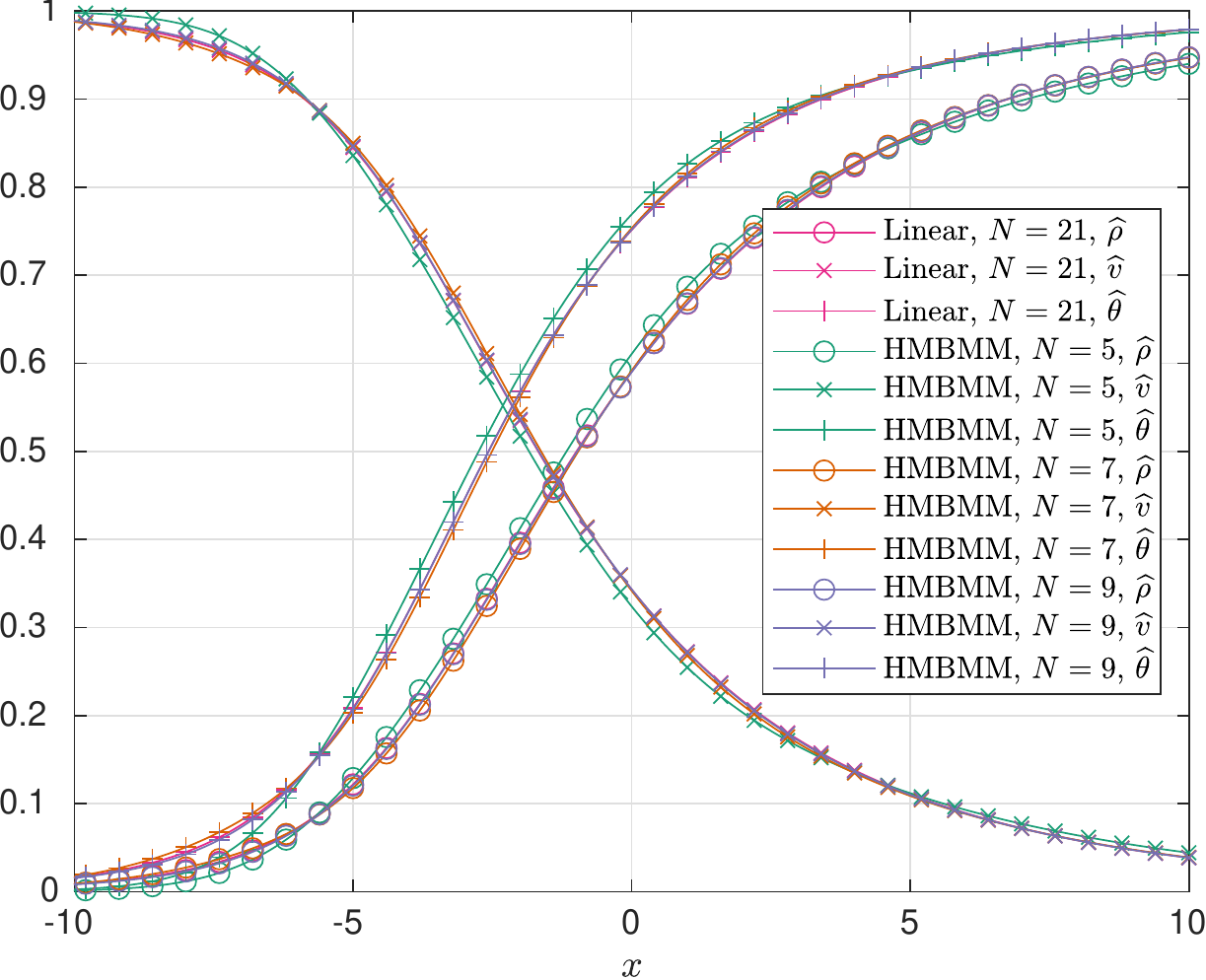}}
\qquad
\subfloat[$\mathit{Ma} = 2.0$]{%
  \label{fig:New_Ma2}
  \includegraphics[width=.45\textwidth,clip]{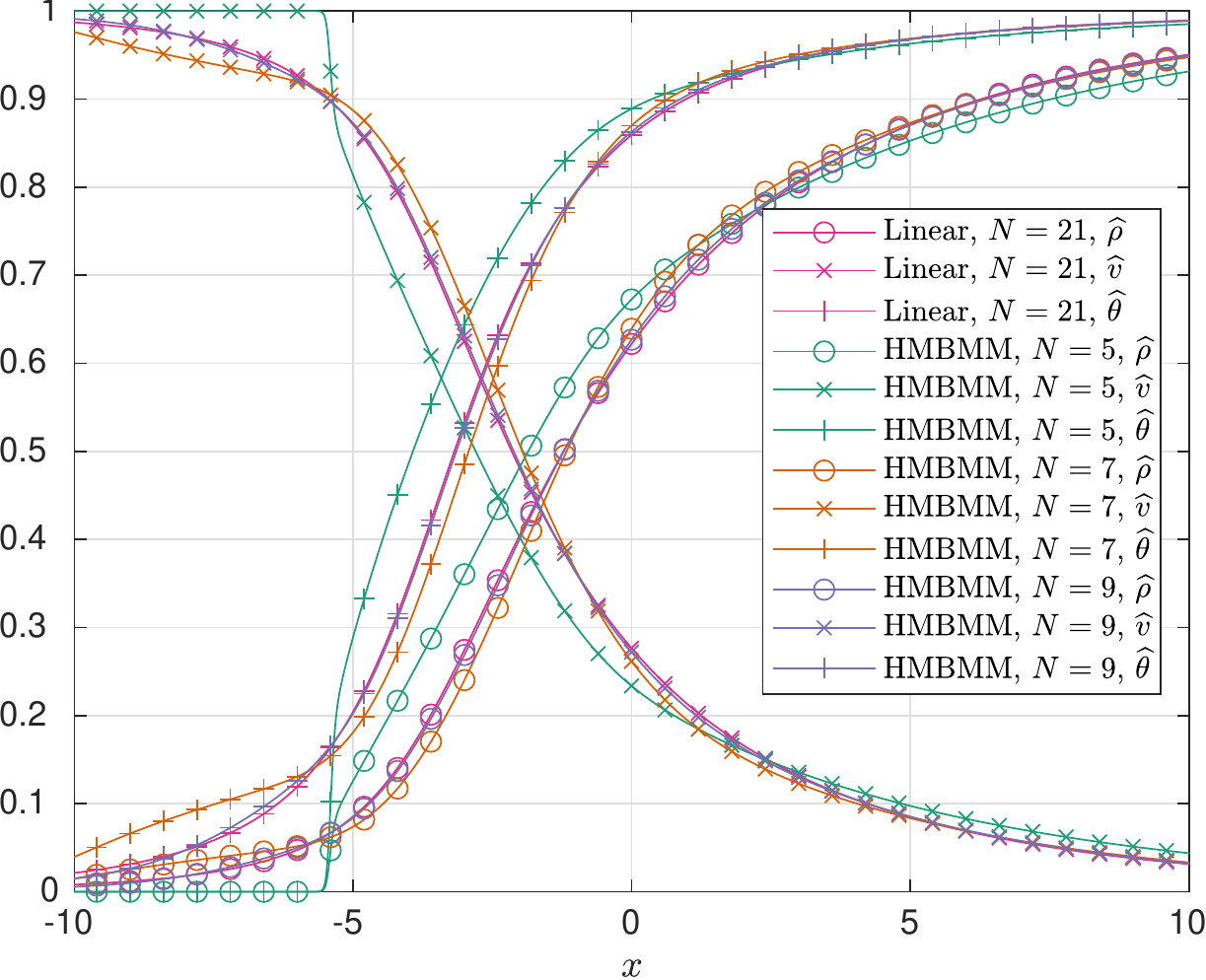}}
  \caption{Numerical results for $\mathit{Ma} = 1.4$ and $\mathit{Ma} = 2.0$}
\label{fig:New}
\end{figure}

We are going to show some numerical results for shock structure computations
using HMBMM. Again we consider the Boltzmann-BGK equation with initial
condition \eqref{eq:init}. Numerical results for Mach numbers $1.4$ and $2.0$
are plotted in Figure \ref{fig:New}.  For $\mathit{Ma} = 1.4$, the quality of
the solution is equally good as the hyperbolic moment method shown in Figure
\ref{fig:Hyp_Ma1.4}. For $\mathit{Ma} = 2.0$, the simulations for all the three
cases $N=5,7,9$ are stable, and the results of $N=9$ almost coincide with the
reference solution. More importantly, the convergence of the numerical solution
as $N$ gets larger can be clearly observed from the numerical results.

The numerical results show that HMBMM does not solve the subshock problem,
since in front of the shock wave, when the distribution function is close to
the Maxwellian, the value of $\Theta$ is close to the temperature $\theta$.
Therefore the characteristic speeds of HMBMM are almost the same as those of
Grad's moment method or the hyperbolic moment method in this region.
Nevertheless, for HMBMM, we can increase the number of moments to resolve the
subshock issue.

\begin{figure}[!ht]
\centering
\subfloat[$\mathit{Ma} = 1.4$]{%
  \label{fig:Theta_Ma1.4}
  \includegraphics[width=.45\textwidth]{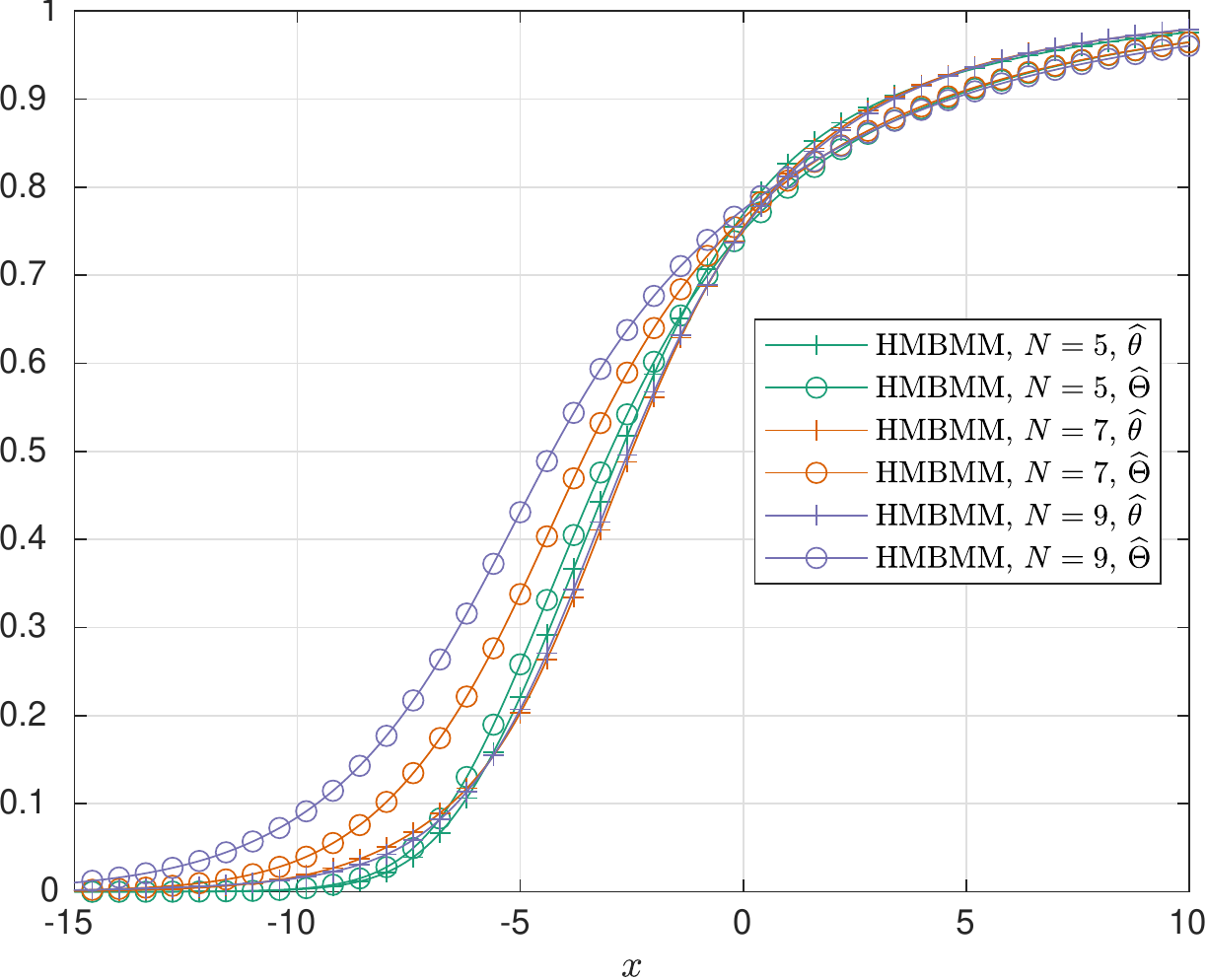}}
\qquad
\subfloat[$\mathit{Ma} = 2.0$]{%
  \label{fig:Theta_Ma2}
  \includegraphics[width=.45\textwidth,clip]{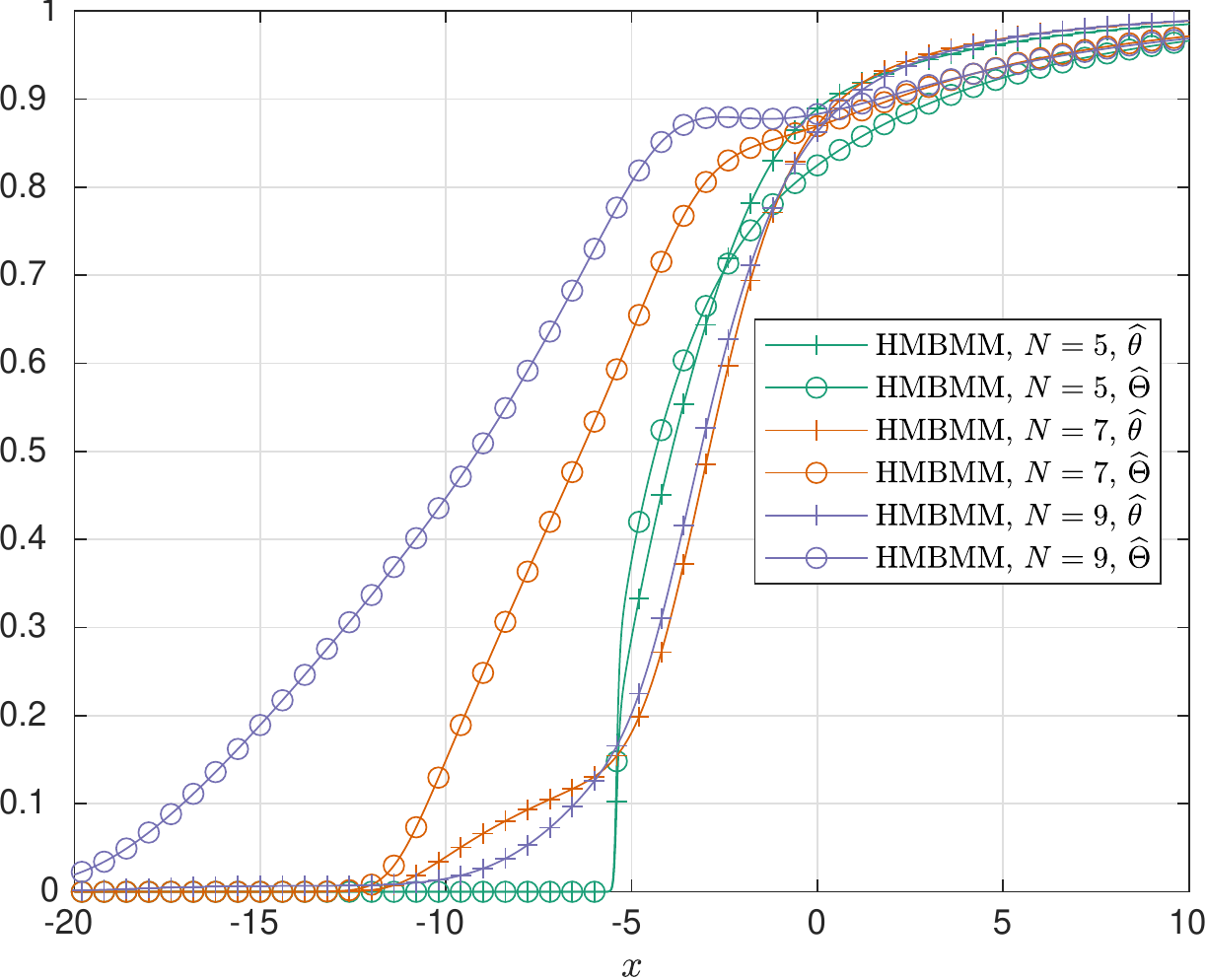}}
  \caption{Comparison between $\theta$ and $\Theta$ in one-dimensional HMBMM}
\label{fig:Theta}
\end{figure}

To better understand the difference between the hyperbolic moment method and
the highest-moment based moment method, we show the difference between $\theta$
and $\Theta$ in Figure \ref{fig:Theta}, where $\widehat{\Theta}$ is defined by
\begin{displaymath}
\widehat{\Theta}(x) =
  \frac{\Theta(x) - \theta_0(-\infty)}{\theta_0(+\infty) - \theta_0(-\infty)}.
\end{displaymath}
It can be observed that for most part of the shock wave, the value of $\Theta$
is greater than the temperature $\theta$, which better fits the tail of the
distribution function. In fact, if the solution of the shock structure takes
the form \eqref{eq:Mott-Smith} everywhere, we should expect that $\Theta$ tends
to $\theta_0(+\infty)$ everywhere as $N$ approaches infinity, which guarantees
the convergence. Such a trend is validated by our numerical results, where
$\Theta$ increases as $N$ increases. Figure \ref{fig:Theta_Ma2} also explains
why the temperature profile of $N=7$ shows significant deviation from the
reference solution in the front part of the shock wave with $\mathit{Ma} = 2$:
when $N=7$, the parameter $\Theta$ is given by the sixth moment, using which
one tends to underestimate the decay rate at the tail of the distribution
function, leading to a relatively poor quality of approximation using
\eqref{eq:new_ansatz}.

\section{Highest-moment-based moment method: Three-dimensional case} \label{sec:3d}
The methodology for the one-dimensional case can be generalized to the
three-dimensional case without much difficulty, although the derivation is
expected to be more tedious. Below we will only brief the idea of the
generalization and provide the final moment equations, followed by some
numerical tests to verify its capability in computing the high-speed shock
structures.

\subsection{Ansatz and the moment equations}
In three-dimensional case, we can again adopt the ansatz of Grad's moment
method and only change the scaling parameter in the basis functions.
Specifically, we assume
\begin{equation} \label{eq:3d_ansatz}
\begin{split}
f(\bx,\bxi,t) &= \sum_{l=0}^{L-1} \sum_{m=-l}^l \sum_{n=0}^{N_l-1}
  f_{lmn}(\bx,t) [\Theta(\bx,t)]^{-(l+2n)/2} \times \\
& \quad p_{lmn} \left( \frac{\bxi - \bv(\bx,t)}{\sqrt{\Theta(\bx,t)}} \right)
  \cdot \frac{1}{[2\pi \Theta(\bx,t)]^{3/2}} \exp \left(
    -\frac{|\bxi - \bv(\bx,t)|^2}{2\Theta(\bx,t)}
  \right),
\end{split}
\end{equation}
where the parameters $L$ and $(N_0, N_1, \cdots, N_{L-1})$ specify the moments
included in the moment system, and $p_{lmn}$ are orthogonal polynomials in the
three-dimensional space. This choice of orthogonal polynomials follows
\cite{Kumar1966, Cai2015}, which is based on the spherical coordinates in the
three-dimensional space. The degree of the polynomial $p_{lmn}$ is $l+2n$, and
the expressions are provided in Appendix \ref{sec:poly}. Since $\bv(\bx,t)$ is
the velocity, the coefficients satisfy
\begin{displaymath}
f_{1m0} = 0, \qquad m = -1,0,1.
\end{displaymath}
In \cite{Cai2018}, the authors suggest choosing $N_0 \geqslant N_1 \geqslant
\cdots \geqslant N_{L-1}$ so that the wall boundary conditions can be
formulated appropriately. Some special choices of the parameters and the
corresponding number of moments are given below:
\begin{itemize}
\item $L = 2$, $N_0 = 2$, $N_1 = 1$: $5$ moments.
\item $L = 3$, $N_0 = 2$, $N_1 = 1$, $N_2 = 1$: $10$ moments.
\item $L = 3$, $N_0 = N_1 = 2$, $N_2 = 1$: $13$ moments.
\item $L = 4$, $N_0 = N_1 = 2$, $N_2 = N_3 = 1$: $20$ moments.
\item $L = 4$, $N_0 = 3$, $N_1 = N_2 = 2$, $N_3 = 1$: $26$ moments.
\end{itemize}
The key factor in this method is the choice of $\Theta$. The derivation is
parallel to the one-dimensional case, which takes into account the conditions
C1--C3. To fulfill Condition C3, we define $\Theta$ based on the following
moment:
\begin{equation} \label{eq:highest_moment}
\int_{\mathbb{R}^3} |\bxi - \bv|^{2K} f(\bxi) \,\mathrm{d}\bxi,
\end{equation}
where $K$ should be chosen as large as possible to best capture the decay of
the tail. According to Condition C2, we need that the above quantity can be
expressed by the moments included in the system. By the definition of $p_{lmn}$
given in \eqref{eq:p_lmn}, we can derive that
\begin{displaymath}
p_{00n}(\bc) = \sqrt{\frac{(2n)!!}{(2n+1)!!}}
  \sum_{i=0}^n (-1)^i \begin{pmatrix} n+1/2 \\ n-i \end{pmatrix}
  \frac{|\bc|^{2i}}{(2i)!!}, \qquad n = 0,1,\cdots,N_0-1.
\end{displaymath}
Therefore using the variables $f_{lmn}$ appearing in \eqref{eq:highest_moment},
we can construct the moments \eqref{eq:highest_moment} for all $K =
0,1,\cdots,N_0-1$. Thus the greatest choice of $K$ is $N_0-1$. The expression 
of $\Theta$ should be formulated following Condition C1. For Maxwellians, the
moment \eqref{eq:highest_moment} is
\begin{equation}
\int_{\mathbb{R}^3} |\bxi - \bv|^{2(N_0-1)}
  \frac{\rho}{(2\pi\theta)^{3/2}} \exp \left( -\frac{|\bxi - \bv|^2}{2\theta} \right)
  \,\mathrm{d}\bxi = (2N_0-1)!! \rho \theta^{2(N_0-1)}.
\end{equation}
This inspires us to define $\Theta$ by
\begin{equation} \label{eq:3d_Theta}
\Theta = \left( \frac{1}{(2N_0-1)!! \rho} \int_{\mathbb{R}^3}
  |\bxi - \bv|^{2(N_0-1)} f(\bxi) \,\mathrm{d}\bxi \right)^{1/(2N_0-2)}.
\end{equation}
By inserting \eqref{eq:3d_ansatz} into \eqref{eq:3d_Theta}, we obtain another
constraint of the coefficients:
\begin{equation} \label{eq:3d_constraint}
\sum_{n=1}^{N_0-1} I_n \Theta^{N_0-1-n} f_{00n} = 0,
\end{equation}
where
\begin{displaymath}
I_n = \int_{\mathbb{R}^3} |\bv|^{2(N_0-1)} p_{00n}(\bc) \cdot \frac{1}{(2\pi)^{3/2}}
  \exp \left( -\frac{|\bc|^2}{2} \right) \,\mathrm{d}\bc,
\end{displaymath}
which satisfies
\begin{displaymath}
I_0 = (2N_0-1)!!, \qquad I_n = (n-N_0) \sqrt{\frac{1}{(N_0-1)(N_0-1/2)}} I_{n-1},
  \quad \forall n > 0.
\end{displaymath}
Similar to the one-dimensional case, one can also prove that for the sum of any
number of any Maxwellians, the ansatz \eqref{eq:3d_ansatz} converges as $N
\rightarrow \infty$. Here
\begin{displaymath}
N = \min\{l+2(N_l-1) \mid l = 0,1,\cdots,L-1\},
\end{displaymath}
which is the maximum degree of polynomials fully included in the ansatz. Thus
Condition C3 is again fulfilled.

Based on this ansatz, the derivation of moment equations is similar to the 1D
case. The general form of the final equations is
\begin{equation} \label{eq:3d_system}
\begin{split}
& \frac{\mathrm{d}f_{lmn}}{\mathrm{d}t} + S_{lmn} + T_{lmn} = \\
& \qquad \frac{\Theta}{\mu}
  \sum_{l_1=0}^{L-1} \sum_{n_1=0}^{N_{l_1}-1}
  \sum_{l_2=0}^{L-1} \sum_{n_2=0}^{N_{l_2}-1}
  \sum_{\substack{m_1 = -l_1,\cdots,l_1 \\ m_2 = -l_2, \cdots, l_2 \\ m_1 + m_2 = m}}
  A_{lmn}^{l_1 m_1 n_1, l_2 m_2 n_2} \Theta^{(l+l_1-l_2)/2+(n-n_1-n_2)}
  f_{l_1 m_1 n_1} f_{l_2 m_2 n_2}.
\end{split}
\end{equation}
Here $S_{lmn}$ includes material derivatives of $\bv$ and $\Theta$, and
$T_{lmn}$ includes spatial derivatives of the coefficients. The form of the
right-hand side comes from the quadratic form of the Boltzmann collision
operator \eqref{eq:quadratic}, and the coefficients $A_{lmn}^{l_1 m_1 n_1, l_2
m_2 n_2}$ specifies the collision kernel. The precise form of the equations
will be given in Appendix \ref{sec:STA}.

The properties such as hyperbolicity and asymptotic limits can again be derived
in the similar way to the one-dimensional case. Instead of carrying out the
derivation with lengthy equations, here we just summarize the results briefly.
The hyperbolicity of the equations can be observed by writing the equations in
the following form:
\begin{displaymath}
\mathbf{D} \frac{\partial \bw}{\partial t} + 
  \sum_{j=1}^3 \mathbf{M}_j \mathbf{D} \frac{\partial \bw}{\partial x_j}
= \boldsymbol{\mathcal{S}},
\end{displaymath}
and it can be demonstrated that each $\mathbf{M}_j$ is symmetric, which implies
the hyperbolicity. In fact, the system is \emph{symmetric hyperbolic}, which
can be observed by multiplying the above equation by $\mathbf{D}^T$. The
asymptotic limits can be studied by Chapman-Enskog expansion. Actually,
for systems with $5, 10, 13, 20$ moments, since $N_0=2$, we can see from
\eqref{eq:3d_Theta} that $\Theta = \theta$, and thus the ansatz
\eqref{eq:3d_ansatz} reduces to Grad's ansatz, so that the corresponding moment
system is identical to the hyperbolic moment equations with the same number of
moments. This also indicates that the Euler limit and the Navier-Stokes limit
can be preserved for sufficient number of moments. However, for the quadratic
collision operators, preserving higher-order limits such as Burnett and
super-Burnett is much more non-trivial just like in Grad's moment methods. The
balance-law form of the equations can also be derived by choosing moments
$M_{lmn} = \langle \overline{p_{lmn}(\bxi)} f(\bxi) \rangle$, where we can find
several equations including non-conservative terms to enforce the
hyperbolicity.

\subsection{Numerical method}
To carry out the simulation of HMBMM, we take the idea of the approach proposed
in \cite{Cai2018} to carry out the simulation. For simplicity, we assume that a
uniform grid with cell size $\Delta x$ is adopted, and at the $k$th time step,
the distribution function on the $j$th grid is denoted by $F^{j,k}(\bxi)$,
which is defined as
\begin{displaymath}
F^{j,k}(\bxi) = \sum_{l=0}^{L-1} \sum_{m=-l}^l \sum_{n=0}^{N_l-1}
  f_{lmn}^{j,k} \Psi^{j,k}(\bxi),
\end{displaymath}
where
\begin{displaymath}
\Psi^{j,k}(\bxi) = (\Theta^{j,k})^{-(l+2n)/2} \cdot
  p_{lmn}\left( \frac{\bxi - \bv^{j,k}}{\sqrt{\Theta^{j,k}}} \right)
  \frac{1}{[2\pi \Theta^{j,k}]^{3/2}}
  \exp \left( -\frac{|\bxi - \bv^{j,k}|^2}{2\Theta^{j,k}} \right).
\end{displaymath}
Our discretization is based on the finite volume scheme:
\begin{equation} \label{eq:intermediate}
F_*^{j,k+1}(\bxi) = F^{j,k}(\bxi) - \frac{\Delta t}{\Delta x}
  \left[ G_+^{j,k}(\bxi) - G_-^{j,k}(\bxi) \right]
  + \Delta t \sum_{l=0}^{L-1} \sum_{m=-l}^l \sum_{n=0}^{N_l-1}
  Q_{lmn}^{j,k} \Psi^{j,k}(\bxi),
\end{equation}
where $Q_{lmn}^{j,k}$ is the right-hand side of \eqref{eq:3d_system} with
$f_{l_1 m_1 n_1}$ and $f_{l_2 m_2 n_2}$ replaced by $f_{l_1 m_1 n_1}^{j,k}$ and
$f_{l_2 m_2 n_2}^{j,k}$, respectively, and $G_+^{j,k}(\bxi)$ and
$G_-^{j,k}(\bxi)$ are the numerical fluxes. To define the numerical fluxes, we
introduce the projection operator $\mathcal{P}^{j,k}$, so that for any
distribution function $g(\bxi)$, the function $\mathcal{P}^{j,k} g$ has the form
\begin{equation} \label{eq:projection}
\mathcal{P}^{j,k} g(\bxi) = \sum_{l=0}^{L-1} \sum_{m=-l}^l \sum_{n=0}^{N_l-1}
  \tilde{g}_{lmn} \Psi_{lmn}^{j,k}(\bxi),
\end{equation}
which satisfies
\begin{displaymath}
\left\langle
  \overline{p_{lmn}\left( \frac{\bxi - \bv^{j,k}}{\sqrt{\Theta^{j,k}}} \right)} g(\bxi)
\right\rangle = 
\left\langle
  \overline{p_{lmn}\left( \frac{\bxi - \bv^{j,k}}{\sqrt{\Theta^{j,k}}} \right)}
  \mathcal{P}^{j,k} g(\bxi)
\right\rangle
\end{displaymath}
for all $l = 0,1,\cdots,L-1$, $m = -l,\cdots,l$ and $n = 0,1,\cdots,N_l-1$.
This operator can be regarded as ``series truncation'' for the expansion of $g$
using basis functions $\Psi_{lmn}^{j,k}$. With this operator, we can write down
the numerical fluxes following the HLL scheme:
\begin{align*}
G_+^{j,k}(\bxi) &= \mathcal{P}^{j,k} \left(
  \frac{\lambda_R^{j+1/2,k} \xi_1 F^{j,k}(\bxi)
    - \lambda_L^{j+1/2,k} \xi_1 \mathcal{P}^{j,k} F^{j+1,k}(\bxi)
    + \lambda_R^{j+1/2,k} \lambda_L^{j+1/2,k}
      [\mathcal{P}^{j,k} F^{j+1,k}(\bxi) - F^{j,k}(\bxi)]
  }{\lambda_R^{j+1/2,k} - \lambda_L^{j+1/2,k}}
\right), \\
G_-^{j,k}(\bxi) &= \mathcal{P}^{j,k} \left(
  \frac{\lambda_R^{j-1/2,k} \xi_1 \mathcal{P}^{j,k} F^{j-1,k}(\bxi)
    - \lambda_L^{j-1/2,k} \xi_1 F^{j,k}(\bxi)
    + \lambda_R^{j-1/2,k} \lambda_L^{j-1/2,k}
      [F^{j,k}(\bxi) - \mathcal{P}^{j,k} F^{j-1,k}(\bxi)]
  }{\lambda_R^{j-1/2,k} - \lambda_L^{j-1/2,k}}
\right).
\end{align*}
Here we apply the operator $\mathcal{P}^{j,k}$ to $F^{j\pm 1,k}$ before
multiplying it by $\xi_1$, as follows the derivation of the equations detailed
in Section \ref{sec:system}. Note that $\mathcal{P}^{j,k} F^{j,k} = F^{j,k}$,
which is why we omit this operator in front of $F^{j,k}$. Multiplying $\xi_1$
by a function with the form \eqref{eq:projection} can be implemented by the
recursion relation of the basis functions:
\begin{displaymath}
\begin{split}
& \xi_1 \sum_{l=0}^{L-1} \sum_{m=-l}^l \sum_{n=0}^{N_l-1}
  \tilde{g}_{lmn} \Psi_{lmn}^{j,k}(\bxi) \\
={} & \sum_{l=0}^{L-1} \sum_{m=-l}^l \sum_{n=0}^{N_l-1}
  \Bigg(
    \sqrt{\frac{(l-m-1)(l-m)(n+l+1/2)}{2(2l-1)(2l+1)}} \Theta^{j,k} \tilde{g}_{l-1,m+1,n} -
    \sqrt{\frac{(l-m-1)(l-m)(n+1)}{2(2l-1)(2l+1)}} \tilde{g}_{l-1,m+1,n+1} \\
& \qquad - \sqrt{\frac{(l+m-1)(l+m)(n+l+1/2)}{2(2l-1)(2l+1)}} \Theta^{j,k} \tilde{g}_{l-1,m-1,n}
  + \sqrt{\frac{(l+m-1)(l+m)(n+1)}{2(2l-1)(2l+1)}} \tilde{g}_{l-1,m-1,n+1} \\
& \qquad - \sqrt{\frac{(l-m+1)(l-m+2)n}{2(2l+1)(2l+3)}} \Theta^{j,k} \tilde{g}_{l+1,m-1,n-1}
  + \sqrt{\frac{(l-m+1)(l-m+2)(n+l+3/2)}{2(2l+1)(2l+3)}} \tilde{g}_{l+1,m-1,n} \\
& \qquad + \sqrt{\frac{(l+m+1)(l+m+2)n}{2(2l+1)(2l+3)}} \Theta^{j,k} \tilde{g}_{l+1,m+1,n-1}
  - \sqrt{\frac{(l+m+1)(l+m+2)(n+l+3/2)}{2(2l+1)(2l+3)}} \tilde{g}_{l+1,m+1,n} \\
& \qquad + v_1^{j,k} \tilde{g}_{lmn} \Bigg) \Psi_{lmn}^{j,k}(\xi).
\end{split}
\end{displaymath}
The maximum characteristic speeds $\lambda_L^{j+1/2, k}$ and
$\lambda_R^{j+1/2,k}$ are computed by
\begin{displaymath}
\lambda_L^{j+1/2,k} = \min \left\{
  0, \lambda_{\min} (\xi_1 \mathcal{P}^{j,k}),
  \lambda_{\min} (\xi_1 \mathcal{P}^{j+1,k})
\right\}, \quad
\lambda_R^{j+1/2,k} = \max \left\{
  0, \lambda_{\max} (\xi_1 \mathcal{P}^{j,k}),
  \lambda_{\max} (\xi_1 \mathcal{P}^{j+1,k})
\right\}.
\end{displaymath}
Here $\lambda_{\min}$ and $\lambda_{\max}$ denote the minimum and maximum
eigenvalues of the operators. Since the operator is invariant up to translation
by $v_1^{j,k}$ and scaling by $\sqrt{\Theta^{j,k}}$, the eigenvalues have the
form
\begin{displaymath}
\lambda_{\min} (\xi_1 \mathcal{P}^{j,k}) = v_1^{j,k} + C_{\min} \sqrt{\Theta^{j,k}}, \quad
\lambda_{\max} (\xi_1 \mathcal{P}^{j,k}) = v_1^{j,k} + C_{\max} \sqrt{\Theta^{j,k}},
\end{displaymath}
and the constants $C_{\min}$ and $C_{\max}$ can be precomputed by the power
method. Finally, note that the expansion of the function $F_*^{j,k+1}(\bxi)$
given in \eqref{eq:intermediate} does not satisfy the constraint
\eqref{eq:3d_constraint}. We need to compute $\bv^{j,k+1}$ and $\Theta^{j,k+1}$
by
\begin{displaymath}
\bv^{j,k+1} = \frac{\langle \bxi F_*^{j,k+1}(\bxi) \rangle}%
  {\langle F_*^{j,k+1}(\bxi) \rangle},
\qquad \Theta^{j,k+1} = \left( \frac{1}{(2N_0-1)!!}
  \frac{\left\langle |\bxi - \bv^{j,k+1}|^{2(N_0-1)} F_*^{j,k+1}(\bxi) \right\rangle}%
    {\langle F_*^{j,k+1}(\bxi) \rangle}
\right)^{1/(2N_0-2)},
\end{displaymath}
with which we can at last evolve the solution by $F^{j,k+1}(\bxi) =
\mathcal{P}^{j,k+1} F_*^{j,k+1}(\bxi)$.

The only component missing in this approach is the implementation of the
operator $\mathcal{P}^{j,k}$ in \eqref{eq:projection}. If the function $g$ is
already expressed in the form of
\begin{displaymath}
g(\bxi) = \sum_{l,m,n} g_{lmn} \Psi_{lmn}^{j,k}(\bxi),
\end{displaymath}
then the operator $\mathcal{P}^{j,k}$ is simply a direct truncation. Otherwise,
the implementation of this operator follows the method introduced in
\cite{Cai2018}. Below we will omit the derivation and just briefly state the
result by taking $\mathcal{P}^{j,k} F^{j+1,k}(\bxi)$ as an example. Suppose
\begin{equation} \label{eq:Pjk}
\mathcal{P}^{j,k} F^{j+1,k}(\bxi) = \sum_{l=0}^{L-1} \sum_{m=-l}^l \sum_{n=0}^{N_l-1}
  \tilde{F}_{lmn}^{j+1,k} \Psi_{lmn}^{j,k}(\bxi).
\end{equation}
Then by defining
\begin{equation} \label{eq:gamma}
\begin{gathered}
V_{-1}^{j,k} = \frac{1}{2} (v_1^{j,k} - \mathrm{i} v_2^{j,k}), \quad
V_0^{j,k} = v_3^{j,k}, \quad
V_1^{j,k} = -\frac{1}{2} (v_1^{j,k} + \mathrm{i} v_2^{j,k}), \\
\gamma_{lm}^{\mu} = \sqrt{ \frac{[l+(2\delta_{1,\mu}-1)m + \delta_{1,\mu}]
  [l-(2\delta_{-1,\mu}-1)m + \delta_{-1,\mu}]}{(2l-1)(2l+1)}},
\end{gathered}
\end{equation}
we can compute the coefficients in \eqref{eq:Pjk} as follows:
\begin{displaymath}
\tilde{F}_{lmn}^{j+1,k} = \sum_{i=0}^{l+2n} \varpi_{lmn}^{(i)},
\end{displaymath}
where
\begin{align*}
\varpi_{lmn}^{(0)} &= F_{lmn}^{j+1,k}, \\
\varpi_{lmn}^{(i)} &= \frac{1}{i} \sqrt{n(n+l+1/2)}
  (\Theta^{j,k} - \Theta^{j+1,k}) \varpi_{l,m,n-1}^{(i-1)} - \frac{\sqrt{2}}{i}
  \sum_{\mu = -1}^1 (V_{\mu}^{j,k} - V_{\mu}^{j+1,k}) \times \\
& \quad \left[ (-1)^{\mu} \sqrt{n+l+1/2} \gamma_{l,m+\mu}^{-\mu}
  \varpi_{l-1,m+\mu,n}^{(k-1)} + \sqrt{n} \gamma_{-l-1,m+\mu}^{-\mu}
  \varpi_{l+1,m+\mu,n-1}^{(k-1)} \right], \quad i > 0.
\end{align*}
Details of the derivation can be found in \cite{Cai2018}.

In applications, we supplement the above first-order method by Heun's method
and linear reconstruction with MC limiter to achieve second order of accuracy,
which is relatively straightforward and will not be introduced. The numerical
tests will be introduced in the next subsection.

\subsection{Numerical tests}
Again we use the problem of shock structure to test HMBMM. For
three-dimensional physics, we set the initial condition to be
\begin{equation} \label{eq:3d_init}
f(x,\bxi,0) = \frac{n_0(x)}{(2\pi\theta_0(x))^{3/2}} \exp \left(
  -\frac{|\bxi - \bv_0(x)|^2}{2\theta_0(x)}
\right),
\end{equation}
where
\begin{gather*}
n_0(x) = \left\{ \begin{array}{@{}ll}
  1, & \text{if } x < 0, \\
  \dfrac{4 \mathit{Ma}^2}{\mathit{Ma}^2 + 3}, & \text{if } x > 0,
\end{array} \right. \qquad
\bv_0(x) = \left\{ \begin{array}{@{}ll}
  (\sqrt{5/3} \mathit{Ma},0,0)^T, & \text{if } x < 0, \\
  \left( \sqrt{\dfrac{5}{3}} \dfrac{\mathit{Ma}^2+3}{4\mathit{Ma}}, 0, 0 \right)^T,
    & \text{if } x > 0,
\end{array} \right. \\[5pt]
\theta_0(x) = \left\{ \begin{array}{@{}ll}
  1, & \text{if } x < 0, \\
  \dfrac{(5\mathit{Ma}^2-1)(\mathit{Ma}^2+3)}{16\mathit{Ma}^2}, & \text{if } x > 0.
\end{array} \right.
\end{gather*}
Note that here the spatial variable $x$ remains to be one-dimensional, so that
the advection term in the Boltzmann equation \eqref{eq:Boltzmann} reduces to
$\xi_1 \partial_x f$. The collision term is given by the inverse-power-law
intermolecular potential, and we choose the power so that the viscosity index
is $\omega = 0.72$, which is often used to simulate the argon gas. More details
about the collision term can be found in Appendix \ref{sec:STA}, from which one
can see that one unit length in the problem setting equals the mean free path
for the flow in front of the shock wave. Here we consider two Mach numbers
$\mathit{Ma} = 3.8$ and $\mathit{Ma} = 6.5$, for which the temperature ratios
are $5.375$ and $14.07$, respectively. Both are larger than $2$ so that methods
based on Grad's ansatz fail to converge.

\begin{figure}[!ht]
\centering
\subfloat[Density, velocity and temperature]{%
  \includegraphics[width=.45\textwidth]{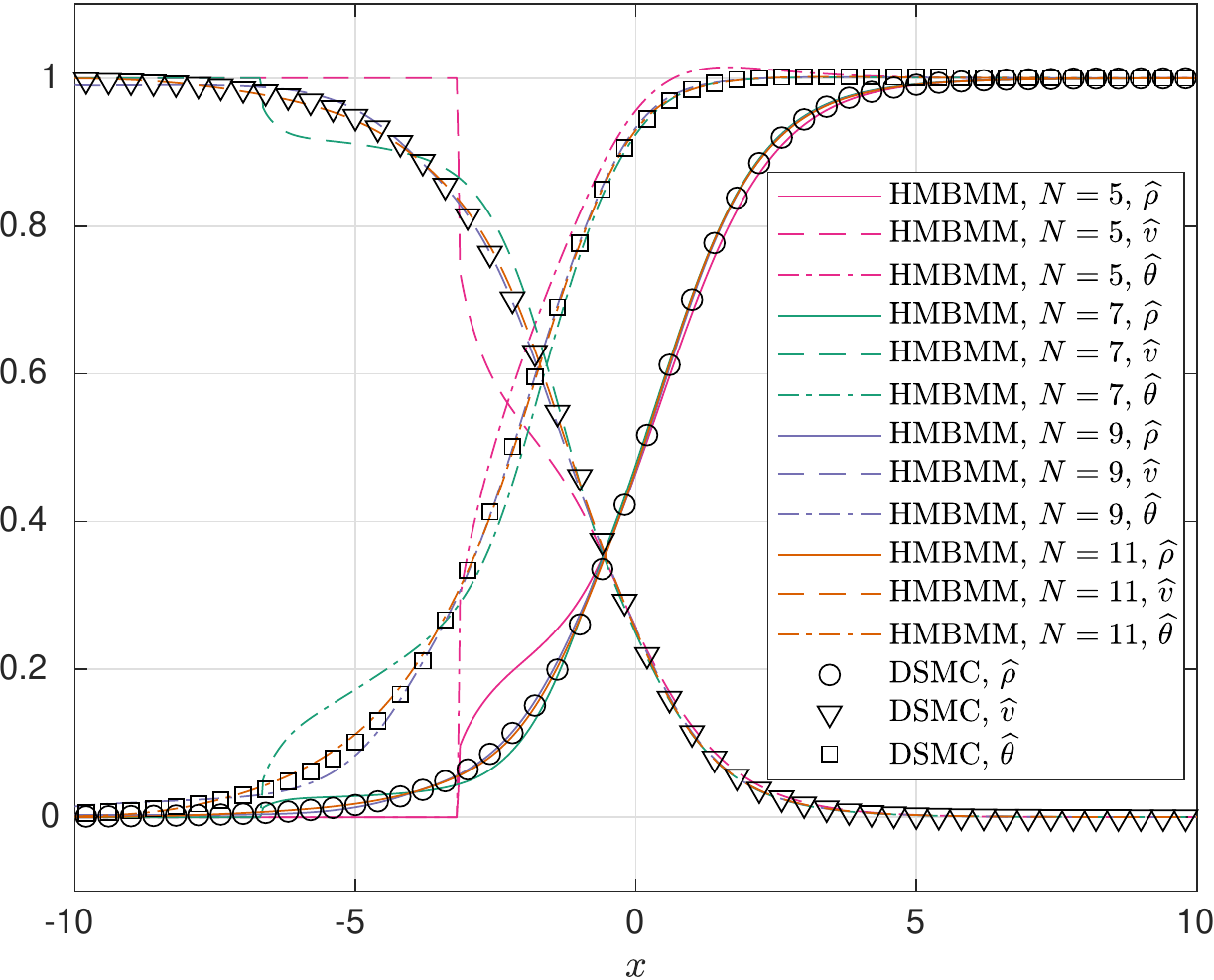}}
\qquad
\subfloat[Comparison between $\theta$ and $\Theta$]{%
  \label{fig:Theta_Ma3.8}
  \includegraphics[width=.45\textwidth,clip]{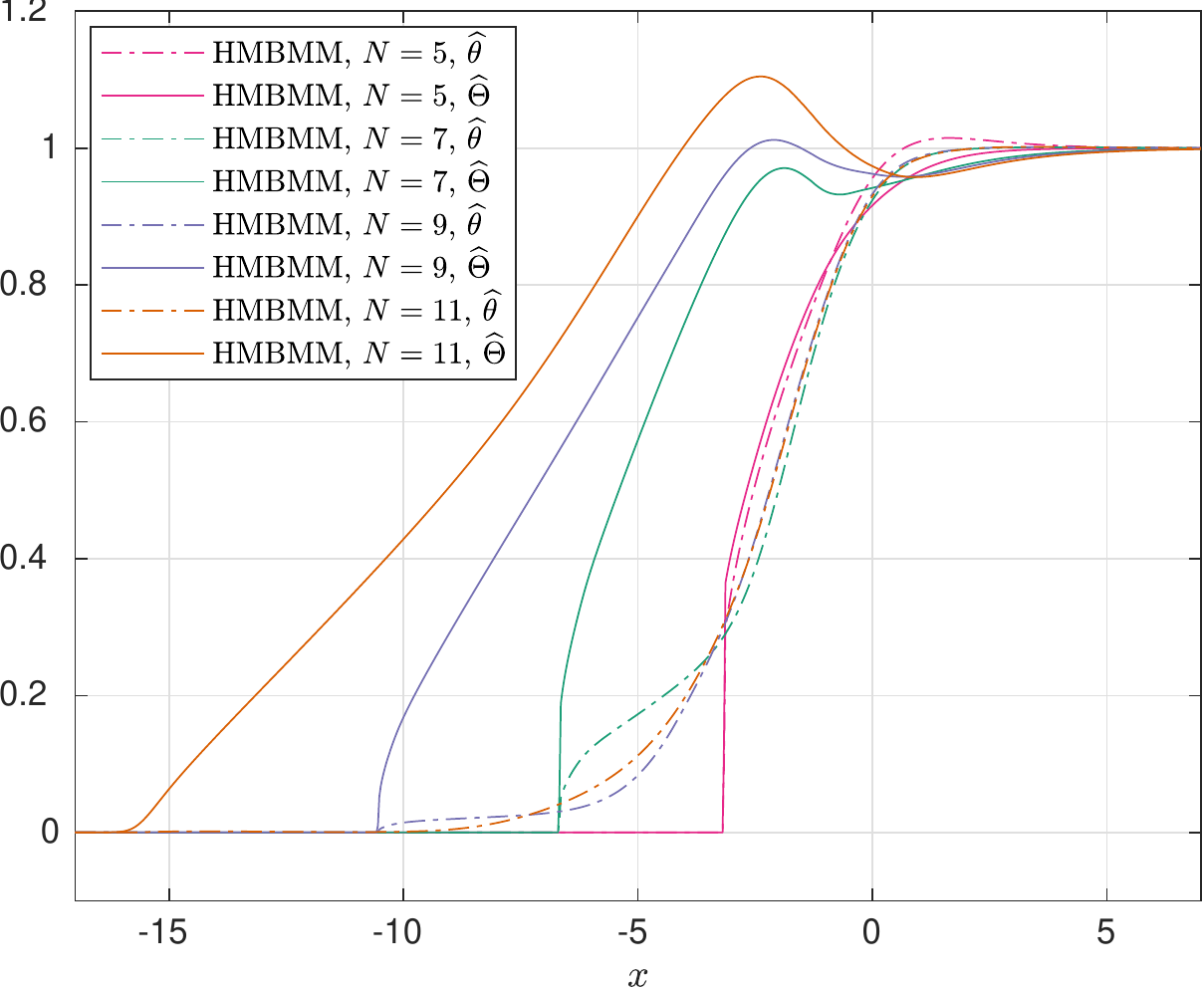}}
  \caption{Numerical results for Mach number $3.8$}
\label{fig:Ma3.8}
\end{figure}

In our simulation, we adopt a second-order method implemented by linear
reconstruction with the minmod limiter, so that the discontinuities (subshock)
can be observed more clearly. A uniform grid with 2000 cells covering the range
$[-30, 30]$ is used. The DSMC solution, computed using Bird's DSMC code
\cite{Bird1963, Bird1994}, is provided as a reference. In the ansatz
\eqref{eq:3d_ansatz}, we choose $L$ and $N_l$ such that all polynomials
$p_{lmn}$ of degree less than $N$ is included. Specifically, we set $L = N$ and
$N_l = \lfloor (N+1-l)/2 \rfloor$. For $\mathit{Ma} = 3.8$, the results are
shown in Figure \ref{fig:Ma3.8}. The left panel shows the curves for normalized
density, velocity and temperature, where one can see that when $N=5,7$, the
subshock still exists. When $N=9$, all the three curves for density, velocity,
and temperature already have good agreement with the reference solution.
However, by looking at Figure \ref{fig:Theta_Ma3.8}, one can still observe the
discontinuity around $x = -10$, while for $N=11$, the subshock is truly
removed. In general, as the number of moments increases, the subshock moves in
the direction of the front of the shock wave, and eventually disappears.
Meanwhile, the discontinuity in the conservative quantities becomes smaller as
$N$ increases. When the discontinuity is sufficiently small, we may consider
the solution acceptable even if the unphysical subshock still exists.

\begin{figure}[!ht]
\centering
\subfloat[Density, velocity and temperature]{%
  \includegraphics[width=.45\textwidth]{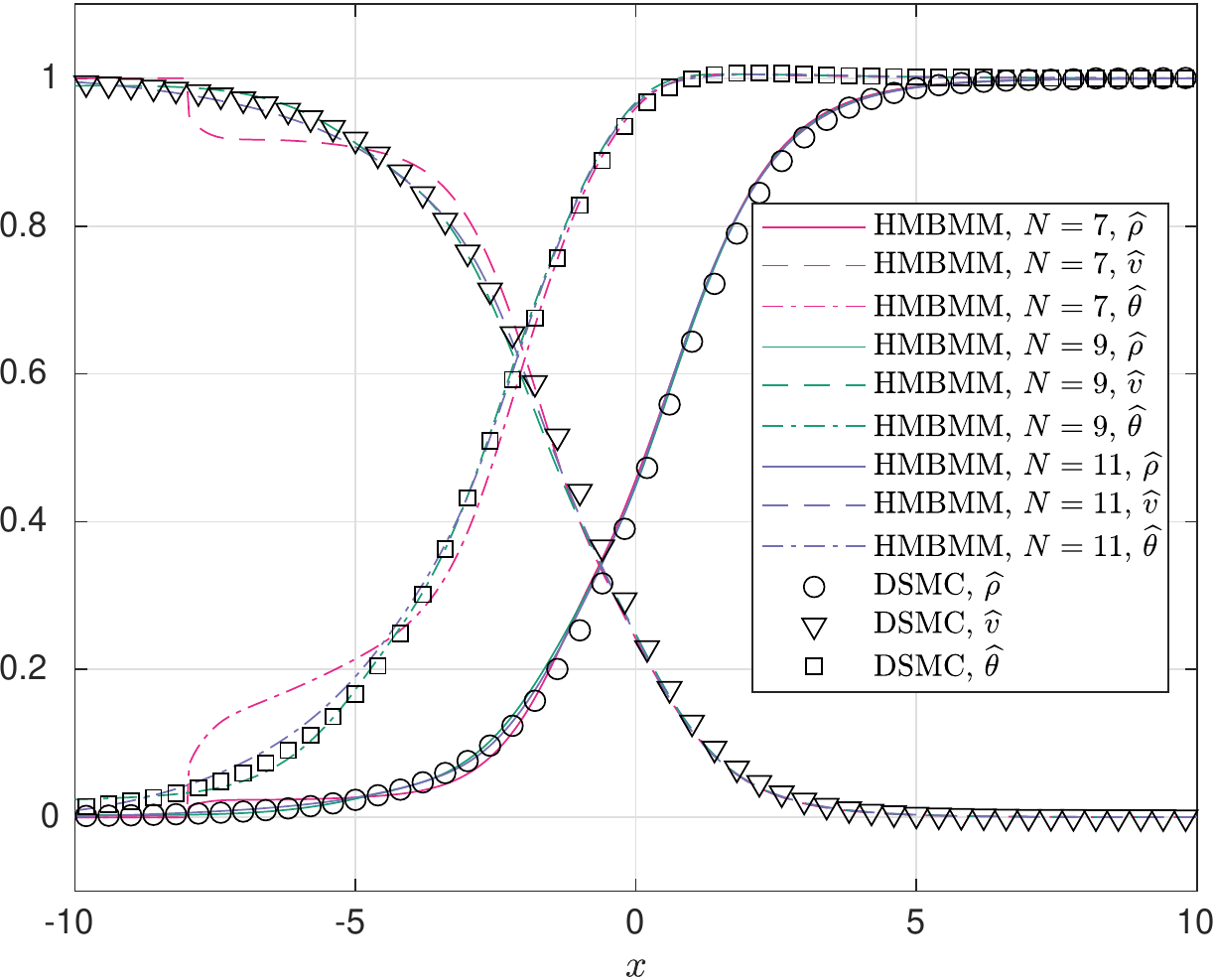}}
\qquad
\subfloat[Comparison between $\theta$ and $\Theta$]{%
  \label{fig:Theta_Ma6.5}
  \includegraphics[width=.45\textwidth,clip]{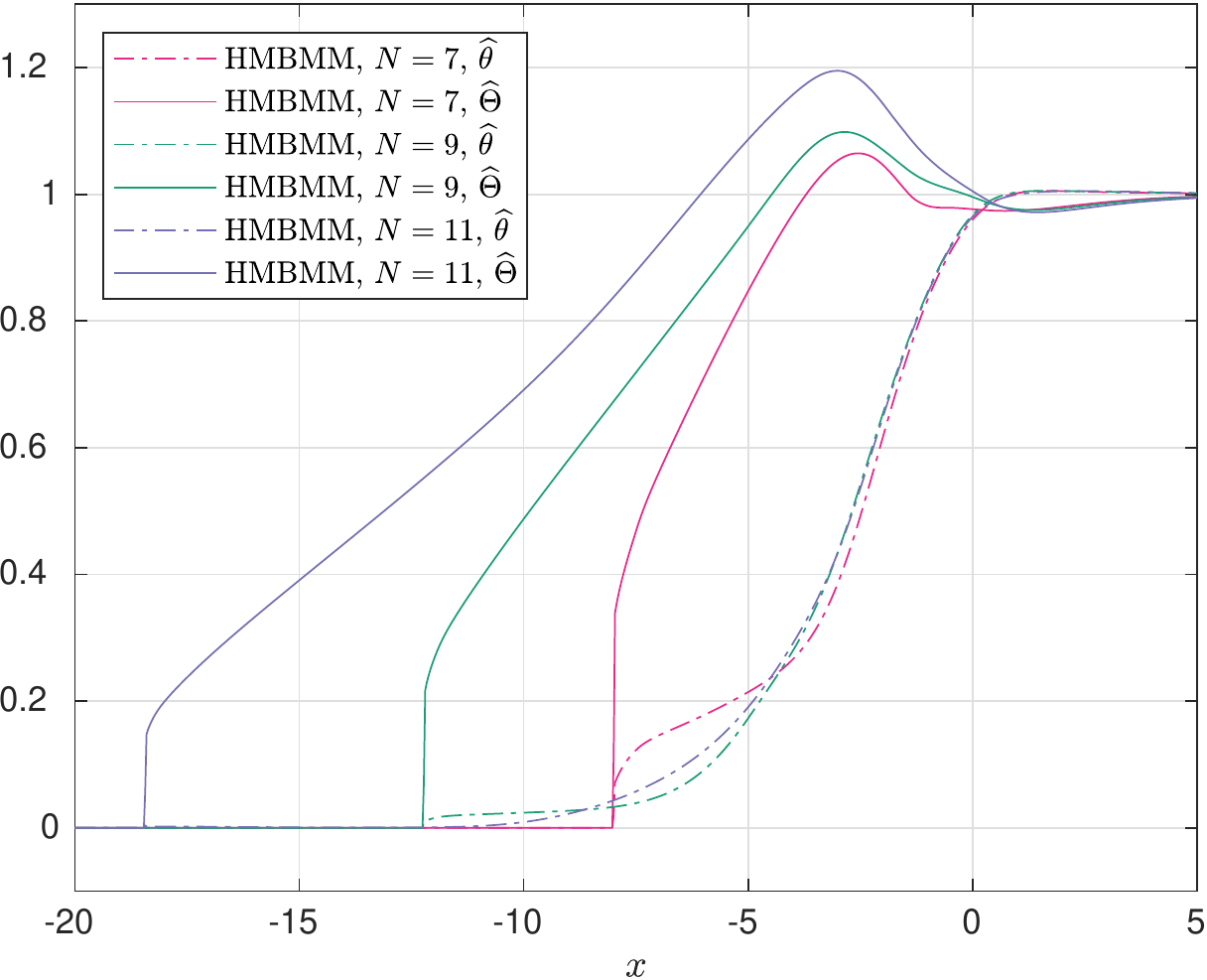}}
  \caption{Numerical results for Mach number $6.5$}
\label{fig:Ma6.5}
\end{figure}

Similar behavior can be observed for $\mathit{Ma} = 6.5$, for which we plot the
results in Figure \ref{fig:Ma6.5}. For $N=11$, although the subshock still
exists, the location of the subshock is quite far from the center of the shock,
so that the jumps in the first few moments are nearly negligible. Again the
convergence is clearly seen as $N$ increases.

\begin{figure}
\centering
\subfloat[Mach number $3.8$]{%
  \includegraphics[width=.47\textwidth,clip]{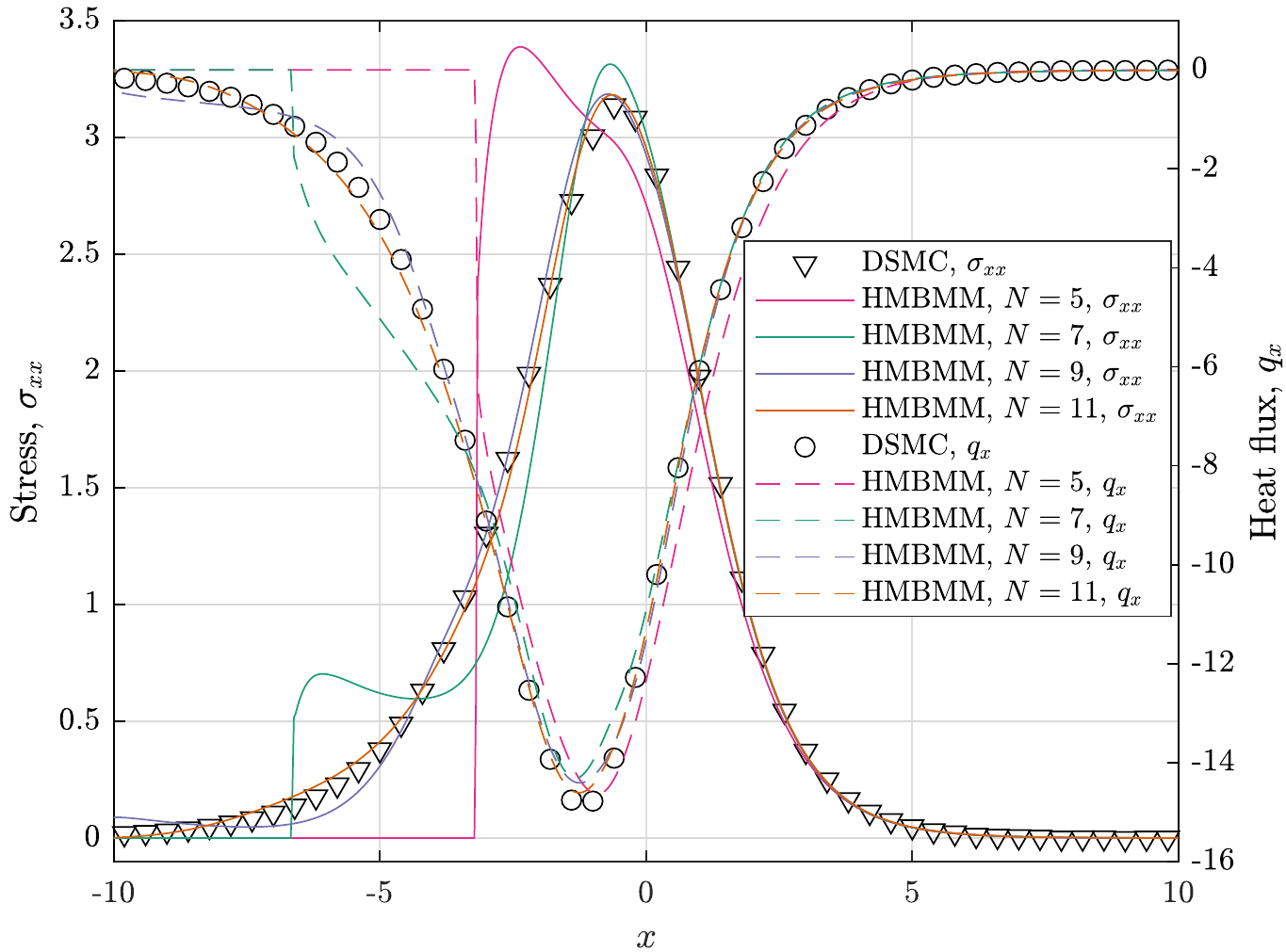}}
\quad
\subfloat[Mach number $6.5$]{%
  \includegraphics[width=.47\textwidth,clip]{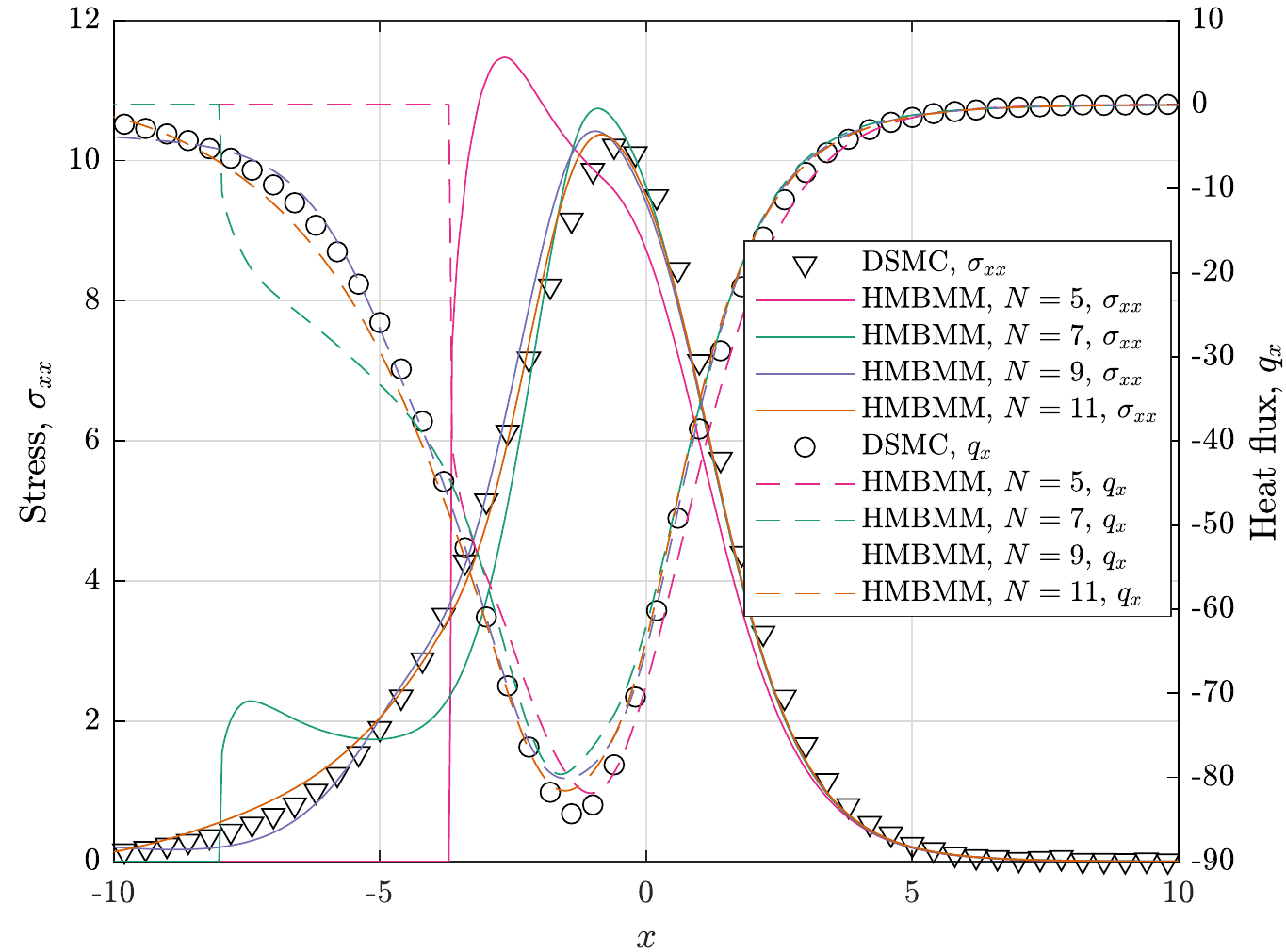}}
\caption{Profiles of $\sigma_{xx}$ and $q_x$}
\label{fig:sigma_q}
\end{figure}

Some non-equilibrium variables, including the stress $\sigma_{xx}$ and the heat
flux $q_x$, are plotted Figure \ref{fig:sigma_q}, again with the DSMC results
as references. These quantities are defined by
\begin{displaymath}
  \sigma_{xx} = \int_{\mathbb{R}^3} (\xi_1 - v_1)^2 f(\bxi) \,\mathrm{d}\bxi - \rho \theta,
  \qquad
  q_x = \frac{1}{2} \int_{\mathbb{R}^3} |\bxi - \bv|^2 (\xi_1 - v_1) f(\bxi) \,\mathrm{d}\bxi.
\end{displaymath}
It can be observed that the values for these quantities are greater for larger
Mach numbers, and the locations of the discontinuities in these quantities are
the same as the locations of the subshocks in equilibrium quantities. In both
cases, the graphs support the convergence to the smooth solutions.

\begin{figure}[!ht]
\centering
\subfloat[$\mathit{Ma} = 3.8$]{%
  \includegraphics[width=.45\textwidth]{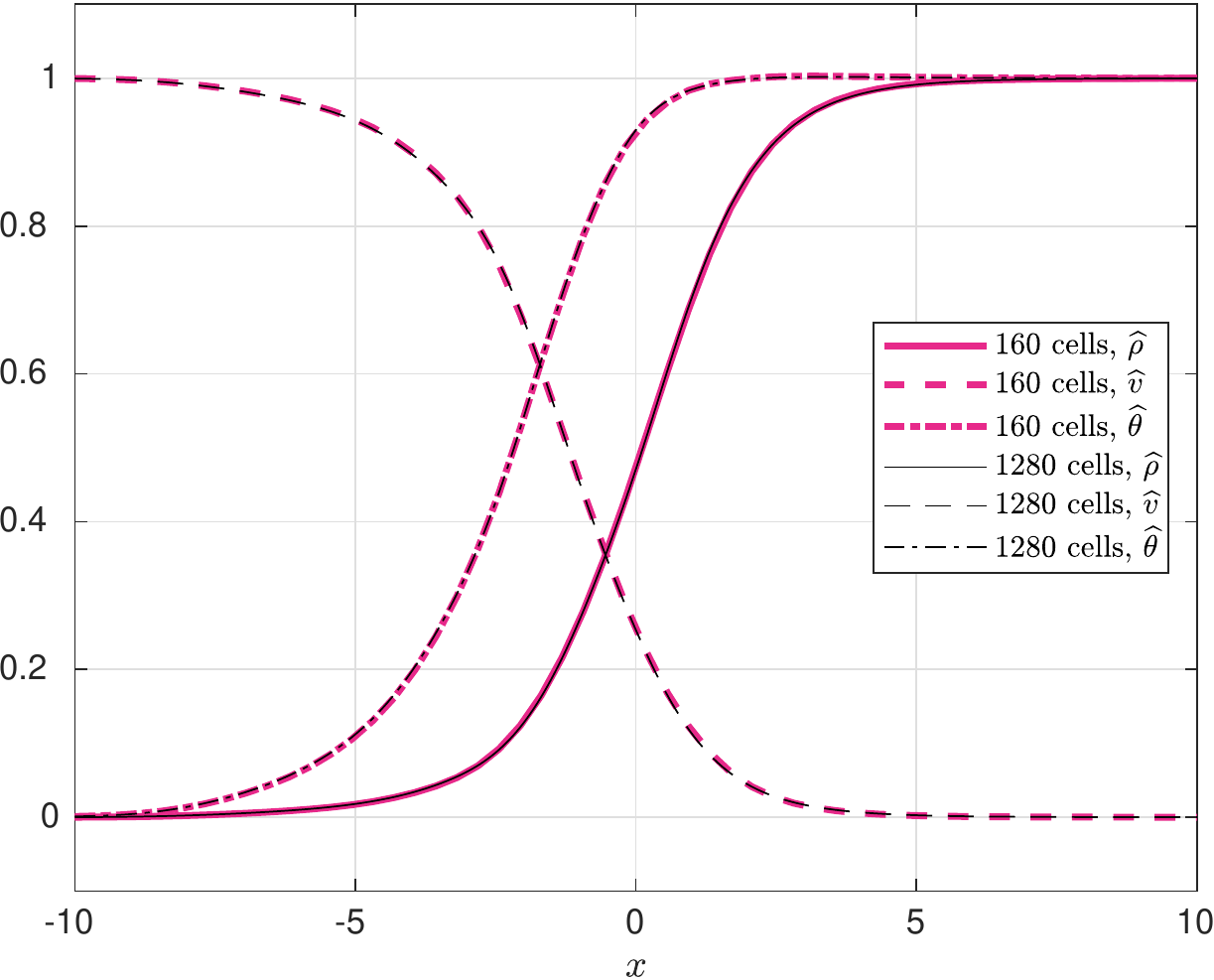}}
\qquad
\subfloat[$\mathit{Ma} = 6.5$]{%
  \includegraphics[width=.45\textwidth,clip]{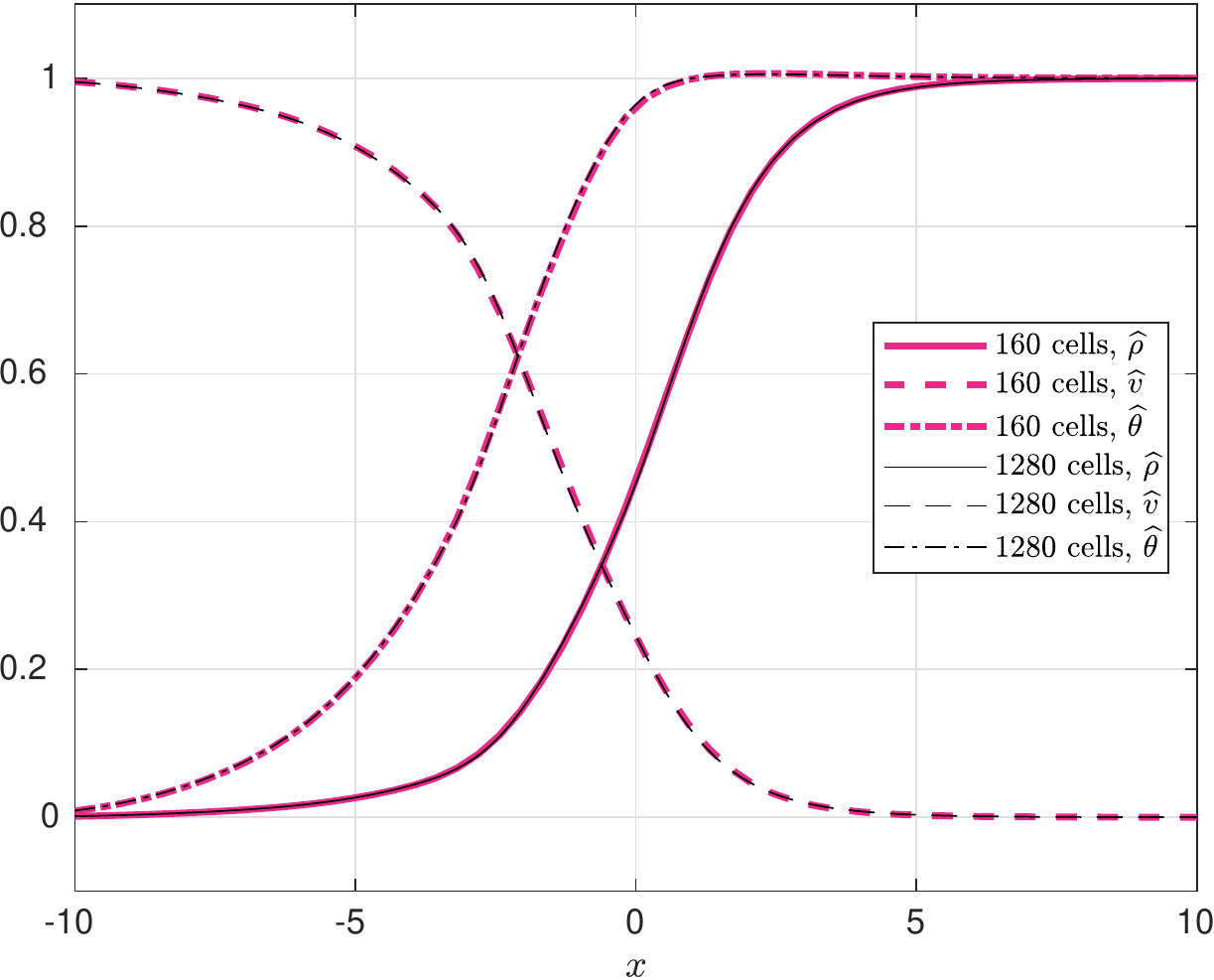}}
  \caption{Numerical results for different grid sizes}
\label{fig:comparison}
\end{figure}

We will now test the numerical efficiency of HMBMM. To this aim, we would like
to reduce the number of grid cells while keeping the numerical solution
acceptable. For both Mach numbers, we consider $N = 11$ and run the codes for
grids with $160$ cells and $1280$ cells. The numerical results are plotted in
Figure \ref{fig:comparison}. One can see that the two solutions are nearly
identical, meaning that for $N=11$ and both Mach numbers, it is sufficient to
use $160$ grid cells in our simulation. The machine for the computation has the
CPU model ``Intel\textsuperscript{\textregistered} Core\texttrademark{}
i7-7600U'' with clock frequency 2.8GHz. The code is run on a single thread with
no parallelization. Some run-time data are given in Table \ref{tab:data}. Note
that our code is still to be optimized (it includes some duplicate computations
of the numerical fluxes) and the efficiency is expected to be further improved
by applying higher-order methods. We consider this computational cost as
competitive since the quadratic collision operator is applied in the
computation.

\begin{table}[!ht]
\centering
\caption{Run-time data for shock structure computation with $160$ grid cells}
\label{tab:data}
\begin{tabular}{cccc}
\hline
& $\mathit{Ma} = 3.8$ & & $\mathit{Ma} = 6.5$ \\
\hline
Number of time steps & 4167 & & 6796 \\
Average time step & 0.012 & & 0.0074 \\
Total computational time & 1227.5s & & 2065.4s \\
Average computational time per time step & 0.295s & & 0.304s \\
\hline
\end{tabular}
\end{table}

%% file: article_application.tex
\section{Applications beyond the plane shock structure problem} \label{sec:application}
As mentioned previously, such an approach is not specially designed for the
shock structure problem. We would therefore like to study the performance of
the method in other problems. In this section, four more examples are presented.
In the first two examples, we select problems where large temperature
ratios do not exist in the initial condition, but emerge after the evolution of
the flow. In the third example, the large temperature ratio is caused
by the wall boundary conditions, and in the fourth example, a two-dimensional
shock structure problem is studied to show the similar behavior as the
one-dimensional case. As described in Appendix \ref{sec:STA}, in all the
simulations, the Knudsen number is $1.0$, meaning that the unit length equals
the mean free path of the gas with unit density and temperature in equilibrium.

\subsection{Colliding flows}
\begin{figure}[!ht]
\includegraphics[width=.45\textwidth]{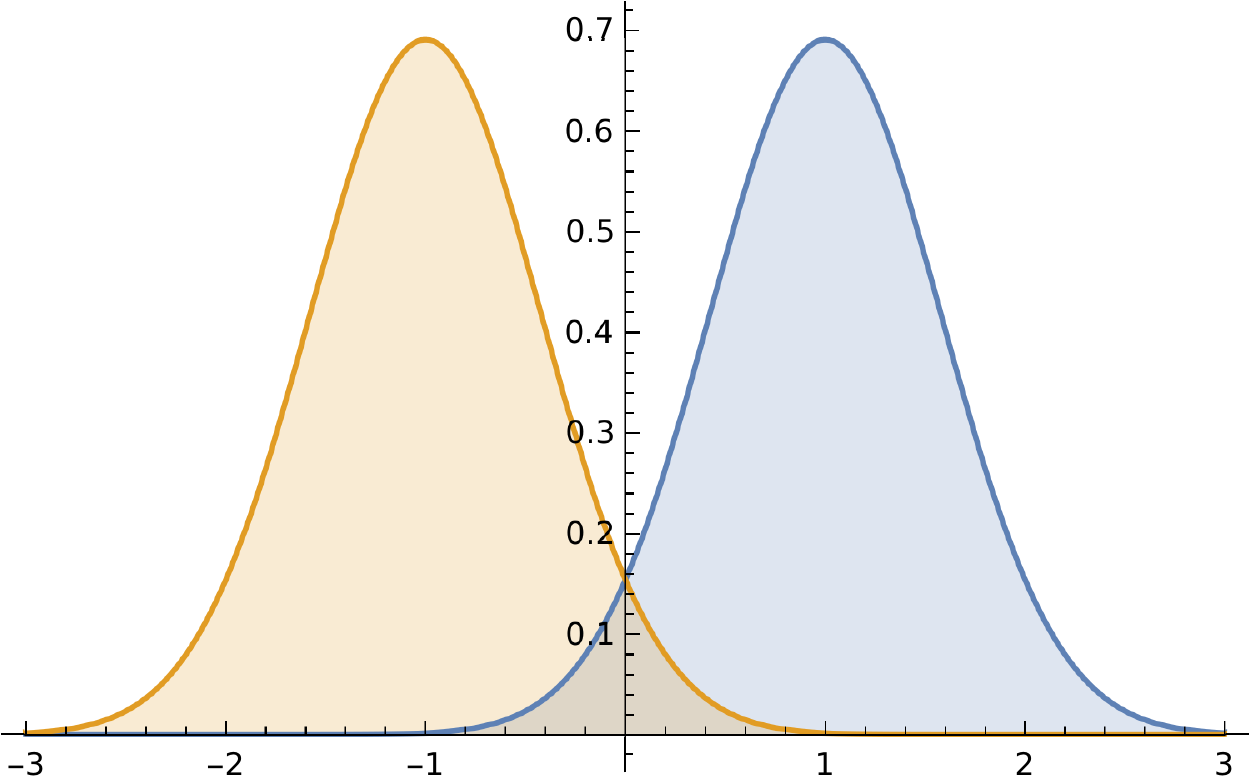}
\caption{One-dimensional marginal distribution functions in the initial
  condition}
\label{fig:init}
\end{figure}

We consider the one-dimensional flow whose initial condition is
\eqref{eq:3d_init} with 
\begin{displaymath}
\rho_0(x) = 1, \qquad \bv_0(x) = \left\{ \begin{array}{ll}
  (1,0,0)^T, & \text{if } x < 0, \\
  (-1,0,0)^T, & \text{if } x > 0,
\end{array} \right. \qquad \theta_0(x) = 1/3.
\end{displaymath}
The two distribution functions in the initial data are plotted in Figure
\ref{fig:init}, which shows that the two Maxwellians have the same width, while
there centers are separated. It can be expected that the fusion of these two
Maxwellians can lead to a larger temperature that does not exist in the initial
value.

\begin{figure}[!ht]
\centering
\subfloat[$\rho$, $t=7.5$]{%
  \includegraphics[width=.45\textwidth]{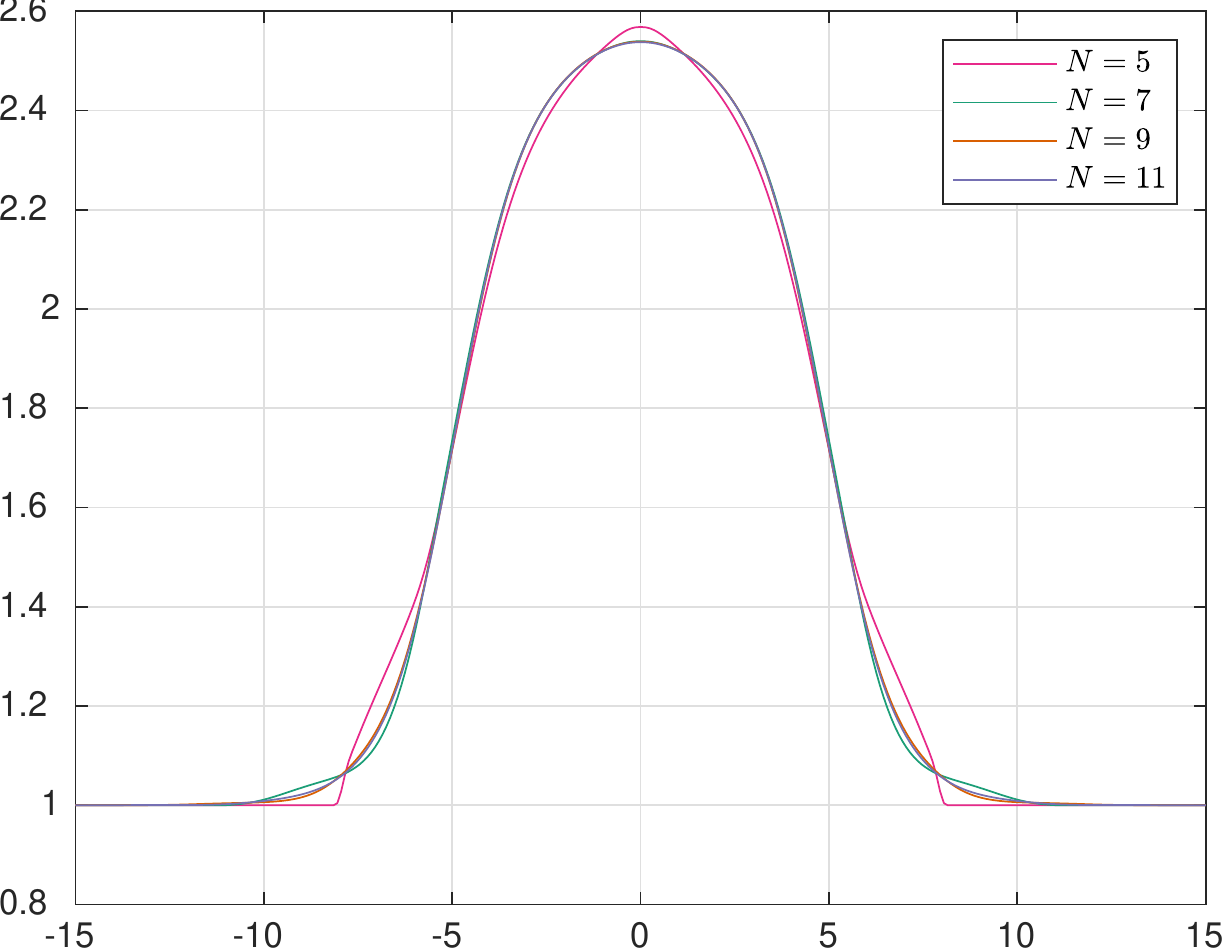}}
\qquad
\subfloat[$\theta$ and $\Theta$, $t=7.5$]{%
  \includegraphics[width=.45\textwidth,clip]{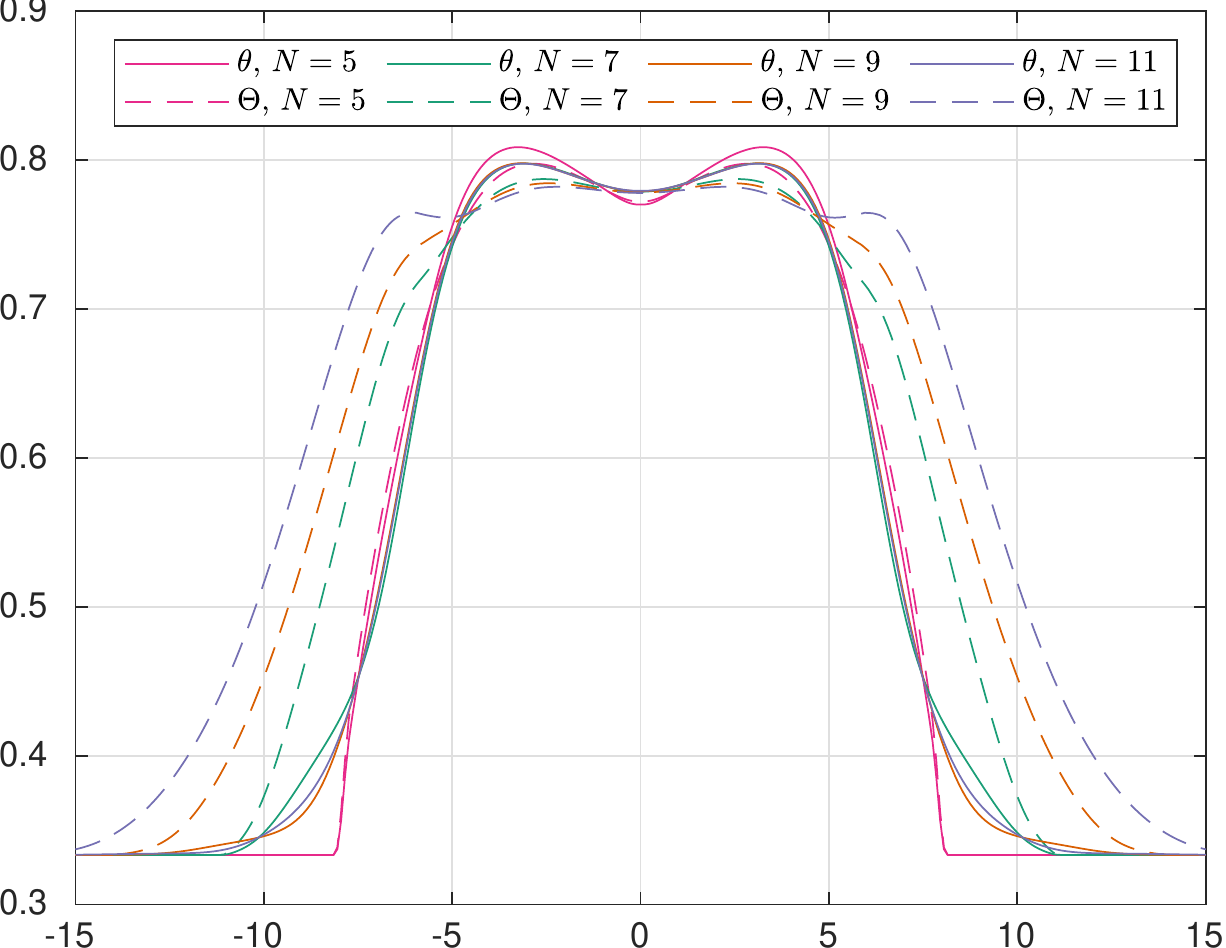}} \\
\subfloat[$\rho$, $t=15$]{%
  \includegraphics[width=.45\textwidth]{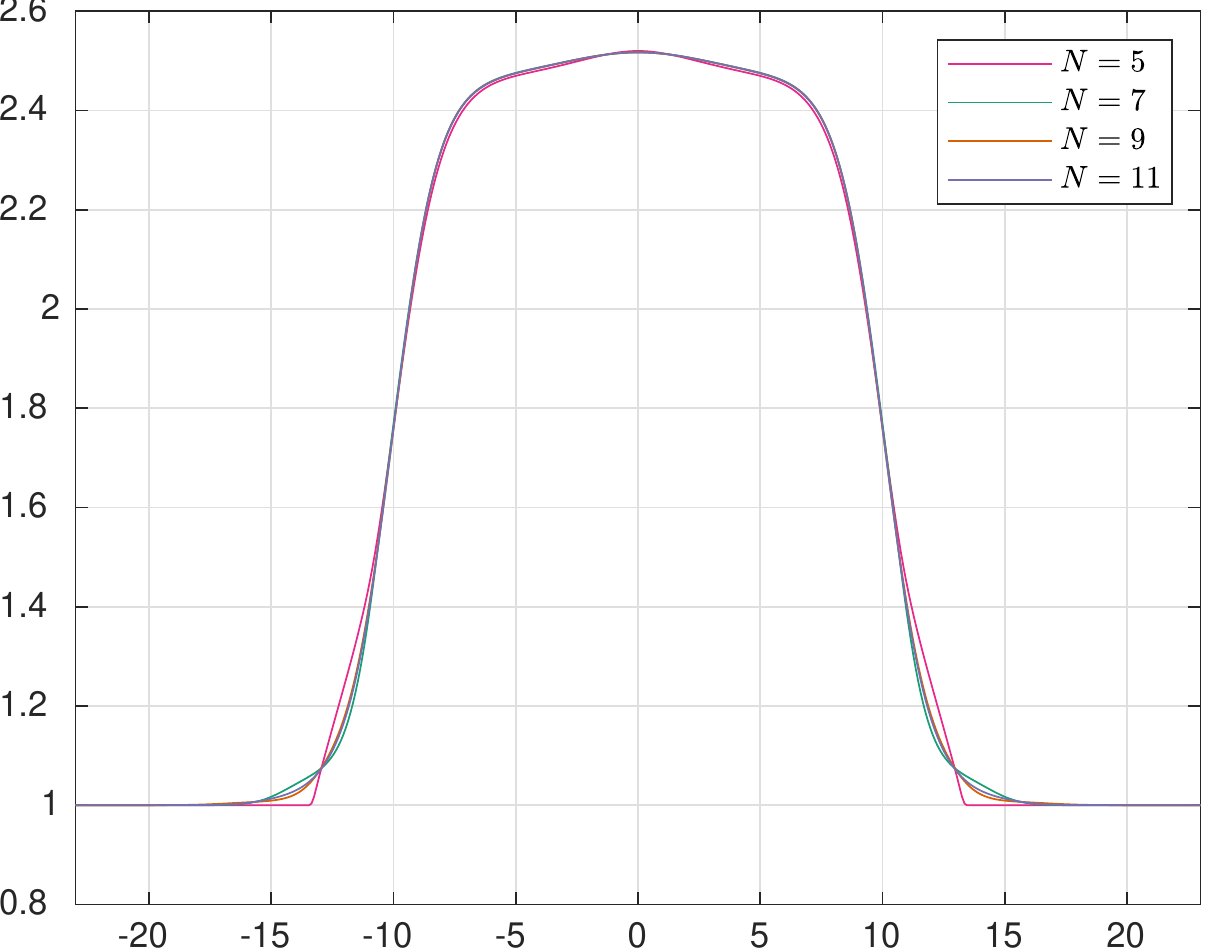}}
\qquad
\subfloat[$\theta$ and $\Theta$, $t=15$]{%
  \includegraphics[width=.45\textwidth,clip]{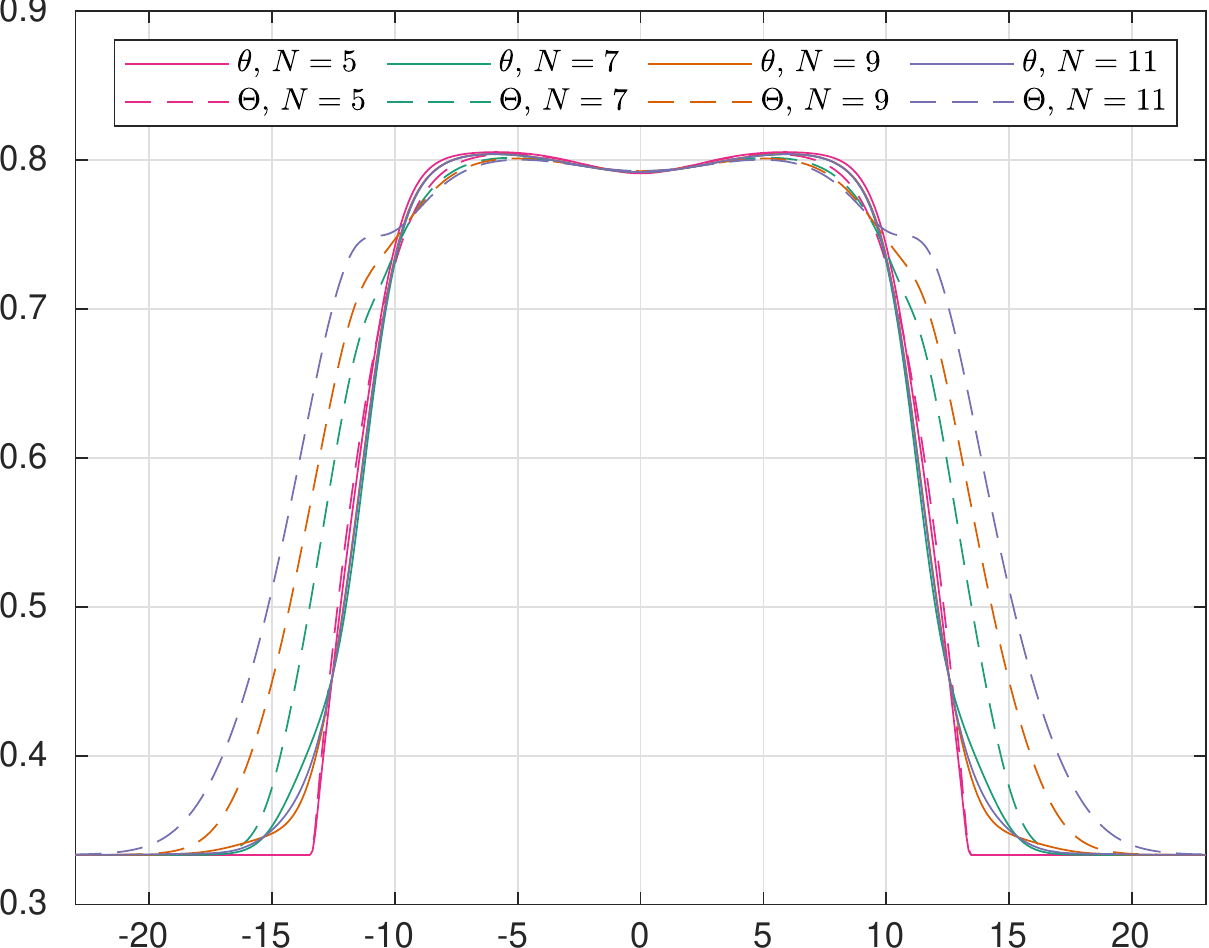}}
\caption{Numerical results for the colliding flow}
\label{fig:ex1}
\end{figure}

In our computation, we set the domain to be $[-30, 30]$, discretized by a
uniform grid with $600$ cells. A second-order scheme is again applied in our
numerical simulation. We present the numerical solutions for $N = 5,7,9$, which
are plotted in Figure \ref{fig:ex1}. It is seen that a much higher temperature
emerges in the middle of the flow such that the temperature ratio is well above
$2.0$, which again fails the Grad-type methods. While for HMBMM, the
simulations are stable, and the results look convergent as $M$ increases.
Again, inside the shocks, the scaling factor $\Theta$ is mostly greater the
temperature, so that the tail of the distribution function is better captured,
as also stablizes the simulation. While $N=5$ (35-moment theory) cannot well
capture the flow structure on both ends of the domain, we can safely increase
$M$ to improve the quality of the solution. Note that for different $M$, the
variable $\Theta$ stands for different moments, as explains why $\Theta$ does
not look convergent in the figures. In the central part of the domain, the
values of $\theta$ and $\Theta$ are relatively close to each other, since the
higher density in this region causes higher collision rate, driving the
distribution functions closer to Maxwellian, for which $\theta$ and $\Theta$
are equal.

The choice of $N$ in applications depends on the requirement of the accuracy.
For this example, $N = 6$ can already provide a good approximation. Note that
it is important to keep $N$ small in simulations since the computational cost
for the quadratic collision term usually increases quickly with respect to $N$.
Being able to use a small $N$ can signficantly accelerate the simulation.

\subsection{Cross-regime flows}
\begin{figure}[!ht]
\centering
\subfloat[$\rho$, $t=2.5$]{%
  \includegraphics[width=.45\textwidth]{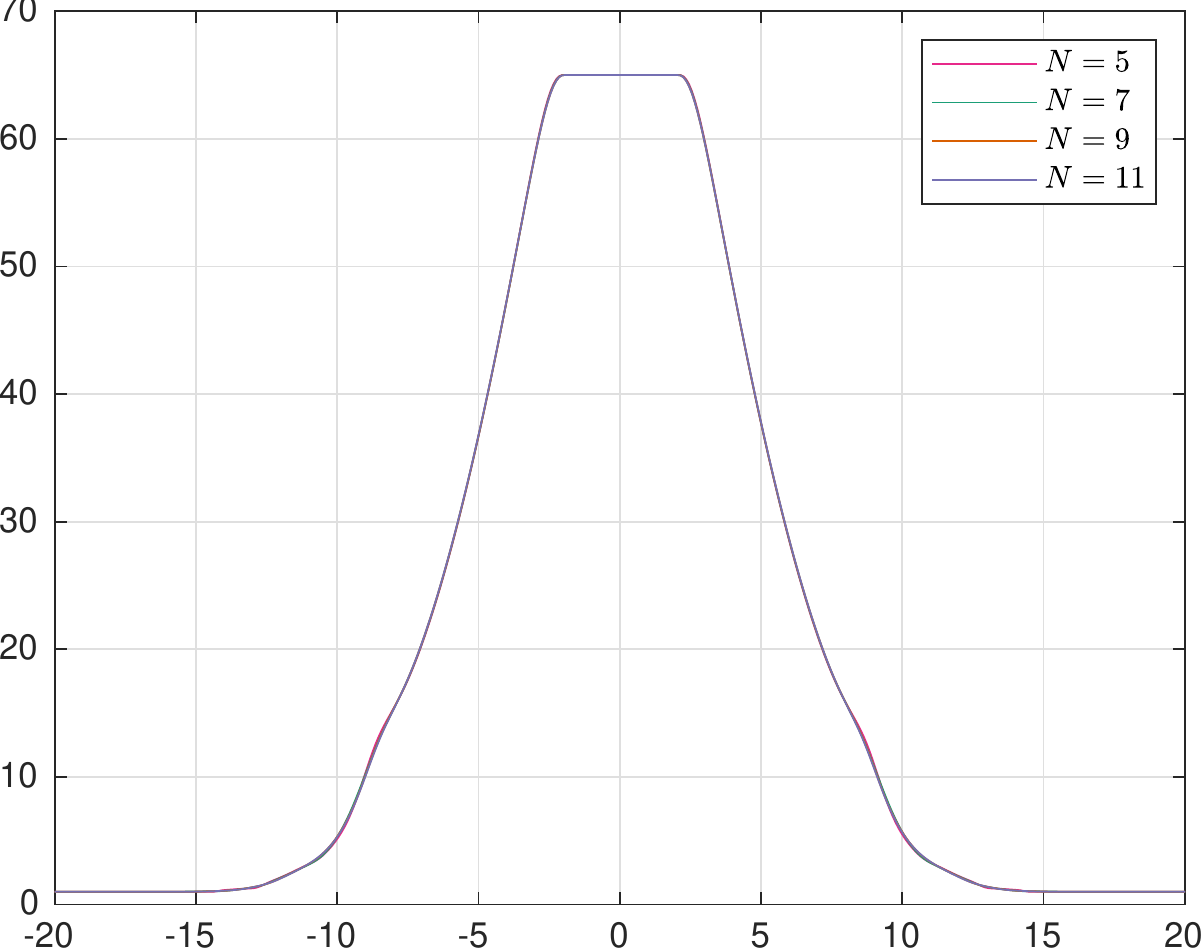}}
\qquad
\subfloat[$\theta$ and $\Theta$, $t=2.5$]{%
  \includegraphics[width=.45\textwidth,clip]{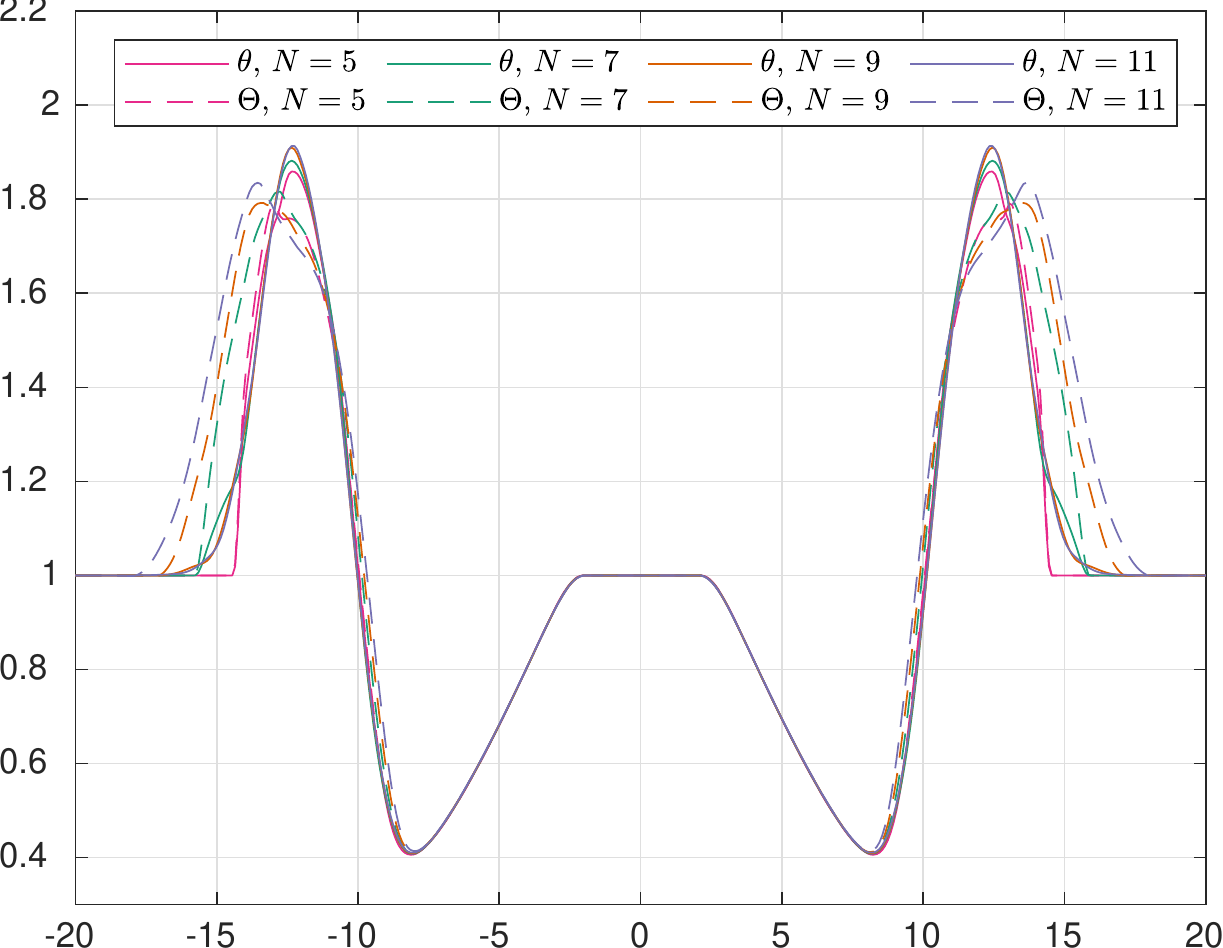}} \\
\subfloat[$\rho$, $t=5$]{%
  \includegraphics[width=.45\textwidth]{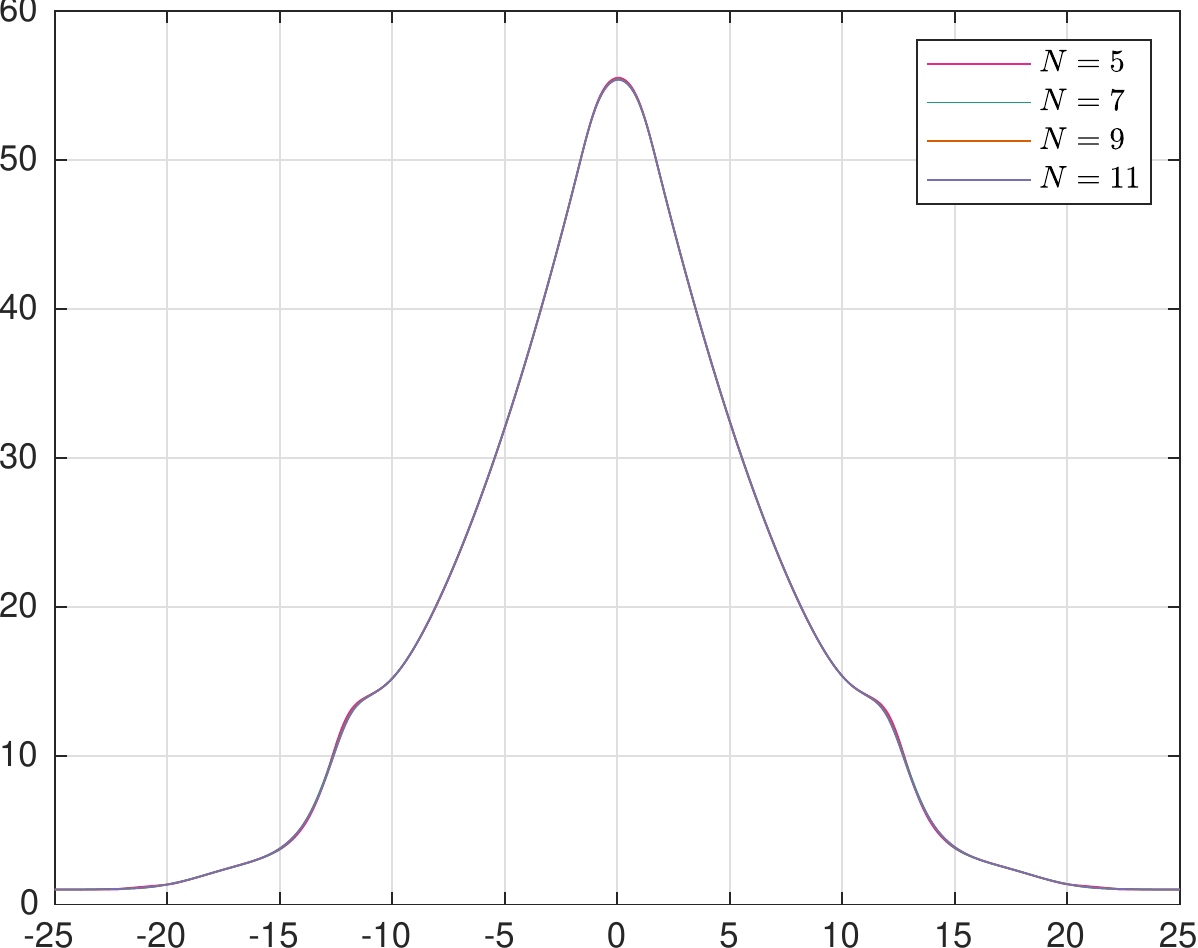}}
\qquad
\subfloat[$\theta$ and $\Theta$, $t=5$]{%
  \includegraphics[width=.45\textwidth,clip]{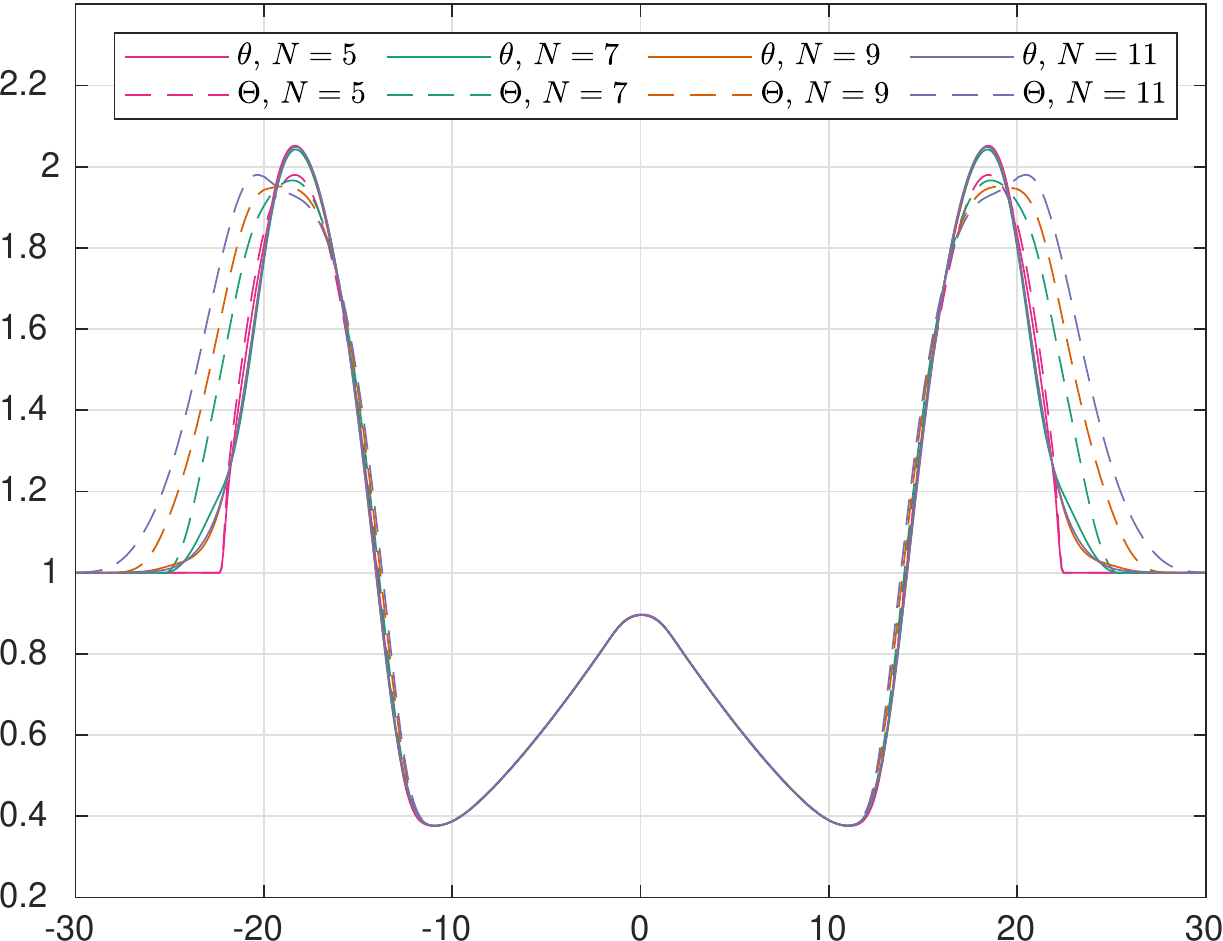}}
\caption{Numerical results for the cross-regime flow}
\label{fig:ex2}
\end{figure}

Now we consider the shock tube problem with a large density ratio, for which
the initial condition is \eqref{eq:3d_init} with
\begin{displaymath}
\rho_0(x) = \left\{ \begin{array}{ll}
  65, & \text{if } |x| < 6, \\
  1, & \text{if } |x| > 6,
\end{array} \right.
\qquad \bv_0(x) = 0, \qquad \theta_0(x) = 1.
\end{displaymath}
In this example, while the Knusden number for $|x| > 6$ is $1.0$, the central
area $[-6,6]$ has the Knudsen number $1/65 \approx 0.015$. Thus this problem
can be considered as the interaction of the slip flow and the transitional
flow. We again set the computationl domain to be $[-30, 30]$ and divide it into
$600$ uniform grid cells. The numerical results at $t = 2.5$ and $t = 5$ are
presented in Figure \ref{fig:ex2}.

For this problem, the high temperature ratio is generated by the high density
ratio in the initial data. Therefore if we use the discrete velocity model to
discretize the velocity space, we need to set a wide velocity domain to capture
the high-temperature distribution functions and set a sufficiently small grid
size to capture the low-temperature distribution functions. However, the
nonlinearity in HMBMM can well adapt the ansatz to fit the above characters of
the problem. The general feature of the numerical solution is similar to the
previous example, such as the convergence with respect to $M$ and the relation
between $\theta$ and $\Theta$, while this example better shows that HMBMM can
simultaneously enjoy the advantages of the hydrodynamic equations and the
spectral method at the same time: in the near-continuum regime, since the
distribution function is close to Maxwellian, we need only a small number of
moments even if the variation of the macroscopic quantities may be large; in
the transitional regime, the method behaves like an adaptive spectral method
which can efficiently capture the distribution function.

\subsection{Fourier flow} \label{sec:Fourier}
In \cite{Cai2020}, the authors proposed another model problem that suffers from
the loss of convergence for Grad's moment method, which is the heat transfer
between two parallel plates, known as the Fourier flow. Mathematically, this
problem can be formulated by the Boltzmann equation
\begin{displaymath}
\frac{\partial f}{\partial t} +
  \xi_1 \frac{\partial f}{\partial x} = \mathcal{S}[f], \qquad
  x \in (-L/2, L/2), \quad \bxi \in \mathbb{R}^3
\end{displaymath}
with boundary conditions
\begin{displaymath}
  f(-L/2, \bxi, t) = \frac{\rho_{\mathrm{left}}(t)}{(2\pi \theta_{\mathrm{left}})^{3/2}}
    \exp \left( -\frac{|\bxi|^2}{2 \theta_{\mathrm{left}}} \right)
  \text{ if } \xi_1 > 0, \quad
  f(L/2, \bxi, t) = \frac{\rho_{\mathrm{right}}(t)}{(2\pi \theta_{\mathrm{right}})^{3/2}}
    \exp \left( -\frac{|\bxi|^2}{2 \theta_{\mathrm{right}}} \right)
  \text{ if } \xi_1 < 0.
\end{displaymath}
Here $\theta_{\mathrm{left}}$ and $\theta_{\mathrm{right}}$ are the
temperatures of the left and right plates, respectively, and
$\rho_{\mathrm{left}}(t)$ and $\rho_{\mathrm{right}}(t)$ are chosen such that
\begin{displaymath}
  \int_{\mathbb{R}^3} \xi_1 f(-L/2, \bxi, t) \,\mathrm{d}\bxi =
  \int_{\mathbb{R}^3} \xi_1 f(L/2, \bxi, t) \,\mathrm{d}\bxi = 0.
\end{displaymath}
Such boundary conditions are known as diffuse reflection, where the
distribution function of ingoing particles is given by a Maxwellian, whose
center and variance are given by the velocity and temperature of the wall. For
moment equations, the implementation of such boundary conditions follows the
method introduced in \cite{Cai2018}, and its general idea is sketched in
Appendix \ref{sec:diff_bc}.

In our test, we choose $L = 5$ so that the effective Knudsen number is $0.2$
since we use the mean free path as the unit length. When
$\theta_{\mathrm{left}}$ and $\theta_{\mathrm{right}}$ are close, the problem
can be well approximated by linearizing the equations about a global
Maxwellian. When linearized, the ansatz of HMBMM \eqref{eq:new_ansatz} becomes
identical to Grad's moment methods \eqref{eq:ansatz}, and the results for
Grad's moment method in this linearized setting have been reported in some
previous works (e.g. \cite{Torrilhon2015}). Therefore, here we choose a
relatively large temperature ratio by setting
\begin{displaymath}
  \theta_{\mathrm{left}} = 1, \qquad \theta_{\mathrm{right}} = 5.
\end{displaymath}
According to \cite{Cai2020}, the steady state of this problem has a temperature
ratio exceeding $2.0$, so that Grad's moment method is expected to diverge. Our
simulation starts with the initial condition
\begin{displaymath}
  f(x,\bxi,0) = \frac{1}{\sqrt{5\pi}} \exp \left( -\frac{|\bxi|^2}{5} \right),
    \qquad \forall x \in (-1,1),
\end{displaymath}
and we evolve the solution until the steady state is reached. A uniform grid
with $200$ cells is used for the spatial discretization.

\begin{figure}[!ht]
\centering
\subfloat[$\rho$ and $\theta$]{%
  \includegraphics[height=.35\textwidth]{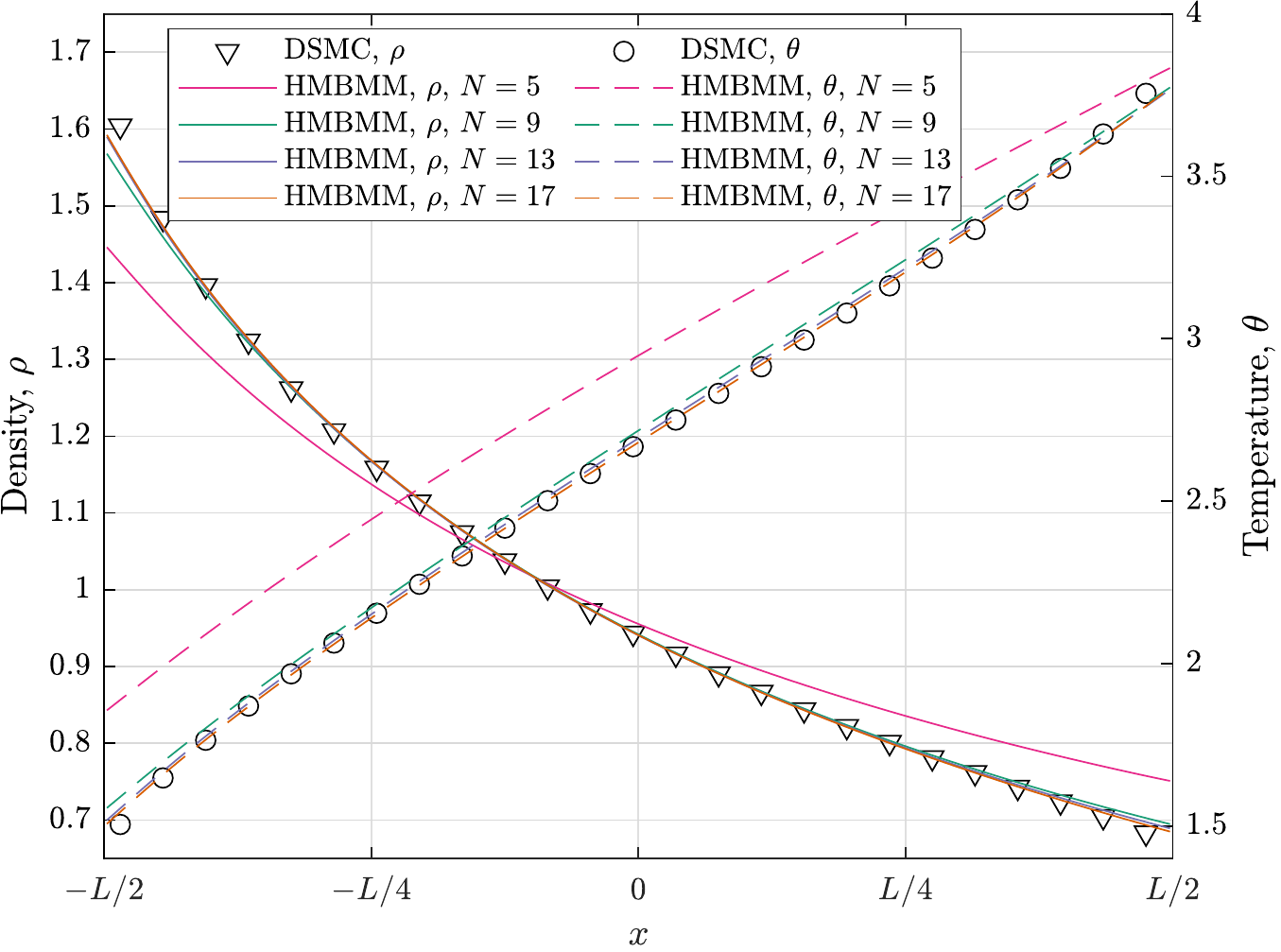}}
\quad
\subfloat[$\sigma_{xx}$ and $q_x$]{%
  \includegraphics[height=.35\textwidth,clip]{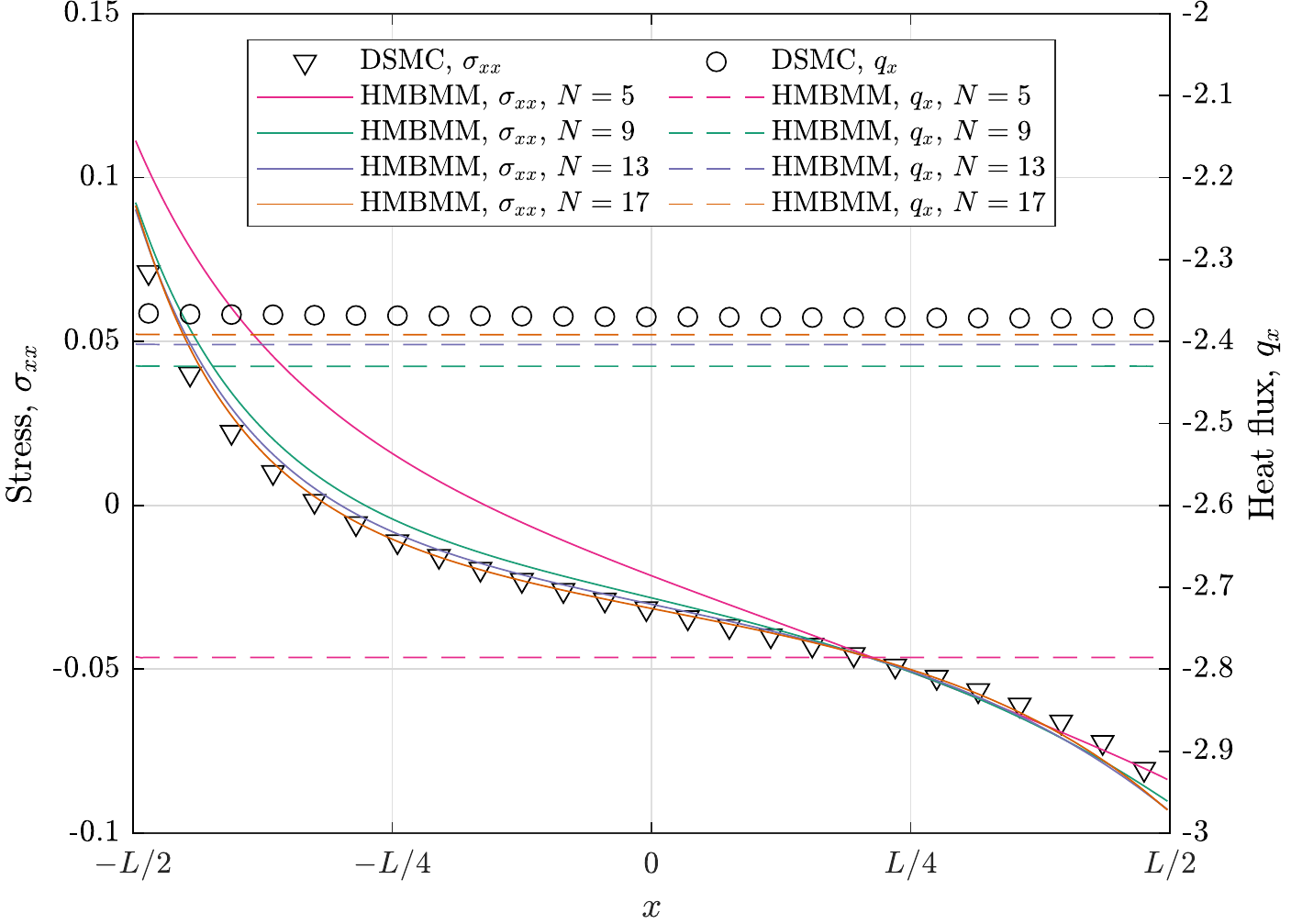}}
\caption{Numerical results for the Fourier flow}
\label{fig:ex3}
\end{figure}

Some numerical results with comparisons to the DSMC results are given in Figure
\ref{fig:ex3}. The temperature profiles confirm that the example involves
temperature ratio greater than $2.0$, so that Grad's moment method would
fail to converge. In fact, as reported in \cite{Cai2020}, the steady-state
solution of Grad's moment method cannot be found in most cases. As for our
method, due to the wall boundary conditions, the distribution functions on both
ends of the domain are expected to be discontinuous, and thus we have used
larger values of $N$ ($5$, $9$, $13$ and $17$) to show the convergence. It can
be seen that under such a Knudsen number, the moment system with $N = 9$ can
already provide quite satisfactory results, and the curves for $N = 13$ and $N =
17$ are almost identical for equilibrium variables $\rho$ and $\theta$, while
some tiny differences can still be observed in non-equilibrium quantities.

\begin{figure}[!ht]
  \centering
  \includegraphics[width=.45\textwidth]{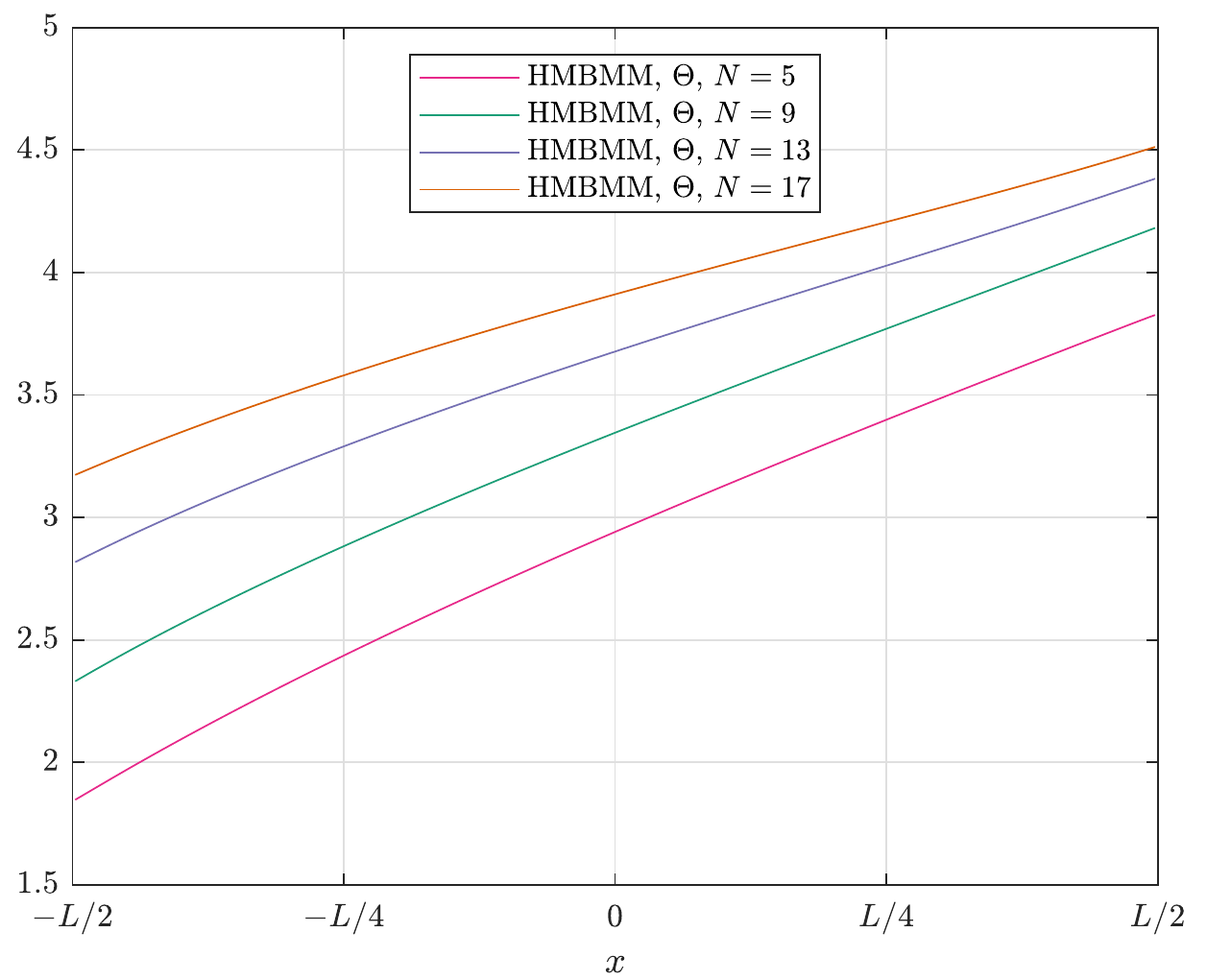}
  \caption{Parameter $\Theta$ for the Fourier flow}
  \label{fig:ex3_Theta}
\end{figure}

The values of $\Theta$ are shown in Figure \ref{fig:ex3_Theta}. According to
the boundary conditions, it is expected that when $N$ approaches infinity, the
value of $\Theta$ should tend to $\max(\theta_{\mathrm{left}},
\theta_{\mathrm{right}}) = 5.0$ everywhere. The reason is that the fat tail
provided by the solid walls is expected to spread all over the domain, as is
proven in \cite{Cai2020} for the BGK collision model. Such a tendancy is
observed in Figure \ref{fig:ex3_Theta}, where we can see an obvious increase of
$\Theta$ on all the points in the domain when $N$ increases. When the minimum
value of $\Theta$ exceeds $2.5$, we can expect that the convergence of the
expansion \eqref{eq:3d_ansatz} can be achieved for all the distribution
functions.

In most cases, for problems with wall boundary conditions, moment methods
perform well mainly for the situation with relatively small Knudsen numbers due
to the slow convergence rate of spectral expansions for discontinuous
functions. We therefore suggest the use of this approach for effective Knudsen
numbers below $0.5$, so that the region with strong discontinuity is
restricted.

\subsection{Hypersonic flow past a flat plate}
\begin{figure}[!ht]
  \centering
  \includegraphics[width=.5\textwidth]{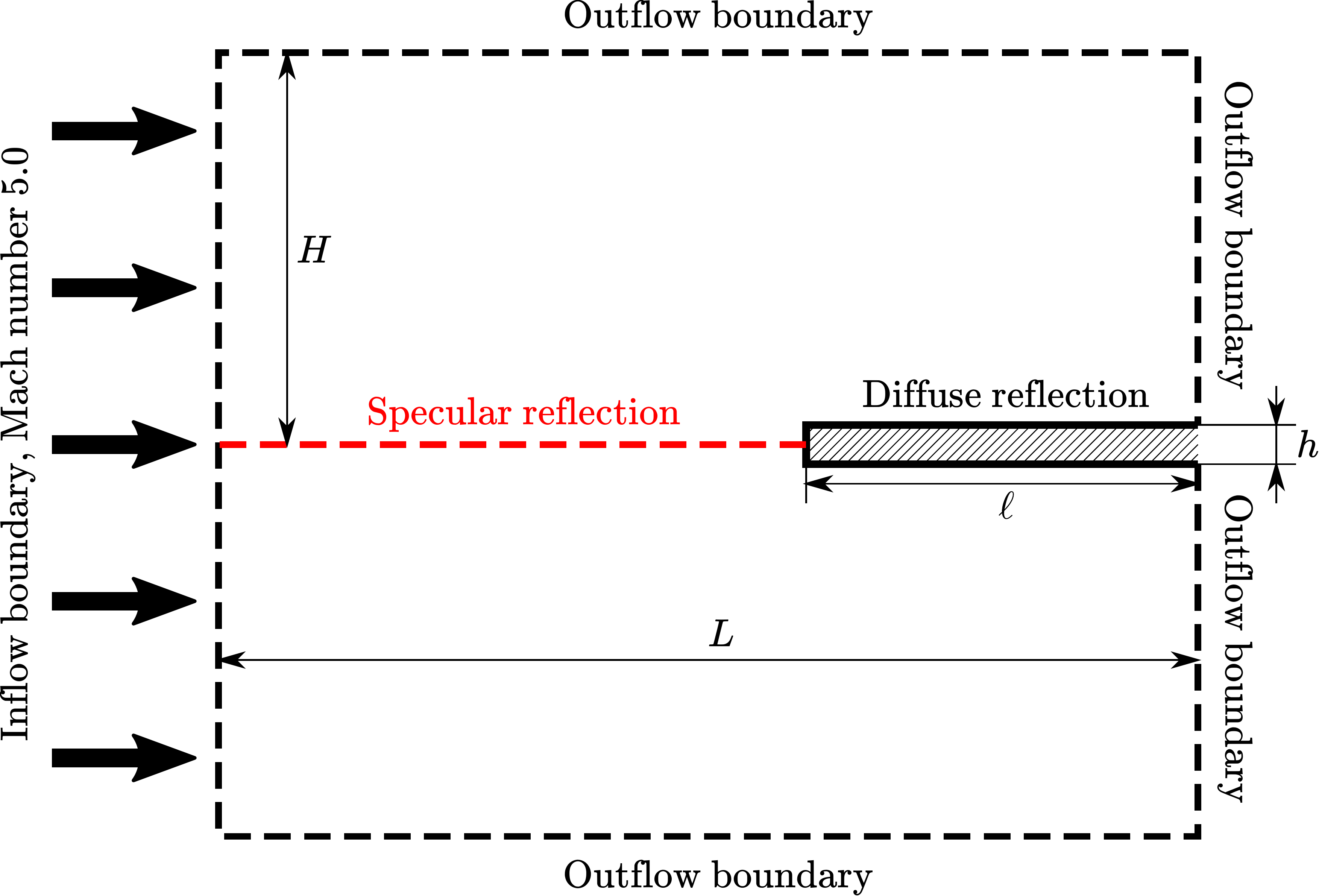}
  \caption{Settings of the two-dimensional shock problem}
  \label{fig:2d_shock}
\end{figure}

In this example, we test the performance of HMBMM in a two-dimensional shock
problem. We consider the case where the incoming flow with Mach number $5.0$
hits a flat plate, generating a bow shock in front of the plate. The setting of
the problem is illustrated in Figure \ref{fig:2d_shock}, where we assume that
the incoming flow is parallel to the plate, so that we can impose the specular
reflection boundary condition as illustrated in Figure \ref{fig:2d_shock}, and
thus only half of the domain needs to be simulated. In our simulation, we set
the thickness of the plate to be $h = 1$, and the size of the computational
domain is given by $L = 25$ and $H = 20$. The length of the plate inside the
computational domain is given by $\ell = 10$. The incoming flow has density
$1.0$ and temperature $1.0$, and inflow and outflow boundary conditions are
imposed on the outer boundaries of the computational domain as indicated in
Figure \ref{fig:2d_shock}. The surface of the plate is assumed to be diffusive
with temperature $\theta_{\mathrm{wall}} = 1$. We refer the readers to Section
\ref{sec:Fourier} for the description of the diffuse reflection boundary
conditions.

\begin{figure}
  \centering
  \subfloat[$N = 5$]{%
    \includegraphics[width=.95\textwidth]{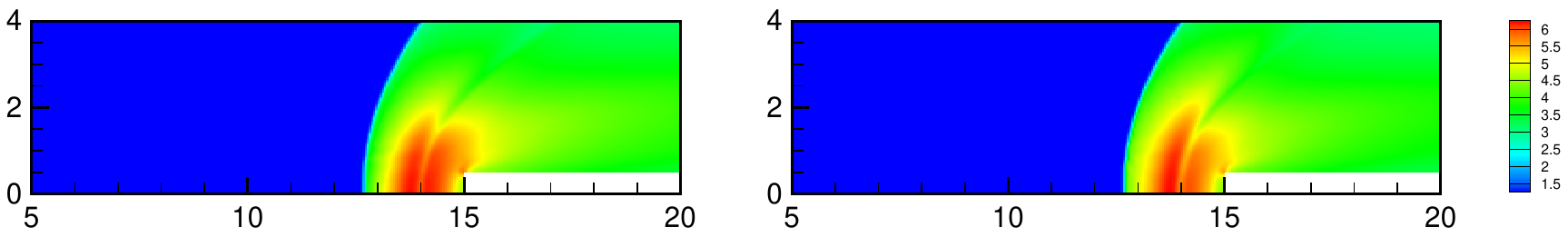}} \\
  \subfloat[$N = 7$]{%
    \includegraphics[width=.95\textwidth]{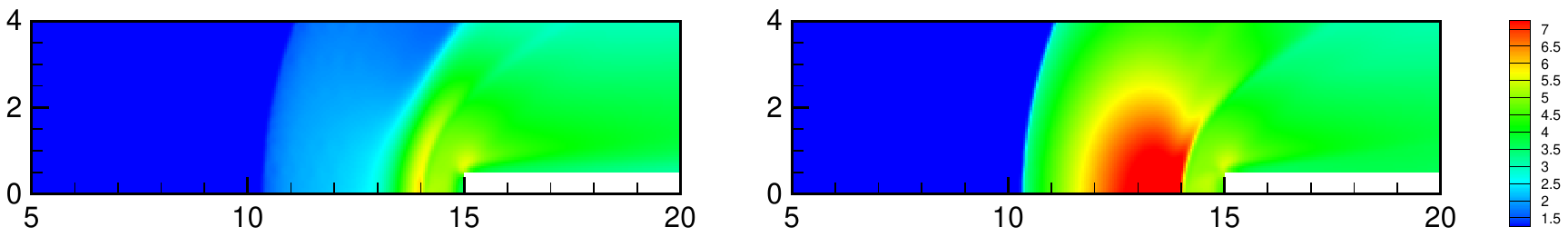}} \\
  \subfloat[$N = 9$]{%
    \includegraphics[width=.95\textwidth]{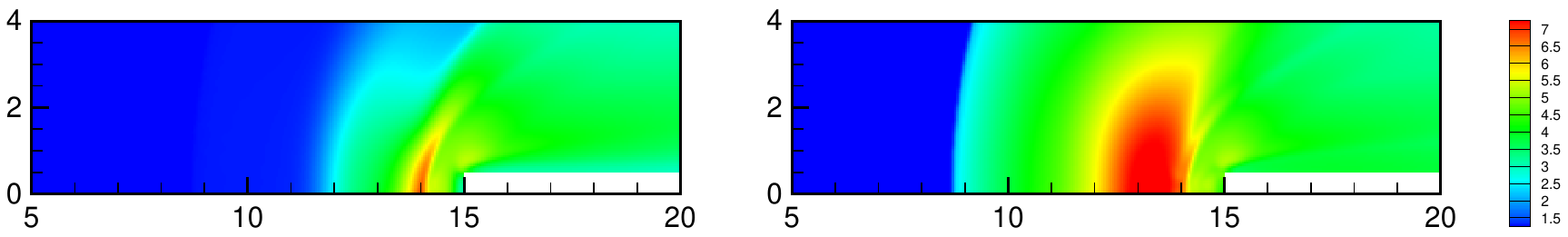}} \\
  \subfloat[$N = 11$]{%
    \includegraphics[width=.95\textwidth]{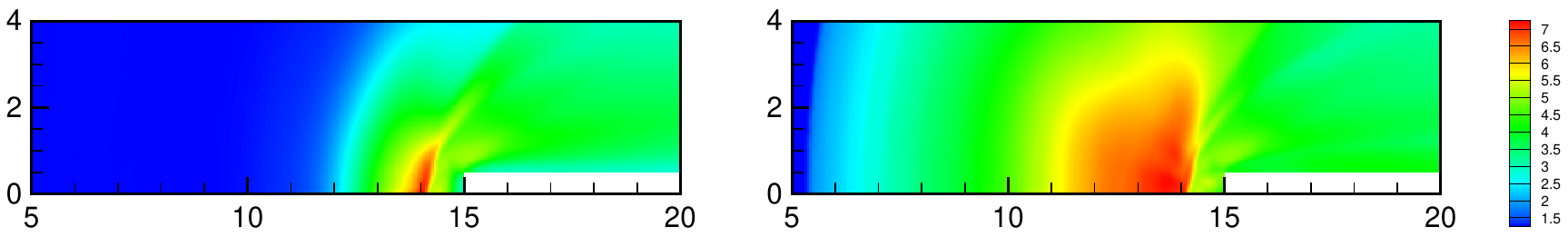}}
  \caption{Numerical solution for the two-dimensional shock problem. Left
    panel: $\theta$; Right panel: $\Theta$.}
  \label{fig:2d_shock_sol}
\end{figure}

In our computation, a uniform grid with cell size $\Delta x = \Delta y = 1/20$
is chosen for the spatial discretization. The numerical solutions for $\theta$
and $\Theta$ are shown in Figure \ref{fig:2d_shock_sol} for $N = 5,7,9,11$,
where only part of the computation domain with interesting flow structure is
plotted. In general, in spite of the large temperature ratio, the trend of
convergence for the temperature plots can be observed, although we expect a
larger $N$ to be used to get accurate flow states for such a large Knudsen
number. Here we would like to draw the readers' attention to the locations of
the subshocks. When $N$ increases, the right panel of Figure
\ref{fig:2d_shock_sol} shows that the subshock gets farther away from the
plate, which agrees with the one-dimensional case. For $N = 5$ and $7$, the
subshock can be clearly observed in the temperature plot, while for $N = 7$,
the temperature jump at the subshock is much weaker. For $N = 9$ and $11$,
although it is known that the subshocks exist, the jumps are already quite
small and hardly observable in the temperature plots.

%% file: article_conclusion.tex
\section{Discussion and conclusion} \label{sec:conclusion}
After witnessing that the shock structure problem disproves a number of moment
hierarchies in the gas kinetic theory, we have built a variation of Grad's
moment method to face its challenge. The novel HMBMM takes into account the
tail of the distribution function, as fixes the convergence issue of Grad's
moment method. Note that although HMBMM is not specially designed for the shock
structure problem, our numerical tests show its capability in predicting the
structures of high-speed shock waves. The HMBMM hierarchy constructs a more
reliable bridge between hydrodynamic models and the kinetic equation, which may 
broaden the usage of the moment methods.

By the above numerical examples, we conclude that the HMBMM is mainly suitable
for high-speed early transitional flows. In this flow regime, Grad's method may
fail due to its convergence problem, and the discrete velocity method (or
Grad's method with linearized ansatz) may require a large number of degrees of
freedom to describe the distribution function. In this case, HMBMM has a chance
to provide reasonable results using a relatively small number of moments. This
is particularly useful when we need to use quadratic collision operator to
capture the flow structure, since the computational cost of the quadratic
collision operator usually increases at least quadratically with respect to the
number of degrees of freedom. For highly nonequilibrium flows where a large
number of degrees of freedom are needed to discretize the distribution
function, the Fourier spectral method \cite{Pareschi1996, Bobylev1997,
Gamba2009, Mouhot2006, Filbet2015, Hu2016} may be more efficient since the time
complexity of such methods is significantly lower. This will happen
when the gas is close to the regime of free molecules or when the shock is very
strong.

Besides, for boundary value problems, due to the discontinuity in the
distribution functions, the moment methods may not be able to well capture the
structure of the boundary layers, which also restricts the use of HMBMM.
Nevertheless, as a simple extension/stabilization of Grad's moment method, it
indicates that the nonlinear moment method without sophisticated closure still
has a chance to work as a numerical solver of the Boltzmann equation, and we
hope the idea of HMBMM can shed some light on the further development of the
moment method. More works on the convergence theories of the moment methods,
especially for boundary value problems, are currently going on.

%% file: article_appendix.tex
\appendix

\section{Definition of the polynomial $p_{lmn}$} \label{sec:poly}
In the ansatz of the three-dimensional distribution functions
\eqref{eq:3d_ansatz}, the polynomials $p_{lmn}$ are three-dimensional
orthogonal polynomials based on spherical coordinates. The definition is
\begin{equation} \label{eq:p_lmn}
p_{lmn}(\bc) = \sqrt{\frac{2^{1-l} \pi^{3/2} n!}{\Gamma(n+l+3/2)}}
  L_n^{(l+1/2)} \left( \frac{|\bc|^2}{2} \right) |\bc|^l
  Y_l^m \left( \frac{\bc}{|\bc|} \right),
\qquad l,n = 0,1,\cdots,\quad m = -l,\cdots,l,
\end{equation}
where $L_n^{(\alpha)}(\cdot)$ is the Laguerre polynomial
\begin{equation}
L_n^{(\alpha)}(x) = \frac{x^{-\alpha} \exp(x)}{n!}
  \frac{\mathrm{d}^n}{\mathrm{d}x^n} [x^{n+\alpha} \exp(-x)],
\end{equation}
and $Y_l^m(\cdot)$ is the spherical harmonics defined for $\boldsymbol{n} =
(\sin \vartheta \cos \varphi, \sin \vartheta \sin \varphi, \cos \vartheta)^T$:
\begin{equation}
Y_l^m(\boldsymbol{n}) =
  \sqrt{\frac{2l+1}{4\pi} \frac{(l-m)!}{(l+m)!}}
  P_l^m(\cos \vartheta) \exp(\mathrm{i} m \varphi),
\end{equation}
with $P_l^m(\cdot)$ being the associated Legendre functions:
\begin{displaymath}
P_l^m(x) = \frac{(-1)^m}{2^l l!} (1-x^2)^{m/2}
  \frac{\mathrm{d}^{l+m}}{\mathrm{d}x^{l+m}} [(x^2-1)^l].
\end{displaymath}
The polynomials $p_{lmn}(\cdot)$ satisfy the following orthogonality:
\begin{displaymath}
\int_{\mathbb{R}^3} p_{lmn}(\bc) p_{l'm'n'}(\bc) \cdot
  \frac{1}{(2\pi)^{3/2}} \exp \left( -\frac{|\bc|^2}{2} \right)
\,\mathrm{d}\bc = \delta_{ll'} \delta_{mm'} \delta_{nn'}.
\end{displaymath}

Note that $p_{lmn}(\cdot)$ is a complex-valued function satisfying
\begin{displaymath}
p_{lmn}(\bc) = (-1)^m \overline{p_{l,-m,n}(\bc)}.
\end{displaymath}
Therefore in order that the distribution function \eqref{eq:3d_ansatz} is real,
the coefficients $f_{lmn}$ must satisfy
\begin{displaymath}
f_{lmn} = (-1)^m \overline{f_{l,-m,n}}.
\end{displaymath}
This means that we only need to store about half of the coefficients appearing
in \eqref{eq:3d_ansatz} during the computation, but all the coefficients
(except the ones with $m = 0$) should be treated as complex numbers. One option
to avoid complex numbers is to use real spherical harmonics defined using the
real and imaginary parts of the complex spherical harmonics. However, such
transformation will significantly complicate the calculation when deriving
moment equations. Here we preserve the complex form as in some previous works
of the author \cite{Cai2015, Cai2019}, as is convenient for both derivation and
numerical implementation.

\section{Definition of HMBMM equations} \label{sec:STA}
Here we provide supplementary clarifications of the three-dimensional equations
\eqref{eq:3d_system}. We will first define $S_{lmn}$ and $T_{lmn}$ on the
left-hand side. The definition of $S_{lmn}$ is
\begin{equation}
  \label{eq:Slmn}
  \begin{split}
  S_{lmn}& = - \sqrt{n(n+l+1/2)}\frac{\partial \Theta}{\partial t}f_{l,m,n-1} \\
  & +\sqrt{2} \sum_{\mu = -1}^{1}\frac{\partial V_{\mu}}{\partial t}\left[
    (-1)^{\mu}\sqrt{n + l + 1/2}\gamma_{l,
      m+\mu}^{-\mu}f_{l-1,m+\mu,n} - \sqrt{n}\gamma_{-l-1,
      m+\mu}^{-\mu}f_{l+1,m+\mu,n-1} \right],
  \end{split}
\end{equation}
where $\gamma_{lm}^{\mu}$ and $V_{\mu}$ are defined in \eqref{eq:gamma}. The
definition of $T_{lmn}$ requires to introduce the following complex
differential operators:
\begin{displaymath}
\frac{\partial}{\partial X_{-1}} =
  \frac{\partial}{\partial x_1} + \mathrm{i} \frac{\partial}{\partial x_2},
\qquad \frac{\partial}{\partial X_0} = \frac{\partial}{\partial x_3},
\qquad \frac{\partial}{\partial X_1} =
  -\frac{\partial}{\partial x_1} + \mathrm{i} \frac{\partial}{\partial x_2},
\end{displaymath}
which can be used to define
\begin{displaymath}
  \begin{split}
  F_{lmn\mu} & = \frac{\partial f_{lmn}}{\partial X_{\mu}}
    + \sqrt{2} \sum_{\nu = -1}^{1}\frac{\partial V_{\nu}}{\partial X_{\mu}}\left[
    (-1)^{\nu}\sqrt{n + l + 1/2}\gamma_{l,m+\nu}^{-\nu}f_{l-1,m+\nu,n}
    - \sqrt{n}\gamma_{-l-1,m+\nu}^{-\nu}f_{l+1,m+\nu,n-1} \right] \\
  & - \sqrt{n(n+l+1/2)}\frac{\mathrm{d} \Theta}{\partial X_{\mu}}f_{l,m,n-1},
    \qquad l = 0,1,\cdots,L-1, \quad m = -l,\cdots,l, \quad n = 0,1,\cdots,N_l-1.
  \end{split}
\end{displaymath}
If the indices of $F_{lmn\mu}$ are not in the range specified in the above
equation, we regard its value as zero. Now we are ready to write down the
formulas for $T_{lmn}$:
\begin{equation}
  \label{eq:cove_coe}
  \begin{aligned}
    & T_{lmn} = \sum_{\mu=-1}^{1} \bigg(
      \frac{1}{2^{|\mu|}} \Big[ \sqrt{2(n+l) + 1}
        \gamma_{l,m-\mu}^{\mu} \Theta F_{l-1,m-\mu,n,\mu}  
        -\sqrt{2(n+1)} \gamma_{l,m-\mu}^{\mu} F_{l-1,m-\mu,n+1,\mu} \\
    & \qquad + (-1)^{\mu}\gamma_{-l-1,m-\mu}^{\mu}
      (\sqrt{2(n+l)+3}F_{l+1,m-\mu,n,\mu} -\sqrt{2n}\Theta
      F_{l+1,m-\mu,n-1,\mu}) \Big] \bigg).
  \end{aligned}
\end{equation}
The derivation of the above equations can be found in \cite{Cai2018}.

The right-hand side of the equation \eqref{eq:3d_system} is determined by the
collision kernel $B(\cdot,\cdot)$ in \eqref{eq:quadratic}. For
inverse-power-law intermolecular potentials, the coefficients $A_{lmn}^{l_1 m_1
n_1, l_2 m_2 n_2}$ are all constants that can be precomputed, and $\mu = \mu_0
\Theta^{\omega}$ with $\omega$ being a constant between $1/2$ and $1$. The
computation of the coefficients $A_{lmn}^{l_1 m_1 n_1, l_2 m_2 n_2}$ can be
found in \cite{Cai2019}. When simulating the shock structure, we choose to
scale the coefficients such that
\begin{displaymath}
A_{200}^{200,000} + A_{200}^{000,200} = -1,
\end{displaymath}
and we set
\begin{displaymath}
\mu_0 = \sqrt{\frac{\pi}{2}} \frac{15}{(5-2\omega)(7-2\omega)},
\end{displaymath}
so that the mean free path of the particles in the equilibrium state in front
of the shock wave equals $1$.

\section{Diffuse reflection boundary condtions} \label{sec:diff_bc}
In this appendix, we provide the sketch of the derivation of the diffuse
reflection boundary conditions for the moment system. The details of the
derivation is nearly identical to the procedure described in \cite{Cai2018},
except that the scaling factor $\theta$ (called $RT$ in \cite{Cai2018}) needs
to be replaced by $\Theta$ for HMBMM.

To ensure that the number of boundary conditions agrees with the number of
characteristics pointing into the domain, we need that $N_0 \geqslant N_1
\geqslant \cdots \geqslant N_l$ in \eqref{eq:3d_ansatz}. Besides, for
convenience, we assume that the solid wall has zero velocity. Under these
assumptions, the derivation of the boundary conditions obeys the following
rules:
\begin{itemize}
  \item Let $\bn$ be the outer unit normal vector on the boundary point $\bx_B$.
    Then the boundary conditions have the form
    \begin{equation} \label{eq:bc}
      \int_{\bxi \cdot \bn < 0} p(\bxi) f(\bx_B, \bxi, t) \,\mathrm{d}\bxi =
      \int_{\bxi \cdot \bn < 0} p(\bxi)
        \frac{\rho_{\mathrm{wall}}(t)}{(2\pi \theta_{\mathrm{wall}})^{3/2}}
        \exp \left( -\frac{|\bxi|^2}{2 \theta_{\mathrm{wall}}} \right)
      \,\mathrm{d}\bxi,
    \end{equation}
    where $\theta_{\mathrm{wall}}$ denotes the temperature of the wall at point
    $x_B$, and $\rho_{\mathrm{wall}}(t)$ is given by
    \begin{equation} \label{eq:rho_wall}
      \rho_{\mathrm{wall}}(t) = \sqrt{\frac{2\pi}{\theta_{\mathrm{wall}}}}
        \int_{\bxi \cdot \bn \geqslant 0} (\bxi \cdot \bn) f(\bx_B, \bxi, t) \,\mathrm{d}\bxi.
    \end{equation}
    In both \eqref{eq:bc} and \eqref{eq:rho_wall}, the distribution function
    $f(\bx_B, \bxi, t)$ is to be replaced by the truncated series
    \eqref{eq:3d_ansatz}, so that equations of the coefficients $f_{lmn}$ can
    be formed by \eqref{eq:bc}.
  \item In \eqref{eq:bc}, the function $p(\bxi)$ must hold the form
    \begin{equation} \label{eq:p}
      p(\bxi) = \sum_{l=0}^{L-1} \sum_{m=-l}^l \sum_{n=0}^{N_l-1}
        C_{lmn} p_{lmn} \left( \frac{\bxi - \bv(\bx_B,t)}{\sqrt{\Theta(\bx_B,t)}} \right)
    \end{equation}
    for some constants $C_{lmn}$. Meanwhile, the polynomial $p(\bxi)$ must satisfy
    \begin{equation} \label{eq:p_sym}
      p(\bxi) = -p(\bxi^*), \qquad \bxi^* = \bxi - 2 (\bxi \cdot \bn) \bn.
    \end{equation}
    This condition means that the polynomial $p$ is symmetric about the wall.
  \item The functions satisfying both \eqref{eq:p} and \eqref{eq:p_sym} form a
    finite dimensional linear space. The full set of boundary conditions at
    $\bx_B$ is given by \eqref{eq:bc} with $p$ chosen as all the bases of this
    finite dimensional space.
\end{itemize}
Clearly $p(\bxi) = \bxi \cdot \bn$ is one of the polynomials satisfying both
\eqref{eq:p} and \eqref{eq:p_sym}. For such a choice of $p$, the boundary
condition \eqref{eq:bc} turns out to be
\begin{displaymath}
  \bv(\bx_B, t) \cdot \bn = 0,
\end{displaymath}
meaning that the normal velocity of the gas is zero on the boundary.